\documentclass[letterpaper,preprint, aps, prd, nofootinbib,showpacs]{revtex4-1}

\usepackage{amsmath, amssymb, amsfonts, graphicx}
\usepackage[subfigure]{graphfig}
\usepackage{subfigure}
\usepackage{epsfig}
\usepackage{epstopdf}

\usepackage[colorinlistoftodos]{todonotes}
\usepackage[colorlinks=true, allcolors=blue]{hyperref}

\usepackage{color}      

\def\ppf{{_\textrm{f}}}
\def\ppi{{_\textrm{i}}}

\usepackage{dcolumn}
\usepackage{bm}
\usepackage{braket}
\usepackage{slashed}
\usepackage{enumitem}
\usepackage{bbm}
\usepackage{float}

\usepackage{cancel}

\def\path{./pdf/}

\begin{document}
\title{Proton momentum and angular momentum decompositions with overlap fermions}

\author{
\texorpdfstring{
Gen Wang$^{1}$,
Yi-Bo Yang$^{2,3,4,5}$,
Jian Liang$^{6,7}$,
Terrence Draper$^{1}$,
and Keh-Fei Liu$^{1}$
{
\vspace*{-0.5cm}
\begin{center}
{
\includegraphics[scale=0.25]{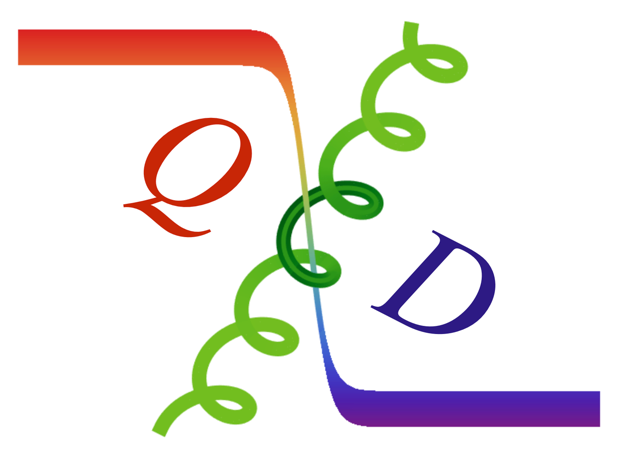}\\
\vspace*{-0.1cm}
($\chi$QCD Collaboration)
}
\end{center}}
}
}

\affiliation{
\scriptsize
$^{1}$\mbox{Department of Physics and Astronomy, University of Kentucky, Lexington, Kentucky 40506, USA} \\
$^{2}$\mbox{International Centre for Theoretical Physics Asia-Pacific, Beijing/Hangzhou, 100019, China}\\
$^{3}$\mbox{CAS Key Laboratory of Theoretical Physics, Institute of Theoretical Physics,}\\ \mbox{Chinese Academy of Sciences, Beijing 100190, China}\\
$^{4}$\mbox{School of Fundamental Physics and Mathematical Sciences, Hangzhou Institute for Advanced Study,}\\ \mbox{ UCAS, Hangzhou 310024, China}\\
$^{5}$\mbox{School of Physical Sciences, University of Chinese Academy of Sciences,
Beijing 100049, China}\\
$^{6}$\mbox{Guangdong Provincial Key Laboratory of Nuclear Science, Institute of Quantum Matter,}\\ \mbox{South China Normal University, Guangzhou 510006, China} \\
$^{7}$\mbox{Guangdong-Hong Kong Joint Laboratory of Quantum Matter, Southern Nuclear Science Computing Center, }\\ \mbox{ South China Normal University, Guangzhou 510006, China} \\
}

\begin{abstract}
We present a calculation of the proton momentum and angular momentum decompositions using overlap fermions on a $2+1$-flavor RBC/UKQCD domain-wall lattice at 0.143 fm with a pion mass of 171 MeV which is close to the physical one.
A complete determination of the momentum and angular momentum fractions carried by up, down, strange and glue inside the proton has been done with valence pion masses varying from 171 to 391 MeV.
We have utilized fast Fourier transform on the stochastic-sandwich method for connected-insertion parts and the cluster-decomposition error reduction technique for disconnected-insertion parts has been used to reduce statistical errors.
The full nonperturbative renormalization and mixing between the quark and glue operators are carried out.
The final results are normalized with the momentum and angular momentum sum rules and reported at the physical valence pion mass at  ${\overline{\rm {MS}}}\, (\mu = 2\ {\rm{GeV}})$.
The renormalized momentum fractions for the quarks and glue are $\langle x \rangle^q = 0.491(20)(23)$ and $\langle x \rangle^g = 0.509(20)(23)$, respectively,
and the renormalized total angular momentum fractions for quarks and glue are $2 J^q = 0.539(22)(44)$ and $2 J^g = 0.461(22)(44)$, respectively.
The quark spin fraction is $\Sigma = 0.405(25)(37)$ from our previous work and the quark orbital angular momentum fraction is deduced from $2 L^q = 2 J^q - \Sigma$ to be $0.134(22)(44)$.
\end{abstract}
\maketitle

{
\section{INTRODUCTION}\label{sec_intro}
A quantitative understanding of the proton spin in terms of its fundamental quark and gluon constituents is an important and challenging question of hadron physics.
Experiments using polarized deep inelastic lepton-nucleon scattering (DIS) processes show that the total helicity contribution from the quarks is just about $25\%-30\%$~\cite{deFlorian:2008mr,deFlorian:2009vb,Blumlein:2010rn,Leader:2010rb,Ball:2013lla,Nocera:2014gqa,Ethier:2017zbq,Deur:2018roz} of the proton spin.
The recent analyses~\cite{deFlorian:2014yva, Nocera:2014gqa} of the high-statistics 2009 STAR~\cite{Djawotho:2013pga} and PHENIX~\cite{Adare:2014hsq} experiments at the Relativistic Heavy Ion Collider (RHIC) showed evidence of nonzero glue helicity in the proton.

Lattice QCD provides the {\it ab initio\/} nonperturbative framework to calculate the spin and momentum moments of quarks and gluons constituents inside the proton directly from the QCD action.
The intrinsic spin carried by each quark flavor was first studied by $\chi$QCD~\cite{Dong:1995rx} in the quenched approximation.
Followup calculations with dynamical fermions, were carried out on multiple lattice spacings and pion masses by $\chi$QCD~\cite{Liang:2018pis}, Extended Twisted Mass Collaboration (ETMC)~\cite{Alexandrou:2017hac,Alexandrou:2017oeh}, and PNDME~\cite{Lin:2018obj},
and they have provided results consistent with experiment with $\Delta u = 0.777(25)(30)$, $\Delta d = - 0.438(18)(30)$ and $\Delta s = - 0.053(8)$ as averaged by the Flavour Lattice Averaging Group (FLAG)~\cite{Aoki:2019cca}.
It is worth noting that the current prediction of $\Delta s$ from lattice QCD is more precise than the phenomenological determinations.
The gluon spin was determined in Ref.~\cite{Yang:2016plb} to be 0.251(47)(16) at the physical pion mass in the ${\overline{\rm {MS}}}$ scheme at $\mu^2 = 10 \ {\rm{ GeV}^2}$.
We further note that the anomalous Ward identity (AWI) was explicitly verified with the overlap fermion which has chiral symmetry~\cite{Liang:2018pis}.
The smallness of the quark spin contribution to the proton spin had been the major source of mystification in the ``proton spin crisis.''
It is now understood that it is due to the fact that the disconnected insertion is unexpectedly large and negative which reduces the positive contribution from the connected insertion~\cite{Liang:2018pis}.
In order to address the angular momentum fractions, a first attempt to fully decompose the proton spin was carried out by the $\chi$QCD collaboration in 2013~\cite{Deka:2013zha} in the quenched approximation and a lot of progress has been made with dynamical fermions for $N_f = 2$~\cite{Alexandrou:2017oeh}, $N_f = 2+1+1$~\cite{Alexandrou:2020sml}, and one preliminary $N_f = 2+1$ result with complete nonperturbative renormalization, mixing and normalization~\cite{Yang:2019dha}.

In this paper, we use the nucleon matrix element of the Belinfante energy-momentum tensor (EMT) to determine the momentum and angular momentum fractions of the up, down, strange, and glue constituents of the nucleon.
The quark orbital angular momentum can be obtained by subtracting the spin component from the total quark angular momentum.
Overlap fermions are used on a $32^3 \times 64$ $2+1$-flavor domain-wall fermion lattice at 0.143 fm with a pion mass of 171 MeV which is close to the physical one.
With a multimass inverter, we are able to simulate several valence pion masses and extrapolate the results to the physical pion mass.
We have utilized fast Fourier transform (FFT) on the stochastic-sandwich method for connected-insertion parts which enabled us to simulate {\cal O}(100) combinations of the initial and final nucleon momenta in the three-point function (3pt) contraction.
With the use of the cluster-decomposition error reduction (CDER) technique~\cite{Liu:2017man}, the statistical errors for all disconnected-insertion parts are greatly reduced.
Since the EMT of each parton species is not separately conserved, we summarize the final momentum and angular momentum fractions by considering nonperturbative renormalization and mixing at ${\overline{\rm {MS}}}(\mu = 2\ {\rm{GeV}})$ and use the momentum and angular momentum conservation sum rules to normalize them.
The numerical approach of this work is based on Ref.~\cite{Wang:2020yqv} with major improvements on the disconnected-insertion parts.

The remaining sections of the paper are organized as follows:
The basic formalism is provided in Sec.~\ref{sec:operator_qg}.
In Sec.~\ref{sec:numerical}, we present the numerical details, such as
details of momentum projection on grid source and FFT on stochastic-sandwich method in Sec.~\ref{sec3:sec_CI_num};
discussions of CDER fits and systematic estimations are in Sec.~\ref{sec3:sec_DI_num};
and a short description of the $z$-expansion fit is in Sec.~\ref{sec3:sec_zexp}.
The details of the 3pt fits and our final results are presented in Sec.~\ref{sec:ana}.
A brief summary is given in Sec.~\ref{sec:summary}.

}

{
\section{Basic formalism}\label{sec:operator_qg}
The nucleon matrix element of the Belinfante EMT can be delineated by four gravitational form factors (GFFs)~\cite{Kobzarev:1962wt,Pagels:1966zza,Ji:1996ek,Ji:1996nm} based on their associated spinor structures as
\begin{eqnarray}\label{eq:sec4_Tmunu}
\begin{aligned}
&\bra{p',s'} \mathcal{T}^{\{\mu \nu\}q,g} \ket{p,s} = \frac{1}{2} \bar{u}(p',s')  \Big{[} T_1(q^2) (\gamma^\mu \bar{p}^\nu + \gamma^\nu \bar{p}^\mu)  \\
&\quad \quad +  \frac{1}{2 m} T_2(q^2) \left( i q_\alpha (\bar{p}^\mu \sigma^{\nu \alpha} + \bar{p}^\nu \sigma^{\mu \alpha}) \right)
+ D(q^2) \frac{q^\mu q^\nu - \eta^{\mu \nu} q^2}{M} 
+ \bar{C}(q^2) M \eta^{\mu \nu}
 \Big{]} ^{q,g} u(p,s),
\end{aligned}
\end{eqnarray}
where $\ket{p,s}$ is the nucleon initial state with momentum $p$ and spin $s$;
$\bra{p',s'}$ is the nucleon final state with momentum $p'$ and spin $s$;
$\bar{u}$ and $u$ are the final and initial nucleon spinors;
$q = p'-p$ is the momentum transfer;
$\bar{p} = (p'+p)/2$ is the averaged initial and final momentum;
and $T_1$, $T_2$, $D$, and ${\bar{C}}$ are the four gravitational form factors.
At the $q^2 \rightarrow 0$ limit, one obtains~\cite{Ji:1996ek} 
\begin{equation}
J^{q,g} = \frac{1}{2} \left[ T_1(0) + T_2(0) \right]^{q,g}, \; \braket{x}^{q,g} = T_1(0)^{q,g},
\end{equation}
in which $J^{q,g}$ is the the total angular momentum fraction for quarks and glue, respectively, and $\braket{x}^{q,g}$ is the second moment of the PDF and is the momentum fraction of the quarks and glue.
We will focus on $\mathcal{T}_{4i}$ which is sufficient for the  evaluation of $T_1(0)$ and $T_1(0)+T_2(0)$.
Following from the conservation of EMT, the momentum and angular momentum are conserved with sum rules
\begin{eqnarray}\label{eq:proton_sum_rules}
\begin{aligned}
&\braket{x}^{q} + \braket{x}^{g} = T_1(0)^q + T_1(0)^g = 1, \\
J^{q} + J^{g} &= \frac{1}{2} \{ [ T_1(0)^q + T_2(0)^q ] + [T_1(0) + T_2(0)]^g \} = \frac{1}{2}. \\
\end{aligned}
\end{eqnarray}
One implication of these two sum rules is that the sum of the $T_2(0)$'s for the quarks and gluons is zero~\cite{Ji:1997pf}, that is,
\begin{eqnarray}
T_2(0)^q + T_2(0)^g = 0.
\end{eqnarray}
The vanishing of total $T_2(0)$, the anomalous gravitomagnetic moment, in the context of a spin-1/2 particle was first derived classically from the post-Newtonian manifestation of the equivalence principle~\cite{Kobzarev:1962wt}. More recently, this has been proven~\cite{Brodsky:2000ii} for composite systems from the light-cone Fock space representation.
$\bar{C}(0)$ is equal to the spatial diagonal part of the stress EMT and is the pressure of the system~\cite{Lorce:2017xzd,Liu:2021gco}.
Due to the conservation of EMT, the total pressure is zero, i.e., $\bar{C}^q(0) + \bar{C}^g(0) = 0$.
This has implication on the confinement from the trace anomaly of the EMT~\cite{Liu:2021gco}.

{
\subsection{Quark and gluon operators}
The Euclidean quark EMT component $\mathcal{T}_{4i}^{q(E)}$ can be written as 
\begin{eqnarray}\label{eq:sec4_quark_E}
\begin{aligned}
\mathcal{T}_{4i}^{q(E)} &= (-1)\frac{i}{4} \sum_f \bar{\psi}_f \left[
      \gamma_4 \overrightarrow{D}_i + \gamma_i \overrightarrow{D}_4
    - \gamma_4  \overleftarrow{D}_i - \gamma_i \overleftarrow{D}_4 \right] \psi_f. \\
\end{aligned}
\end{eqnarray}
The left and right gauge covariant derivatives on the lattice are
\begin{eqnarray}
\begin{aligned}
\overrightarrow{D}_\mu \psi(x) &= \frac{1}{2 a} \left[ U_\mu(x) \psi(x+a_\mu) - U_\mu^\dagger(x - a_\mu) \psi(x-a_\mu) \right] ,\\
\bar{\psi}(x) \overleftarrow{D}_\mu &= \frac{1}{2 a} \left[ \bar{\psi}(x+a_\mu) U_\mu^\dagger(x) - \bar{\psi}(x-a_\mu) U_\mu^\dagger(x - a_\mu) \right] ,\\
\end{aligned}
\end{eqnarray}
with each $\psi$ being a quark field operator on the lattice and each $U$ a gauge link. 
The Euclidean gluon EMT component $\mathcal{T}_{4i}^{g(E)}$ is
\begin{eqnarray}\label{eq:sec4_glue_E}
\begin{aligned}
\mathcal{T}_{4i}^{g(E)} &= (+i) \left[ -\frac{1}{2} \sum_{k=1}^3 2 {\rm{Tr}}^{\rm{color}} [G_{4k} G_{ki} + G_{ik} G_{k4} ] \right], \\
\end{aligned}
\end{eqnarray}
in which $G_{\mu \nu}$ is the Euclidean field-strength tensor
\begin{equation}\label{eq:sec5_glue_field}
G_{\mu \nu}^{(E)}(x) = \frac{1}{8} \left( P_{\mu \nu} (x) - P_{\mu \nu}^\dagger(x) \right),
\end{equation}
with $P_{\mu \nu}$ being the ``cloverleaf" link operator
\begin{eqnarray}
\begin{aligned}
P_{\mu \nu}&= U_\mu(x) U_\nu(x+\mu) U_\mu^\dagger (x+\nu) U_\nu^\dagger(x) \\\
&\quad + U_\nu(x) U_\mu^\dagger(x-\mu+\nu) U_\nu^\dagger(x-\mu) U_\mu(x-\mu)  \\
&\quad + U_\mu^\dagger(x-\mu) U_\nu^\dagger(x-\mu-\nu) U_\mu(x-\mu-\nu) U_\nu(x-\nu) \\
&\quad + U_\nu^\dagger(x-\nu) U_\mu(x-\nu) U_\nu(x-\nu+\mu) U_\mu^\dagger(x) \\
\end{aligned}
\end{eqnarray}
which is built from the hypercubic (HYP) smeared gauge links. The difference between the bare matrix elements and the HYP-smeared matrix elements will be compensated by the nonperturbative renormalization procedure~\cite{Yang:2018bft}. More details of our convention of gamma matrices and operators can be found in Ref.~\cite{Deka:2013zha}.

}

{
\subsection{Three-point correlation functions}
The EMT matrix element can be extracted from the 3pt along with the associated two-point correlation function (2pt) as
\begin{eqnarray}\label{eq:sec5_2pt_con}
G_{\alpha \beta}^{NN} (\vec{p},t) =\sum_{\vec{x}} e^{-i \vec{p} \cdot \vec{x}} \bra{0} T [\chi_\alpha (\vec{x},t) \bar{\chi}_\beta(\vec{0},0) ] \ket{0},
\end{eqnarray}
with
$\chi_\alpha(x) = \epsilon_{abc} u(x)^a_\alpha \left[ {u(x)^b} \widetilde{\mathcal{C}} d(x)^c \right]$ the nucleon interpolation field~\cite{Wilcox:1991cq} and
$\mathcal{C} \equiv \gamma_2 \gamma_4$ the charge conjugation operator with $\widetilde{\mathcal{C}} = \mathcal{C} \gamma_5$.
In the $t \gg 1 $ limit, the unpolarized nucleon 2pt $C_{\rm{2pt}}(\vec{p},t)$ is 
\begin{eqnarray}\label{eq:sec5_2pt_fit}
\begin{aligned}
C_{\rm{2pt}}(\vec{p},t) \equiv {\rm{Tr}} [\Gamma_0 G^{NN} (\vec{p},t)] \xrightarrow[]{ t \gg 1  } \frac{Z_p^2 }{(La)^3} \frac{E_p + m}{E_p} e^{-E_{p} (t-t_0)} + A e^{-E_{p}^1 (t-t_0)}, \\
\end{aligned}
\end{eqnarray}
in which $\Gamma_0=P_+ = \frac{1+\gamma_4}{2}$ is the unpolarized projection for the nucleon, $Z_p^2$ is the spectral weight, $m$ is the nucleon rest mass, $E_p$ and $E_p^1$ are the ground-state energy and first excited-state energy, respectively, and $A$ is the spectral weight associated with the excited-state contamination.
The 3pt of EMT is
\begin{eqnarray}\label{eq:sec2_3pt_lat}
\begin{aligned}
G_{\alpha \beta}^{\mathcal{T}_{4 i}^{q,g} }(t\ppf,\tau,\vec{p}\ppf,\vec{p}\ppi) =
    \sum_{\vec{x}\ppf,\vec{z}} e^{-i \vec{p}\ppf \cdot (\vec{x}\ppf - \vec{z})} e^{ i \vec{p}\ppi \cdot \vec{z}} \times 
    \bra{0} T [\chi_\alpha (\vec{x}\ppf,t\ppf) \mathcal{T}_{4i}^{q,g}(\vec{z},\tau) \bar{\chi}_\beta(\vec{0},0) ] \ket{0}, \\
\end{aligned}
\end{eqnarray}
in which $z = \{ \vec{z}, \tau \}$ is the current position, $x\ppf = \{ \vec{x}\ppf, t\ppf \}$ is the sink position, $\vec{p}\ppf$ is the momentum of the final nucleon, $\vec{p}\ppi$ is the momentum of the initial nucleon, and the momentum transfer is $\vec{q} = \vec{p}\ppf - \vec{p}\ppi$. With the unpolarized/polarized projection for the nucleon, we define $C_{\rm{3pt}}$ as
\begin{eqnarray}\label{eq:sec2_3pt_Tr}
C_{\rm{3pt},\Gamma_\alpha}^{4 i}(t\ppf, \tau,\vec{p}\ppf,\vec{p}\ppi) \equiv {\rm{Tr}} [\Gamma_\alpha G^{\mathcal{T}_{4 i}^{q,g}}(t\ppf,\tau,\vec{p}\ppf,\vec{p}\ppi) ],
\end{eqnarray}
with $\alpha \in \{0,1,2,3\}$, $i \in \{ 1, 2, 3\}$, $\Gamma_0 = \frac{1+\gamma_4}{2}$ the unpolarized projection for nucleon and $\Gamma_k = i\Gamma_0 \gamma_5 \gamma_k$ the polarized projections.
In order to extract $T_{i}$, we take the ratios of 3pt and 2pt functions,
\begin{eqnarray}\label{eq:sec5_3pt_ratio}
\begin{aligned}
R^{4 i}_{\Gamma_\alpha}(t\ppf,\tau,\vec{p}\ppf,\vec{p}\ppi) &\equiv \frac{C_{\rm{3pt},\Gamma_\alpha}^{4 i}(t\ppf, \tau,\vec{p}\ppf,\vec{p}\ppi)}{C_{\rm{2pt}}(\vec{p}\ppf,t\ppf)} \times
    \sqrt{\frac{ C_{\rm{2pt}}(\vec{p}\ppi,t\ppf - \tau)  C_{\rm{2pt}}(\vec{p}\ppf,\tau)  C_{\rm{2pt}}(\vec{p}\ppf,t\ppf) }{ C_{\rm{2pt}}(\vec{p}\ppf,t\ppf-\tau)  C_{\rm {2pt}}(\vec{p}\ppi,\tau)  C_{\rm{2pt}}(\vec{p}\ppi ,t\ppf) }} \\
&\xrightarrow[t\ppf-t \gg 1]{ t\ppf \gg 1  } \frac{a_1 T_1(Q^2) + a_2 T_2(Q^2) + a_3 D(Q^2) }{4 \sqrt{E_{p'} (E_{p'} + m) E_{p} (E_{p} + m)}},
\end{aligned}
\end{eqnarray}
where the $a_i$ are known coefficients which depend on the momentum and energy of the nucleon,
and $Q^2 = (p' - p)^2$ is the momentum transfer squared.

In this paper, we focus on the evaluation of the $T_1$ and $[T_1+T_2]$ form factors with $\mathcal{T}_{4i}$ by choosing specific momentum and polarization projection settings.
We set the initial and final momentum of the nucleon to be the same to target the $T_1$ form factor, 
\begin{eqnarray}\label{eq:sec5_case2}
\begin{aligned}
&R^{     4i }_{\Gamma_0}(t\ppf,\tau,\vec{p},\vec{p})   =  p_i T_1 (0) , \\
\end{aligned}
\end{eqnarray}
with $i \in \{1,2,3\}$.
The following settings are used to calculate the $[T_1 + T_2]$ form factor,
\begin{eqnarray}\label{eq:sec5_case0}
\begin{aligned}
R^{4 i}_{\Gamma_j}(t\ppf,t,\vec{p}, \vec{0})  &= \frac{-i}{4}\sqrt{\frac{E_p+m}{2 E_p}} \epsilon_{i,j,k} p_k [T_1 + T_2] (Q^2) , \\
R^{4 i}_{\Gamma_j}(t\ppf,t,\vec{0}, \vec{p})  &= \frac{-i}{4}\sqrt{\frac{E_p+m}{2 E_p}} \epsilon_{i,j,k} p_k [T_1 + T_2] (Q^2) , \\
R^{4 i}_{\Gamma_j}(t\ppf,t,\vec{p},-\vec{p})  &= \frac{-i}{2} \epsilon_{i,j,k} p_k [T_1 + T_2] (Q^2) , \\
\end{aligned}
\end{eqnarray}
in which the first two momentum settings have either the initial or the final momentum equal to $\vec{0}$, while the third case sets the initial and final momentum of the nucleon in opposite directions which results in larger momentum transfers.

\begin{figure}
  \centering
  \includegraphics[scale=0.50]{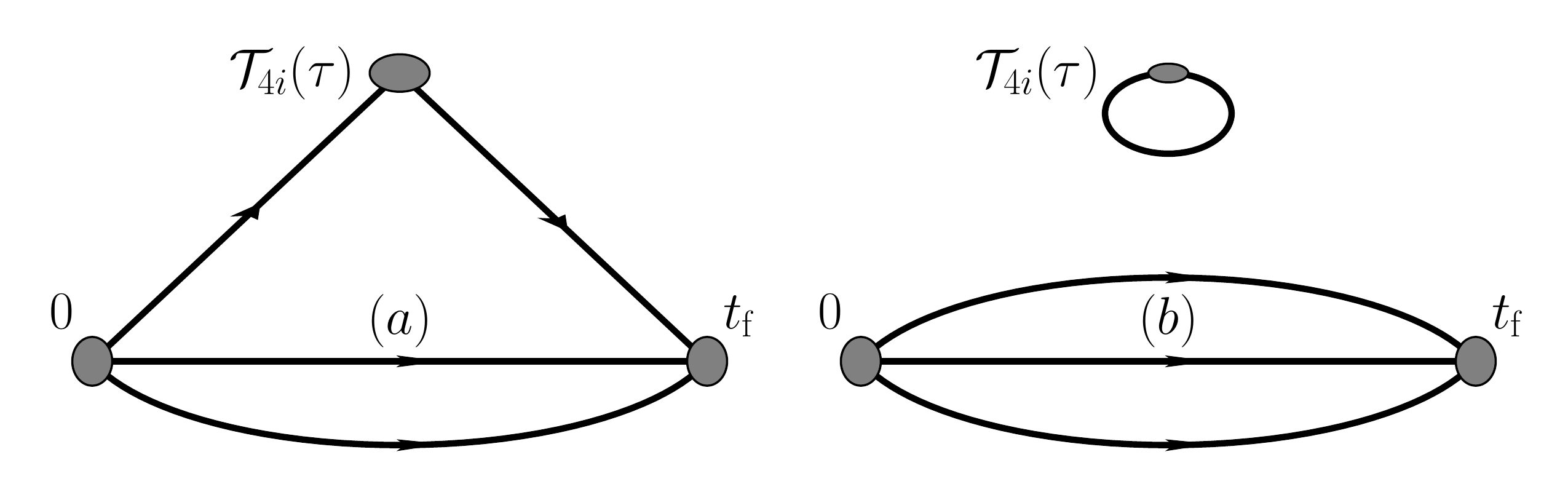}
  \caption{Illustration of the nucleon 3pts with (a) connected insertions (CI) and (b) disconnected insertions (DI).}
  \label{fig:feyn_3pt}
\end{figure}

Using Wick contractions, the evaluation of 3pt in Eq.~(\ref{eq:sec2_3pt_lat}) on the lattice gives two topologically distinct contributions: connected insertions (CI) and disconnected insertions (DI), which are shown in Fig.~\ref{fig:feyn_3pt}.
In the case of CI, the $\psi/\bar{\psi}$ from current $\mathcal{T}_{4i}$ is contracted with the $\bar{\psi}/\psi$ from the source/sink nucleon interpolating field, whereas, in the case of DI, the $\psi/\bar{\psi}$ from the current $\mathcal{T}_{4i}$ is self-contracted at current position $z$ to form a loop.
For the DI case, the gauge-averaged 3pt can be written as
\begin{eqnarray}\label{eq:C3pt_DI_Loop}
\begin{aligned}
&C_{\rm{3pt},{\Gamma_\alpha}}^{4 i}(t\ppf,\tau,\vec{p}\ppf,\vec{p}\ppi)_{\rm{DI}}  = \sum_{\vec{z},\vec{x}\ppf} e^{-i \vec{p}\ppf \cdot \vec{x}\ppf }
  e^{ i \vec{q} \cdot \vec{z} }
    \times \bra{0} {\rm{Tr}}  \left[\Gamma_\alpha \chi (\vec{x}\ppf,t\ppf)  \bar{\chi} (\vec{0},0)   \right]
   \times \left[ \mathcal{T}_{4 i} (\vec{z},\tau) \right] \ket{0} \\
&= \braket{ {\rm{Tr}} [ \Gamma_\alpha G^{NN}(\vec{p}\ppf,t\ppf;U)] \times L^{4 i}[\tau,\vec{q};U]} - \braket{ {\rm{Tr}} [ \Gamma_{\alpha} G^{NN}(\vec{p}\ppf,t\ppf;U)]} \times \braket{ L^{4 i}[\tau,\vec{q};U] } , \\
\end{aligned}
\end{eqnarray}
in which $\braket{\cdots}$ denotes the gauge average and $G^{NN}(\vec{p},t;U)$ is the nucleon propagator under gauge field $U$ and $L^{4i}[\tau,\vec{q};U]$ is the current loop of quark/gluon.
We have subtracted the uncorrelated part of the loop and of the nucleon propagator.
The quark loop $L = L_f^{4 i}[t,\vec{q};U]$ is constructed from the propagator of quark flavor $f$ as
\begin{eqnarray}
\begin{aligned}
L_f^{4 i}[t,\vec{q};U] &= \frac{i}{8 a} \sum_{\vec{z}} e^{i \vec{q} \cdot \vec{z}}  \\
{\rm{Tr}} &\left\{ D^{-1}_{f}(z+a_i,z;U) \gamma_4 U_i(z) -  D^{-1}_{f}(z-a_i,z;U) \gamma_4 U_i^\dagger(z-a_i) \right. \\
 + & D^{-1}_{f}(z,z-a_i;U)   \gamma_4 U_i(z-a_i)     -  D^{-1}_{f}(z,z+a_i;U) \gamma_4 U_i^\dagger(z) \\
 + & D^{-1}_{f}(z + a_4,z;U) \gamma_i U_4(z)   -  D^{-1}_{f}(z-a_4,z;U) \gamma_i U_4^\dagger(z-a_4) \\
 +& \left.  D^{-1}_{f}(z,z-a_4;U) \gamma_i U_4(z-a_4) -  D^{-1}_{f}(z,z+a_4;U) \gamma_i U_4^\dagger(z) \right\} , \\
\end{aligned}
\end{eqnarray}
in which the trace $\rm{Tr}$ is the trace over color and spin, and $D^{-1}_{f}(z+a_\nu,z;U)$ is the quark propagator from point $z$ to point $z+a_\nu$ under gauge field $U$ with flavor $f$.
In the case of the gluon 3pt, only DI contributes as in Eq.~(\ref{eq:C3pt_DI_Loop}) with the current loop $L = L_g^{4 i}[t,\vec{q};U]$ as
\begin{eqnarray}
\begin{aligned}
L_g^{4 i}[t,\vec{q};U] &= (-i) \sum_{\vec{z}} e^{i \vec{q} \cdot \vec{z}}  \times 
 \left[  \sum_{k=1}^3 {\rm{Tr}}^{\rm{color}} [G_{4k}(z) G_{ki}(z) + G_{ik}(z) G_{k4}(z) ] \right],
\end{aligned}
\end{eqnarray}
with the field-strength tensor $G_{\mu \nu}$ defined in Eq.~(\ref{eq:sec5_glue_field}).
}

{
\subsection{Operator renormalization, mixings and normalization}\label{sec_renorm}

Although the total form factors $T_1$, $T_2$, $D$ and $\bar{C}$, such as $T(Q^2) \equiv \sum_{i=u,d,\cdots,g} T^i(Q^2)$, are renormalization and scale invariant, the quark and gluon pieces are not separately scale independent and conserved.
We renormalize our results at ${\overline{{\rm{MS}}}}(\mu =2 \ {\rm {GeV}})$ with a nonperturbative renormalization procedure.
Since we only consider the operator $\mathcal{T}_{4i}$ in this paper, a purely multiplicative and linear mixing renormalization procedure is involved for the $T_1$, $T_2$, and $D$ form factors and their linear combinations such as $T_1(Q^2) + T_2(Q^2)$, namely,
\begin{eqnarray}\label{eq:sec4_renorm} 
\begin{aligned}
{{T}^{u/d,R}_{\rm{CI}}}&=Z^{\overline{\rm{MS}}}_{QQ}(\mu) {{T}^{u/d,L}_{\rm{CI}}}, \\
{{T}^{u/d/s}_{\rm{DI}}}&=Z^{\overline{\rm{MS}}}_{QQ}(\mu) {{T}^{u/d/s,L}_{\rm{DI}}}+\delta Z^{\overline{\rm{MS}}}_{QQ}(\mu)\sum_{q = u,d,s}\!\![{{T}^{q,L}_{\rm{CI}}} + {{T}^{q,L}_{\rm{DI}}}] + Z^{\overline{\rm{MS}}}_{QG}(\mu) {{T}^{g,L}_{\rm{DI}}} , \\
{{T}^{g,R}_{\rm{DI}}}&=Z^{\overline{\rm{MS}}}_{GQ}(\mu) \sum_{q = u,d,s}\!\! [{{T}^{q,L}_{\rm{CI}}} + {{T}^{q,L}_{\rm{DI}}}] + Z^{\overline{\rm{MS}}}_{GG}\, {{T}^{g,L}_{\rm{DI}}} ,
\end{aligned}
\end{eqnarray}
in which ${T^{q/g,L}_{\rm{CI}}}$ and ${T^{q/g,L}_{\rm{DI}}}$ are the CI and DI bare form factors under the lattice regularization, respectively.
Reference.~\cite{Yang:2018nqn} has done a complete calculation of the nonperturbative renormalization constants on the 32ID lattice which are shown in Table~\ref{tab:sec4_32ID_T44}.
More precisely, we calculated the renormalization and mixing coefficients of both the quark and glue operators under the regularization independent momentum subtraction scheme (RI/MOM) scheme nonperturbatively, and then used the perturbative matching (3-loop for the quark operator renormalization and 1-loop for the other cases) to convert the RI/MOM renormalization/mixing coefficients to those under the $\overline{\mathrm{MS}}$ scheme at 2 GeV.
It turns out that the quantum corrections of the glue operator with either quark or gluon external state under the RI/MOM scheme are at a few-percent level with the dimensional regularization used by the $\overline{\mathrm{MS}}$ scheme, but they are sizeable with the lattice regularization we used. Particularly, the value of $Z_{GQ}$ is 0.57 when the lattice spacing is as large as 0.14 fm.
Thus we would like to emphasize here that the nonperturbative renormalization and mixing calculation is essential to obtain reliable momentum and angular momentum fraction results.
\begin{table}[H]
  \centering{
  \begin{tabular}{| c | c | c | c | c | c |  }
    \hline
    Lattice & $Z_{QQ}$ & $\delta Z_{QQ}$ & $Z_{QG}$ & $Z_{GQ}$  & $Z_{GG}$\\
    \hline
    32ID      & 1.25(0)(2) & 0.018(2)(2) & 0.017(17) & 0.57(3)(6) & 1.29(5)(9) \\
    \hline
  \end{tabular}
  \caption{The nonperturbative renormalization and mixing constants on the 32ID lattice.}
  \label{tab:sec4_32ID_T44}
  }
\end{table}

Since there are no conserved EMT operators on the lattice due to the lack of infinitesimal translational and rotational symmetries, we have to normalize the final results with Eq.~(\ref{eq:proton_sum_rules}).
A way of normalizing the momentum and angular momentum fractions is proposed in Ref.~\cite{Deka:2013zha},
in which the normalization constants for quarks and glue $N^{q,L}$ and $N^{g,L}$ satisfy
\begin{eqnarray}\label{eq:nor_condistion1}
\begin{aligned}
N^{q,L} \braket{x}^{q,R} + N^{g,L} \braket{x}^{g,R} &= 1, \\
N^{q,L} J^{q,R} + N^{g,L} J^{g,R} &= \frac{1}{2}, \\
\end{aligned}
\end{eqnarray}
and the normalized quantities are given by
\begin{eqnarray}
\begin{aligned}
\braket{x}^q = N^{q,L} \braket{x}^{q,R} , &\; \braket{x}^g = N^{g,L} \braket{x}^{g,R}, \\
J^q = N^{q,L} J^{q,R}, &\; J^g = N^{g,L} J^{g,R} .
\end{aligned}
\end{eqnarray}
By solving Eq.~(\ref{eq:nor_condistion1}) we get $N^{q,L}$ and $N^{g,L}$ as
\begin{eqnarray}
\begin{aligned}
N^{q,L} &= \frac{-2 J^{g,R}+ \braket{x}^{g,R}}{ 2 J^{q,R} \braket{x}^{g,R} - 2 J^{g,R} \braket{x}^{q,R}}
= \frac{- T_2^{g,R}}{ T_2^{q,R} T_1^{g,R} - T_2^{g,R} T_1^{q,R} }, \\
N^{g,L} &= \frac{ 2 J^{q,R}- \braket{x}^{q,R}}{ 2 J^{q,R} \braket{x}^{g,R} - 2 J^{g,R} \braket{x}^{q,R}}
= \frac{ T_2^{q,R} }{ T_2^{q,R} T_1^{g,R} - T_2^{g,R} T_1^{q,R} }, \\
\end{aligned}
\end{eqnarray}
in which $T_1^{q/g,R}$ and $T_2^{q/g,R}$ are the nucleon form factors from the local current after renormalization.
However, the $T_2$ form factors, which are required in the numerator of the normalization, are small and have almost no signal under our current statistics.
Given the current situation, we assume $N^L \equiv N^{q,L} = N^{g,L}$ and 
use joint fits to get $N^L$ from the momentum and angular momentum fractions sum rules
\begin{eqnarray}\label{eq:nor_condition_Z}
\begin{aligned}
N^L \braket{x}^{q,R} + N^L \braket{x}^{g,R} = 1, \quad
N^L J^{q,R} + N^L\, J^{g,R}  = \frac{1}{2}.
\end{aligned}
\end{eqnarray} 

Note that when the results at several lattice spacings are available, one can compare the continuum limits of the quark and gluon momentum and angular momentum fractions with or without the above normalization, and take the difference as a systematic uncertainty.

}
}

{
\section{Numerical details}\label{sec:numerical}

We use overlap fermions on a $32^3 \times 64$ ensemble (32ID) of HYP smeared $2+1$-flavor domain-wall fermion configurations at $a=0.143\  {\rm{fm}}$ and $m_{\pi} = 172\ {\rm {MeV}}$, generated by RBC/UKQCD with Iwasaki plus the dislocation suppressing determinant ratio (DSDR) gauge action (labeled with ID)~\cite{Boyle:2015exm}.
The effective quark propagator of the massive overlap fermion is the inverse of the operator $(D_c + m)$~\cite{Chiu:1998eu,Liu:2002qu}, where $D_c=  D_{ov}/(1 - D_{ov}/(2\rho))$ is chiral, i.e., $\{D_c, \gamma_5\} = 0$ \cite{Chiu:1998gp}.
In the expression of $D_c$, the overlap Dirac operator $D_{ov}=\rho\big(1+\gamma_5\epsilon(\gamma_5D_w(-\rho))\big)$ is defined through the sign function of $\gamma_5D_w(-\rho)$, where $D_w(-\rho)$ is the Wilson fermion operator with $\rho = -1.5$.
A multimass inverter is used to calculate the propagators on 200 gauge configurations with 6 valence quark masses which correspond to valence pion masses, $173.76(17)$, $232.61(17)$, $261.34(17)$, $287.11(17)$, $325.47(17)$, and $391.11(17) \ {\rm{MeV}}$.
Gauge invariant box smearing~\cite{Allton:1990qg,Liang:2016fgy} with box half size of 1.0 ${\rm{fm}}$ is applied to have a better overlap with the nucleon ground state.
On each of the configurations, three source propagators $D^{-1}(y|\mathcal{G})$ are computed, with $\mathcal{G}$ the smeared $Z_3$-noise grid source~\cite{Dong:1993pk} with $\{2,2,2,2\}$ equally spaced points in the $\{x,y,z,t\}$ directions, respectively.

For the CI, we use the stochastic-sandwich method~\cite{Yang:2015zja, Liang:2018pis} with FFT described below in Sec.~\ref{sec3:sec_CI_FFT} to calculate the 3pt.
Low-mode substitution (LMS) on a grid source has been used to improve signals of the nucleon.
In order to estimate the propagators between current positions and sink positions (at time slice $t'$), we generate $n_{\rm{noi}}$ sets of high-mode propagators $D^{-1}_{H,\rm{noi}}({z}, \eta_j)$ defined in the following Eq.~(\ref{eq:sec3_S_LH}).
Four source-sink separations $t'=7,8,9,10 \ (a) = 1.0, 1.14, 1.29, 1.43 \ {\rm{(fm)}}$ are used to control the excited-state contamination with $n_{\rm{noi}} = {{2,3,4,5}}$, respectively.

For the DI, we use smeared $Z_3$-noise grids to calculate the nucleon correlation functions with the spatial location of the grid chosen randomly on different source time slices. And we repeat the calculation for 16 different source time slices on each configuration to have good statistics.
The gluon operator $L_g^{ 4 i}[\tau,\vec{z};U]$ is constructed on all the current positions $z$ to have full statistics.
The quark loop $L_f^{4 i}[\tau,\vec{q};U]$ with flavor $f$ is calculated based on the point source propagators $D_f^{-1}(y|z)$ with $y = z \pm a_x, a_y \ {\rm{or}} \ a_z$.
The low-mode part of this propagator is calculated exactly using the 900 pairs of low-lying eigenvectors of the overlap Dirac operator.
The high-mode part is estimated with 8 sets of a 4-4-4-2 space-time $Z_4$-noise grid with even-odd dilution.
Each set has a different spatial grid location and an additional time shift.
The 6 valence quark masses used in the construction of the quark loops vary from light quark masses to the strange quark region.
For the strange quark DI contributions, we use bare valence strange quark mass $m_s a = 0.08500$ with the nonperturbative mass renormalization constant~\cite{Liu:2013yxz} $Z_m = 0.8094(26)$ which gives $m_s = 94.83(55) {\rm{MeV}}$. 
This is consistent with our global-fit value $101(3)(6) \ {\rm{MeV}}$ at $2 \ {\rm{GeV}}$ in the ${\overline{\rm{MS}}}$ scheme calculated in~\cite{Yang:2014sea}.
References~\cite{XQCD:2013odc, Gong:2015iir, Yang:2015uis} contain more details regarding the DI calculation.

The total number of propagators we generated is 3 (grid source propagators) + 14 (sink noises propagators) + 16 (grid source propagators for the nucleon correlation functions) + 8 (propagators for quark loops) = 41 on each of the 200 configurations.

{
\subsection{Connected insertions}\label{sec3:sec_CI_num}

{
\subsubsection{Momentum projection on grid source}\label{sec3:sec_CI_mom}
In order to have good signals for the rest of the nucleon correlation functions, we have developed the grid source with $Z_3$ noises~\cite{Dong:1993pk} along with the low-mode substitution (LMS) method~\cite{Li:2010pw, XQCD:2013odc, Yang:2015zja, Liang:2016fgy, Liang:2018pis}.
In addition, for 2pts and 3pts with finite source momenta, we have developed the use of mixed momenta~\cite{Yang:2015zja, Liang:2018pis} to accommodate $Z_3$ noise grid source and momenta.
However, such a mixed-momenta method has worse signals at certain momenta and higher computational cost.
In this section, we will describe our new way of applying momentum projection on a grid source with LMS.

In order to introduce the modification of LMS for source momentum projection, we start with the fact that contributions from high-mode, low-mode, and their mixture parts of the correlation functions can be measured independently using different source positions and statistics for each contribution.
This has been applied to meson~\cite{Neff:2001zr,DeGrand:2004qw,Giusti:2004yp,Blum:2018mom,Aubin:2019usy,Borsanyi:2020mff} and nucleon~\cite{Giusti:2005sx} correlation functions to improve signals.
More specifically, the quark $D^{-1}({y}|{x})$ propagator from ${x}$ to ${y}$ can be split into its high-mode and low-mode parts defined as
\begin{eqnarray}\label{eq:S_LH_src}
\begin{aligned}
D^{-1}(y|x) = D^{-1}_L(y|x) + D^{-1}_H(y|x) ,
  \;\;\;  D^{-1}_L(y|x) &= \sum_{\lambda_i \leq \lambda_c} \frac{1}{\lambda_i +m} v_i(y) v_i^{\dagger}(x),
\end{aligned}
\end{eqnarray}
with $\lambda_i$ the low lying overlap eigenvalue and $v_i$ the eigenvector of the {$i$th} eigenmode of $D_c$.
$\lambda_c$, the highest eigenvalue in LMS, is in the range of twice the pion mass which is much larger than the quark mass $m$ with the number of eigenmodes $n_{\rm v} \sim 1800$ on 32ID.
$D^{-1}_H(y|x)$ is calculated with deflation of the overlap operator using low-mode eigenvectors $v_i$. 
Consider the nucleon correlation function from a point source $x=\{\vec{x},t_0\}$ with finite momentum $\vec{p}$,
\begin{eqnarray}\label{eq:sec3_nucleon_corr}
\begin{aligned}
C_{{N}}(\vec{p},t) =  \sum_{y} e^{-i \vec{p} \cdot (\vec{y} - \vec{x})} \braket{ {\chi}(y) \bar{{\chi}}(x)}, \\
\end{aligned}
\end{eqnarray}
with $y=\{\vec{y},t_1\}$ and $t = t_1 - t_0$. This nucleon correlation function can be split into four contributions as
\begin{eqnarray}\label{eq:sec3_split_low_high}
\begin{aligned}
C_{{N}}(\vec{p},t) =  C_{N,LLL}(\vec{p},t) + C_{N,LLH}(\vec{p},t) + C_{N,LHH}(\vec{p},t) + C_{N,HHH}(\vec{p},t) \\
\end{aligned}
\end{eqnarray}
in which $L$($H$) denotes the low-mode(high-mode) of the propagators involved in contractions.
Since the ensemble average of each contribution is translationally invariant, we can measure each piece independently.
We can focus on increasing the signal-to-noise ratio on the most noisy parts with a different number of sources/statistics for each part.
Combining this idea with LMS under grid source, we first calculate the first three parts with grid source propagator.
The random ${\rm{Z}}_3$ grid source used in LMS is defined as
\begin{eqnarray}\label{eq:sec2_grid_def}
\begin{aligned}
\mathcal{G}(\vec{w}_0) &\equiv \sum_{i}^n \eta_i V(\vec{w}_i), \ \vec{w}_i \in (x_0 + m_x \Delta_x, y_0 + m_y \Delta_y,z_0 + m_z \Delta_z)\\
\end{aligned}
\end{eqnarray}
where $V(\vec{w}_i)$ is the smeared source centered at $\vec{w}_i$, $\eta_i$ is a ${\rm{Z}}_3$ noise on each of the grid points $\vec{w}_i$, $\vec{w}_0 = (x_0,y_0,z_0)$ is the starting point of the grid, $\Delta_{x,y,z} = L/2\  {\rm{or}}\  L/3\  {\rm{or}}\  L/4 \cdots$ is the offset in the spatial direction respectively, $m_{x,y,z} \in \{0,1,\cdots, L_s/ \Delta_{x,y,z}\}$ is the offset number in each direction for each grid point, and $n= \frac{L_s^3}{\Delta_x \Delta_y \Delta_z}$ is the number of grid points of the grid source.
As the Dirac operator is a linear operator, the grid source propagator can be written as
\begin{eqnarray}\label{eq:low_plus_high}
\begin{aligned}
D^{-1}(y|\mathcal{G}(\vec{w}_0)) &= \sum_{i}^n \eta_i D^{-1}(y|\vec{w}_i) = \sum_{i}^n \eta_i D^{-1}_L(y|\vec{w}_i) + D^{-1}_H(y| \mathcal{G}(\vec{w}_0)),\\
\end{aligned}
\end{eqnarray}
with $D^{-1}(y|\vec{w}_i)$ the propagator from each of the grid points $\vec{w}_i$ and $D^{-1}_H(y| \mathcal{G}(\vec{w}_0))$ the high-mode part of the noise grid-source propagator.
As shown in Ref.~\cite{Li:2010pw}, nucleon correlation functions from $D^{-1}(y|\mathcal{G}(\vec{w}_0))$ directly will have worse signals 
and we can approach an intermediate propagator coming from the grid source point $\vec{w}_i$ (noting that the high-mode part is the full noise grid-source propagator) as
\begin{eqnarray}\label{eq:prop_G_LH}
\begin{aligned}
D^{-1}_{\mathcal{G}}(y | \vec{w}_i) &= \eta_i D^{-1}_L(y|\vec{w}_i) + D^{-1}_H(y| \mathcal{G}(\vec{w}_0)).\\
\end{aligned}
\end{eqnarray}
The partial nucleon correlation function constructed from this propagator, but  without the portion with all three quarks in the $H$ modes, is
\begin{eqnarray}\label{eq:sec3_corr_phase}
\begin{aligned}
C_{{N}}^{\mathcal{G}, \vec{w}_i}(\vec{p},t) = \sum_{y} & e^{-i \vec{p} \cdot (\vec{y} - \vec{w}_i)}  \Big\langle C(D^{-1}_{\mathcal{G}}(y | \vec{w}_i), D^{-1}_{\mathcal{G}}(y | \vec{w}_i), D^{-1}_{\mathcal{G}}(y | \vec{w}_i)) \\
  &- C(D^{-1}_H(y| \mathcal{G}(\vec{w}_0)), D^{-1}_H(y| \mathcal{G}(\vec{w}_0)), D^{-1}_H(y| \mathcal{G}(\vec{w}_0))) \Big\rangle \\
\end{aligned}
\end{eqnarray}
where $\braket{...}$ denotes the ensemble average and $C(D^{-1},D^{-1},D^{-1})$ denotes the nucleon contractions with three propagators.
Since the pure high-mode parts in Eq.~(\ref{eq:sec3_corr_phase}) will not give the correct phases for different momenta under grid sources, we have subtracted them out from the correlator.
With gauge invariance and noise averaging ($\braket{\eta_i \eta_j \eta_k}_{Z_3} = \delta_{i,j} \delta_{j,k}$), it is easy to show that
\begin{eqnarray}
\begin{aligned}
C_{{N}}^{\mathcal{G}, \vec{w}_i}(\vec{p},t) &= C_{N,LLL}(\vec{p},t) + C_{N,LLH}(\vec{p},t) + C_{N,LHH}(\vec{p},t),
\end{aligned}
\end{eqnarray}
which gives us the the first three terms in Eq.~(\ref{eq:sec3_split_low_high}).
Furthermore, we shall average the contributions from different $\vec{w}_i$ as $\frac{1}{n} \sum_i C_{{N}}^{\mathcal{G}, \vec{w}_i}(\vec{p},t)$ to have better statistics.
The remaining pure high-mode part $C_{N}^{HHH}(\vec{p},t)$ of Eq.~(\ref{eq:sec3_split_low_high}) could be calculated with a point source high-mode propagator $D_H^{-1}(y|x_P)$ starting from any position $x_P$.
In summary, the new method is
\begin{eqnarray}\label{eq:sec3_corr_phase_final}
\begin{aligned}
C_{{N}}^{\mathcal{G}}(\vec{p},t) = &\frac{1}{n}\sum_{y}  e^{-i \vec{p} \cdot (\vec{y} - \vec{w}_i)}  \Big\langle \sum_i \Big[ C(D^{-1}_{\mathcal{G}}(y | \vec{w}_i), D^{-1}_{\mathcal{G}}(y | \vec{w}_i), D^{-1}_{\mathcal{G}}(y | \vec{w}_i)) \\
  &\quad\quad - C(D^{-1}_H(y| \mathcal{G}(\vec{w}_0)), D^{-1}_H(y| \mathcal{G}(\vec{w}_0)), D^{-1}_H(y| \mathcal{G}(\vec{w}_0))) \Big] \Big\rangle \\
  & + \sum_{y}  e^{-i \vec{p} \cdot (\vec{y} - \vec{x}_P)} \Big\langle C(D_H^{-1}(y|x_P), D_H^{-1}(y|x_P) , D_H^{-1}(y|x_P)) \Big\rangle. \\
\end{aligned}
\end{eqnarray}
In this construction, the momentum projection for the source need not be carried out at the propagator level.
Instead, it is implemented at the correlator level, which saves inversion and contraction time for multiple momenta.

}

{
\subsubsection{FFT on stochastic-sandwich method}\label{sec3:sec_CI_FFT}
In order to approach different current and sink momenta combinations under the stochastic-sandwich method with LMS~\cite{Yang:2015zja, Liang:2018pis},
we utilize the fact that the low and high modes for the propagator $D^{-1}({z}|x\ppf)$ between the current and sink can be well separated into
multiplication of functions of sink position $x\ppf$ and current position $z$.
This facilitates FFT usage on the momenta projection of $\vec{p}\ppf$ and $\vec{p}$ on $x\ppf$ and $z$, respectively~\cite{Wang:2020nbf}.
Such FFT on the stochastic-sandwich method can be applied to the CI part of nucleon 3pts 
\begin{eqnarray}\label{eq:sec3_3pt_contra}
\begin{aligned}
&C_{{\rm{CI}},\Gamma_\alpha}^{\mathcal{O}^{u/d}}(t\ppf, \tau,\vec{p}\ppf,\vec{p}\ppi) = \Big\langle \sum_{\vec{x}\ppf,\vec{z}} e^{-i \vec{p}\ppf \cdot \vec{x}\ppf} e^{ i \vec{q} \cdot \vec{z}}
  \times  {\rm{Tr}} \left[ \Gamma_\alpha \chi(\vec{x}\ppf, t\ppf) \mathcal{O}^{u/d}(z) \bar{\chi}(\vec{0},0)  \right] \Big\rangle \\
\end{aligned}
\end{eqnarray}
which shares the same variables as in Eq.~(\ref{eq:sec2_3pt_lat}) and Eq.~(\ref{eq:sec2_3pt_Tr}) and $\mathcal{O}^{u/d}(z)$ is any local current for an up/down quark.
We use a point source at $(\vec{0},0)$ in the following instead of a grid source for illustrative purpose.
Then the evaluation of the CI part of Eq.~(\ref{eq:sec3_3pt_contra}) for the up/down quark part can be written as
\begin{eqnarray}
\begin{aligned}
&C_{{\rm{CI}}, \Gamma_\alpha}^{\mathcal{O}^{u/d}}(t\ppf, \tau,\vec{p}\ppf,\vec{p}\ppi)
= \sum_{\vec{x}\ppf ,\vec{z}} e^{-i \vec{p}\ppf \cdot \vec{x}\ppf} e^{ i \vec{q} \cdot \vec{z}}
  \times {\rm{Tr}} \left[M^{\mathcal{O}^{u/d}}_\alpha (x\ppf|0) D^{-1}_{u/d}(x\ppf |z) {\mathcal{O}^{u/d}}(z) D_{u/d}^{-1} (z|0) \right]. \\
\end{aligned}
\end{eqnarray}
We can write $M^{\mathcal{O}^{d}}_\alpha$ as
\begin{eqnarray}\label{eq:section2_2pt_PAB1}
\begin{aligned}
&(M^{\mathcal{O}^{d}}_\alpha)_{\beta \beta'}^{b  b'}(y|x)  = \epsilon_{abc} \epsilon_{a' b' c'} \Big( \left[ \underline{D_u^{-1}(y|x)^{ a' a}} \right]_{\beta \beta'} {\rm{Tr}} \left[\Gamma_\alpha D_u^{-1}(y|x)^{ c' c} \right] \\
&\quad \quad \quad + {\rm{Tr}} \left[ \underline{D_u^{-1}(y|0)^{ a' a} \Gamma_\alpha D_u^{-1}(y|x)^{c' c}} \right]_{\beta \beta'} \Big),\\
\end{aligned}
\end{eqnarray}
with the quantity $\underline{Q} \equiv (\widetilde{\mathcal{C}} Q \widetilde{\mathcal{C}}^{-1})^T$ for a matrix $Q$ defined in Dirac space
and $M^{\mathcal{O}^{u}}_\alpha = M_{1,\alpha}^{\mathcal{O}^{u}} + M_{2,\alpha}^{\mathcal{O}^{u}} + M_{3,\alpha}^{\mathcal{O}^{u}} + M_{4,\alpha}^{\mathcal{O}^{u}}$ with
\begin{eqnarray}\nonumber
\begin{aligned}
(M_{1,\alpha}^{\mathcal{O}^{u}})_{\gamma \gamma'}^{ a a'}(y|x) &= \epsilon_{abc} \epsilon_{a' b' c'} \left[ \underline{D_d^{-1}(y|0)^{ b' b}} \right]_{\gamma \gamma'} {\rm{Tr}} \left[\Gamma_\alpha D_u^{-1}(y|0)^{ c' c} \right] , \\
(M_{2,\alpha}^{\mathcal{O}^{u}})_{\gamma \gamma'}^{ a a'}(y|x) &= \epsilon_{abc} \epsilon_{a' b' c'}  \left[ \Gamma_\alpha D_u^{-1}(y|0)^{c' c} \underline{D_d^{-1}(y|0)^{ b' b}} \right]_{\gamma \gamma'} , \\
(M_{3,\alpha}^{\mathcal{O}^{u}})_{\gamma \gamma'}^{ c c'}(y|x) &= \epsilon_{abc} \epsilon_{a' b' c'}  {\rm{Tr}} \left[ \underline{D_d^{-1}(y|0)^{ b' b}} D_u^{-1}(y|0)^{ a' a} \right] \left[\Gamma_\alpha \right]_{\gamma \gamma'} , \\
(M_{4,\alpha}^{\mathcal{O}^{u}})_{\gamma \gamma'}^{ c c'}(y|x) &= \epsilon_{abc} \epsilon_{a' b' c'}  \left[ \underline{D_d^{-1}(y|0)^{ b' b}} D_u^{-1}(y|0)^{ a' a} \Gamma_\alpha \right]_{\gamma \gamma'}. \\
\end{aligned}
\end{eqnarray}

The low-mode part of the propagator $D^{-1}(x\ppf|z)$ between the current and sink is calculated exactly and its high-mode part is calculated with the noise-estimated propagator $D^{-1}_{H,{\rm{noi}}}({z},\eta_j)$ as
\begin{eqnarray}\label{eq:sec3_S_LH}
\begin{aligned}
D^{-1}({z}|{x_{\textrm{f}}})   &= D^{-1}_L({z}|{x_{\textrm{f}}}) + D^{-1}_H({z}|{x_{\textrm{f}}}) , \\
D^{-1}_L({z}|{x_{\textrm{f}}}) &=\sum_{\lambda_i \leq \lambda_c} \frac{1}{\lambda_i +m} v_i({z}) v_i^{\dagger}({x_{\textrm{f}}}),    \\
D^{-1}_H({z}|{x_{\textrm{f}}}) &= \frac{1}{n_{\rm{noi}}} \sum_{j=1}^{n_{\rm{noi}}} D^{-1}_{H,{\rm{noi}}}({z},\eta_j)\eta_j^\dagger({x_{\textrm{f}}}),
\end{aligned}
\end{eqnarray}
in which $\eta_j({x_{\textrm{f}}})$ is a $Z_3$ noise and $n_{\rm{noi}}$ is the number of noises at the sink position $x\ppf$.
Then, we can decompose $C_{\rm{CI}}$ into factorized forms within the sums of the eigenmodes for the low modes and the $n_{\rm{noi}}$ number of noises $\eta_j$ for the high modes,
\begin{eqnarray}\label{eq:sec3_FFT_b2}
\begin{aligned}
C_{{\rm{CI}},\Gamma_\alpha}^{\mathcal{O}^{u/d}}
   =& \langle \sum_{\lambda_i \leq \lambda_c} {\rm{Tr}} [\frac{1}{\lambda_i + m}  G^{L,\mathcal{O}^{u/d}}_i(\vec{q},\tau) F_{i,\alpha}^{L,\mathcal{O}^{u/d}}(\vec{p}\ppf,t\ppf) ] \\
&+\sum_{j=1}^{n_{\rm{noi}}} \frac{1}{n_{\rm{noi}}} {\rm{Tr}}[G_j^{H,\mathcal{O}^{u/d}}(\vec{q},\tau) (F_{j,\alpha}^{H,\mathcal{O}^{u/d}}(\vec{p}\ppf,t\ppf)] \rangle,
\end{aligned}
\end{eqnarray}
where
\begin{eqnarray}\label{eq:sec2_FFT_b3}
\begin{aligned}
G^{L,\mathcal{O}^{u/d}}_i(\vec{q},\tau) &&=&\sum_{\vec{z}} e^{ i \vec{q} \cdot \vec{z}} v_i^\dagger({z}) \mathcal{O}^{u/d} D^{-1}({z}|{0}), \\
F^{L,\mathcal{O}^{u/d}}_{i,\alpha}(\vec{p}\ppf,t\ppf) &&=& \sum_{\vec{x}\ppf}  e^{-i \vec{p}\ppf \cdot \vec{x}\ppf}  M^{u/d}_\alpha(x\ppf|0) v_i({x\ppf}), \\
G^{H,\mathcal{O}^{u/d}}_j(\vec{q},\tau) &&=&\sum_{\vec{z}} e^{ i \vec{q} \cdot \vec{z}} \gamma_5 (D^{-1}_{H,{\rm{noi}}}({z},\eta_j))^\dagger \gamma_5 \mathcal{O}^{u/d} D^{-1}({z}|{0}), \\
F^{H,\mathcal{O}^{u/d}}_{j,\alpha}(\vec{p}\ppf,t\ppf) &&=& \sum_{\vec{x}\ppf} e^{-i \vec{p}\ppf \cdot \vec{x}\ppf} M^{u/d}_\alpha(x\ppf|0) \eta({x\ppf}), \\
\end{aligned}
\end{eqnarray}
in which we have defined $D^{-1} (z|0) = D_u^{-1} (z|0) =  D_d^{-1} (z|0)$ to be the light quark propagator, and used $D^{-1}(x\ppf|z) = \gamma_5 (D^{-1}(z|x\ppf))^\dagger \gamma_5 $ for the high-mode propagator $D^{-1}_{H,{\rm{noi}}}({z},\eta_j)$.
It is also straightforward to replace the point source at $(\vec{0},0)$ with a grid source LMS described in Sec.~(\ref{sec3:sec_CI_mom}).

With these implementations, we can have any combination of $\vec{q}$ and $\vec{p_{\textrm{f}}}$ without much additional cost.
This property is essential for EMT calculations as it enables us to approach different parts of $\mathcal{T}_{\mu \nu}$ which require different nucleon kinematics.
Also the averaging over all equivalent momenta setups gives much higher statistics compared to the traditional stochastic-sandwich method with similar computational cost.

}

}

{
\subsection{Disconnected insertions }\label{sec3:sec_DI_num}

We have applied the CDER technique~\cite{Liu:2017man} to have better control of the statistical uncertainties for the quark and glue DI parts. The associated 3pts are rewritten as
\begin{eqnarray}\label{eq:secDI_3pt_DI_cder}
\begin{aligned}
&C_{{\rm{DI}},\Gamma_\alpha}^{\mathcal{O}}(t\ppf, \tau,\vec{p}\ppf,\vec{p}\ppi; R) = \Big\langle \sum_{\vec{x}\ppf,|\vec{r}| < R} e^{-i \vec{p}\ppf \cdot \vec{x}\ppf} e^{ i \vec{q} \cdot (\vec{x}\ppf + \vec{r})}
   {\rm{Tr}} \left[ \Gamma_\alpha \chi(\vec{x}\ppf, t\ppf) \mathcal{O}(\vec{x}\ppf + \vec{r},\tau) \bar{\chi}(\vec{\mathcal{G}}, 0)  \right] \Big\rangle_{\rm{DI}} \\
& = \sum_{\vec{x}\ppf,|\vec{r}| < R}  { \Big\langle \rm{Tr}}  \left[
    e^{-i \vec{p}\ppf \cdot \vec{x}\ppf} e^{ i \vec{q} \cdot (\vec{x}\ppf + \vec{r})} \Gamma_\alpha
    G^{NN}(x\ppf; \mathcal{G}) \mathcal{O}(\vec{x}\ppf + \vec{r},\tau) \right] \Big\rangle_{\rm{s}},\\
\end{aligned}
\end{eqnarray}
with $G^{NN} (x\ppf;\mathcal{G})$ the grid source nucleon propagator with LMS from position $\mathcal{G} \equiv (\vec{\mathcal{G}},0)$ to $x\ppf \equiv (\vec{x}\ppf,t\ppf)$.
The cutoff $R$ is the distance between the current position and the sink position and
$\langle \mathcal{O}(z) \mathcal{O}(x\ppf)\rangle_{\rm{s}} \equiv \langle \mathcal{O}(z) \mathcal{O}(x\ppf)\rangle - \langle \mathcal{O}(z) \rangle \langle \mathcal{O}(x\ppf)\rangle$
is the vacuum-subtracted correlation function.
It is shown~\cite{Araki:1962zhd} that, under the assumptions of translation invariance, stability of the vacuum, existence of a lowest nonzero mass and local commutativity, $\langle \mathcal{O}(z) \mathcal{O}(x\ppf)\rangle_s$ satisfies
\begin{eqnarray}\label{eq:secDI_cder_M}
\langle \mathcal{O}(z) \mathcal{O}(x\ppf)\rangle_s \leq A r^{-\frac{2}{3}} e^{-M r}
\end{eqnarray}
for large enough spacelike distance $r=|\vec{z} - \vec{x\ppf}|$,
with $M$ the smallest nonzero inverse correlation length for the correlator and $A$ a constant.
This exponential falloff in distance $r$ is known as the {\it{cluster decomposition theorem}}~\cite{Araki:1962zhd,Strocchi:1975xz}.
As demonstrated in Ref.~\cite{Liu:2017man} and utilized in Ref.~\cite{Liang:2018pis,Yang:2018bft}, the signal of the summed correlation function in Eq.~(\ref{eq:secDI_3pt_DI_cder}) for the DI will saturate at some $R$ which is larger than the corresponding correlation length.
But the noise will keep growing as the two operators fluctuate independently due to the fact that the variance of the two disconnected operators has a vacuum insertion.
Examples of the correlators in Eq.~(\ref{eq:secDI_3pt_DI_cder}) as a function of the current-sink separation $r$ for the source-sink separation 4 (in lattice unit) for the glue DI and strange quark DI are shown in the left panels of Fig~\ref{fig:DI_cder_basic_0}.
The correlators fall off exponentially as expected and go to zero at around $1.5 \ {\rm{fm}}$.
In view of cluster decomposition behavior in Eq.~(\ref{eq:secDI_cder_M}), we plot the effective mass $M$ as a function of distance $r$ in the middle panels of Fig~\ref{fig:DI_cder_basic_0} and the fitted values of $M$ and $A$ are shown in the legend.
We find that the glue and quark DI for $T_1(0)$ and $[T_1+T_2]$ form factors have different correlation lengths ($1/M$) and they are treated separately during CDER fits.

\begin{figure}[htbp]
  \centering
  \includegraphics[page=1,width = 0.32 \textwidth,height=0.26 \textwidth]{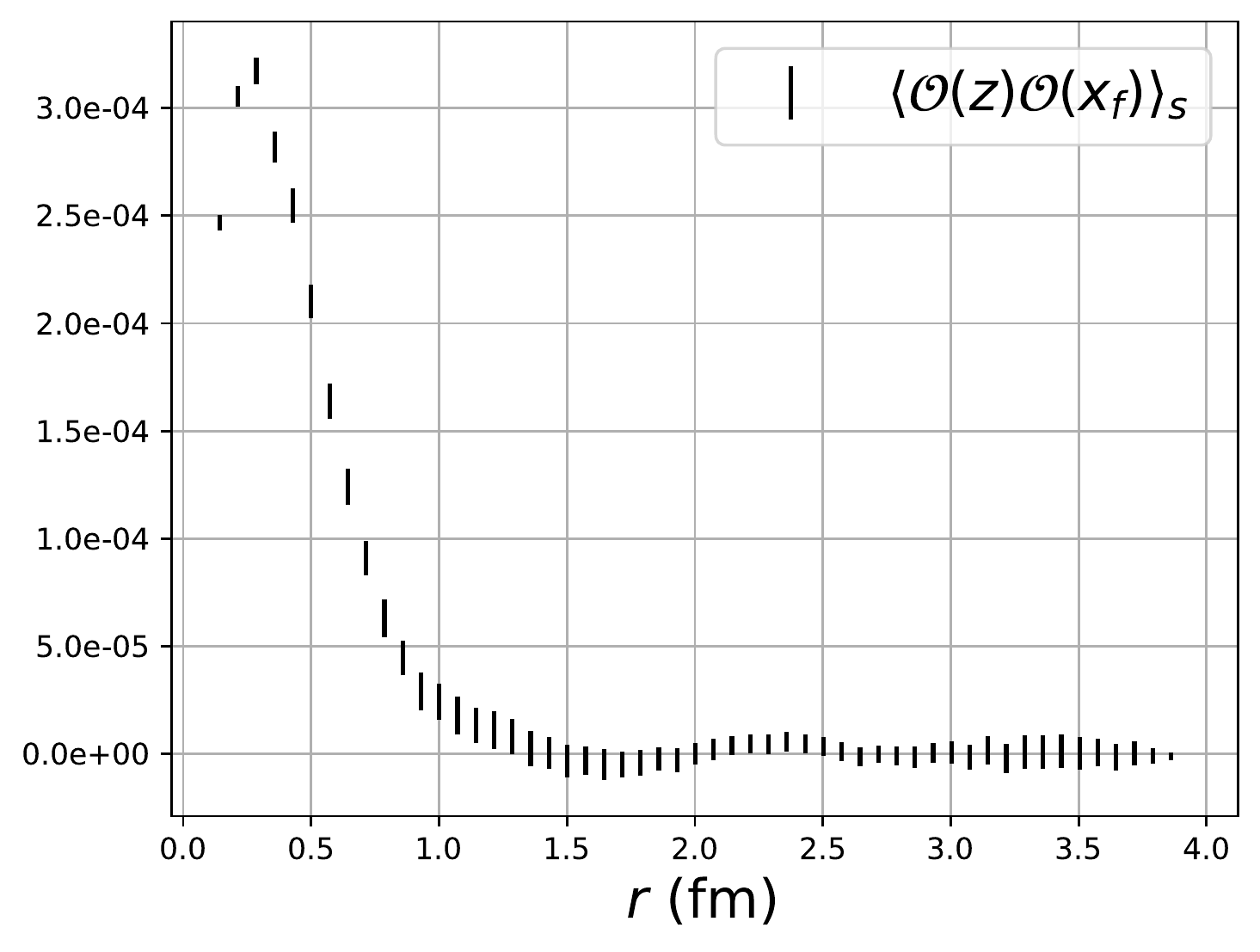}
  \includegraphics[page=1,width = 0.32 \textwidth,height=0.26 \textwidth]{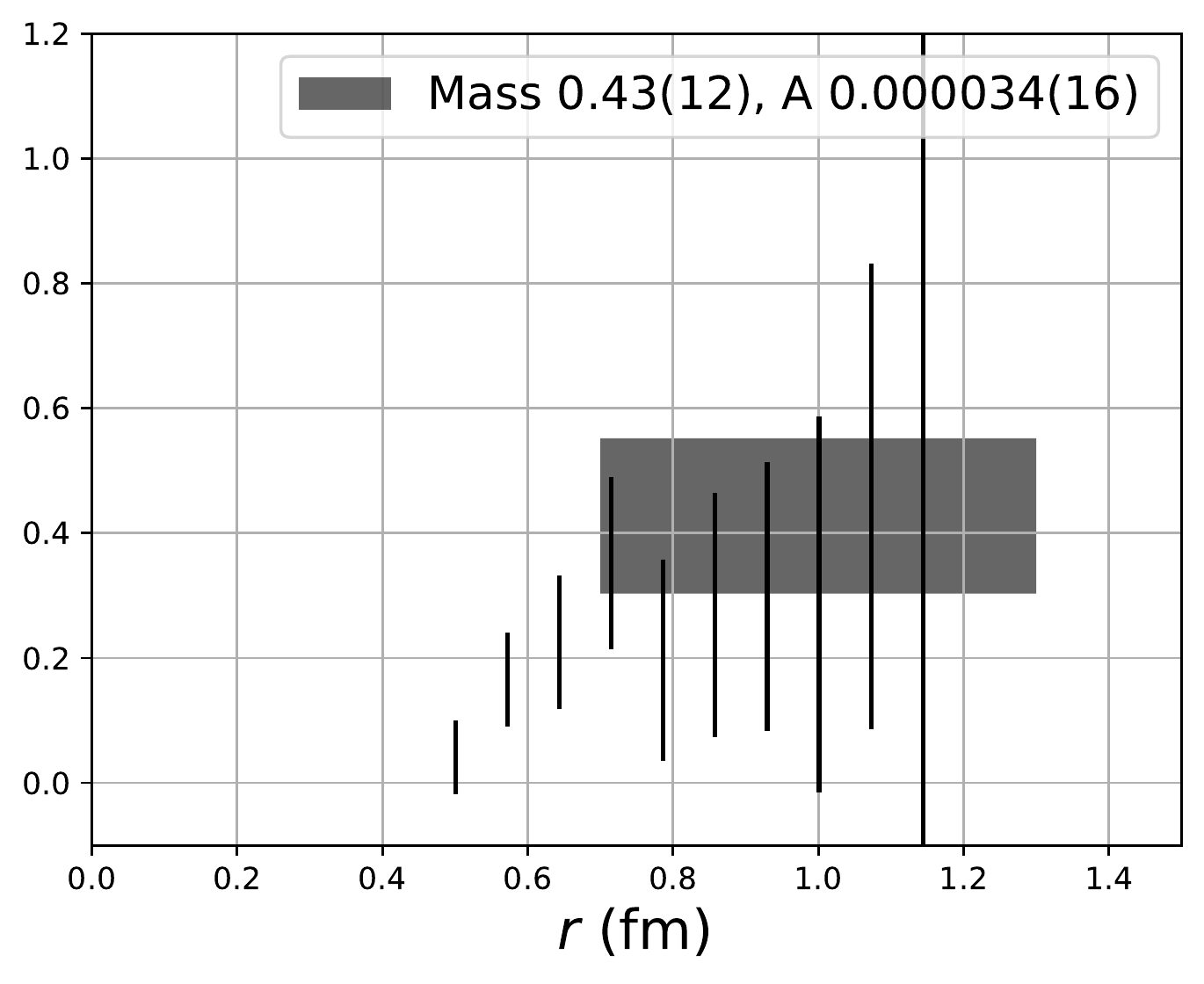}
  \includegraphics[page=1,width = 0.32 \textwidth,height=0.26 \textwidth]{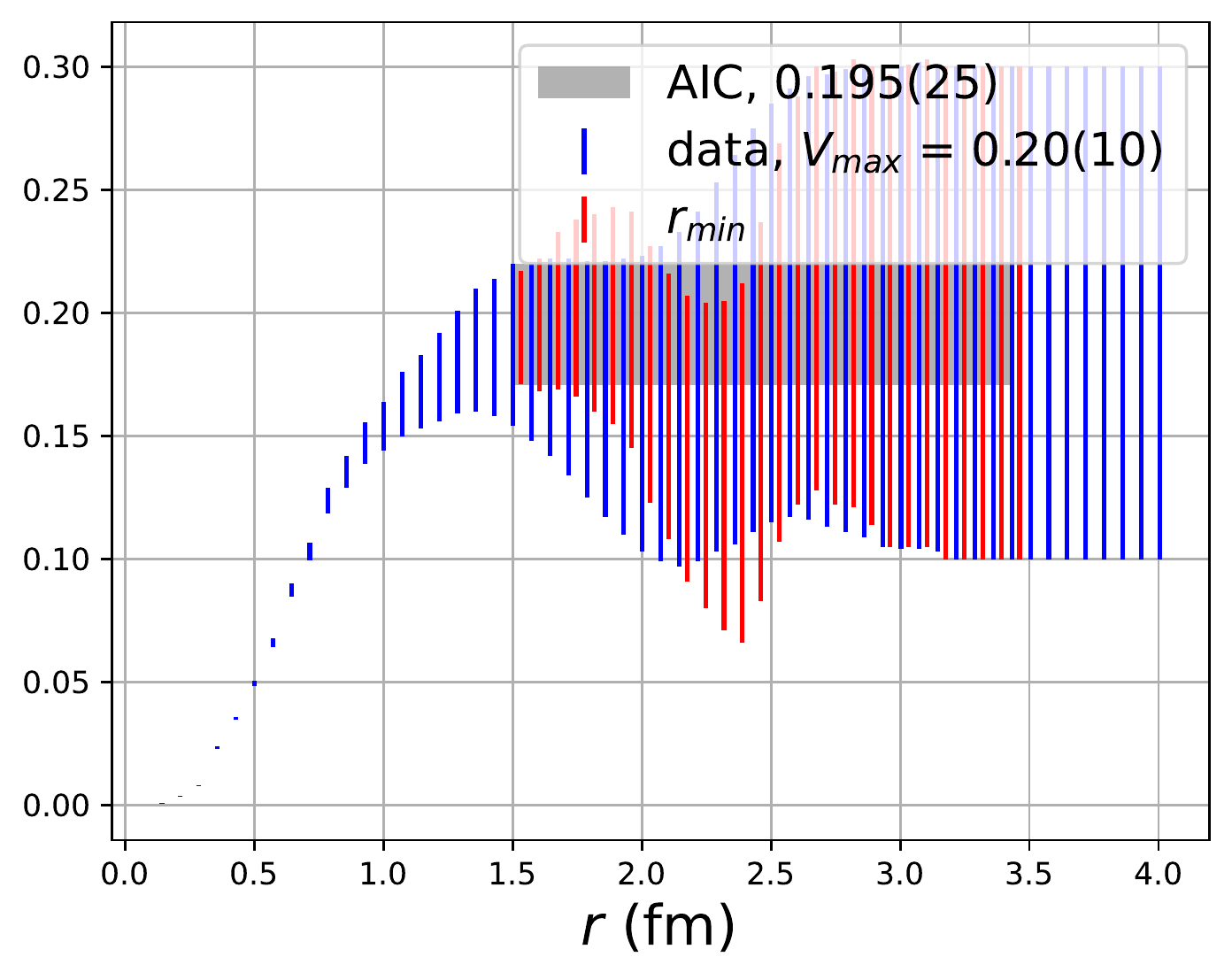} \\
  \includegraphics[page=1,width = 0.32 \textwidth,height=0.26 \textwidth]{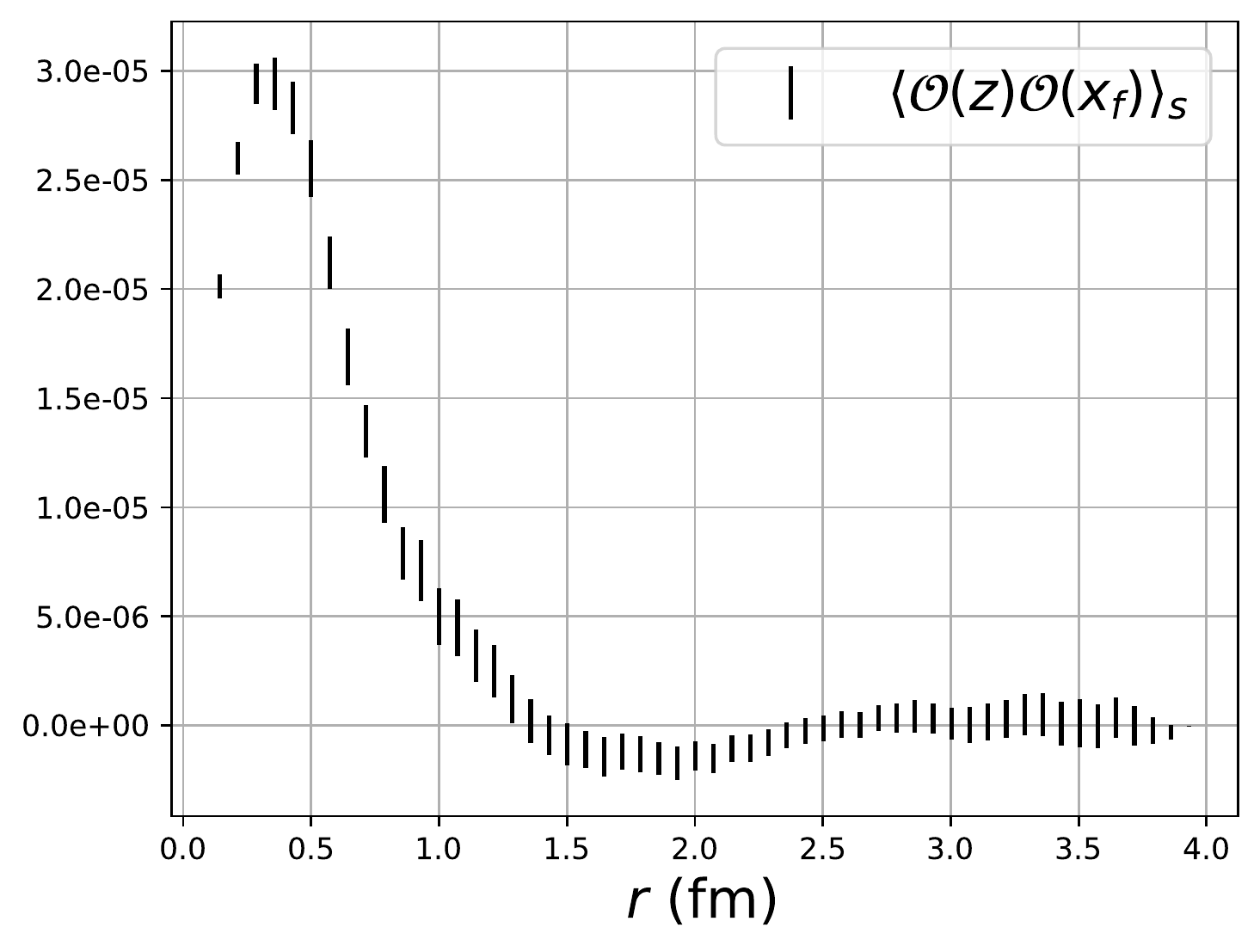}
  \includegraphics[page=1,width = 0.32 \textwidth,height=0.26 \textwidth]{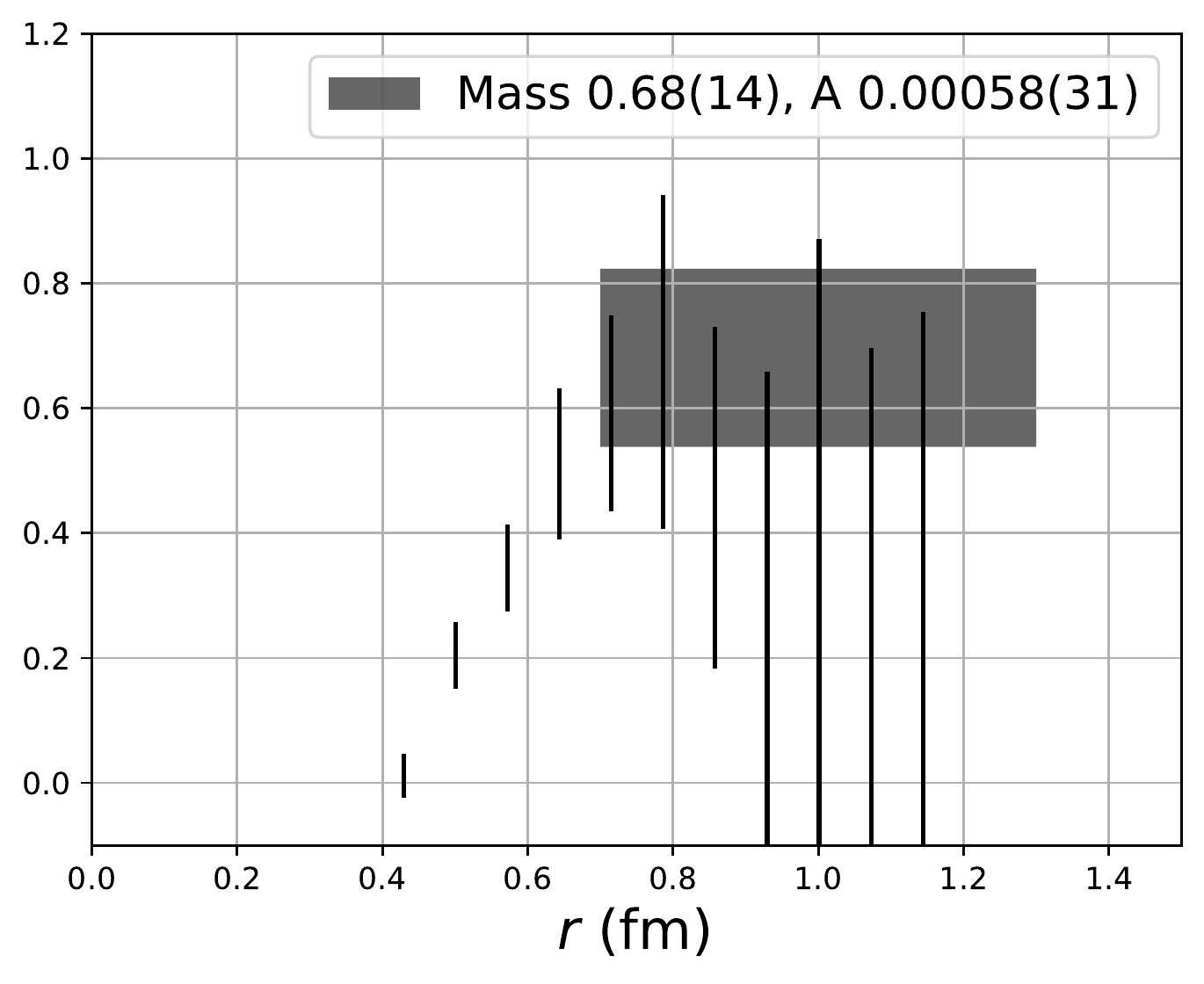}
  \includegraphics[page=1,width = 0.32 \textwidth,height=0.26 \textwidth]{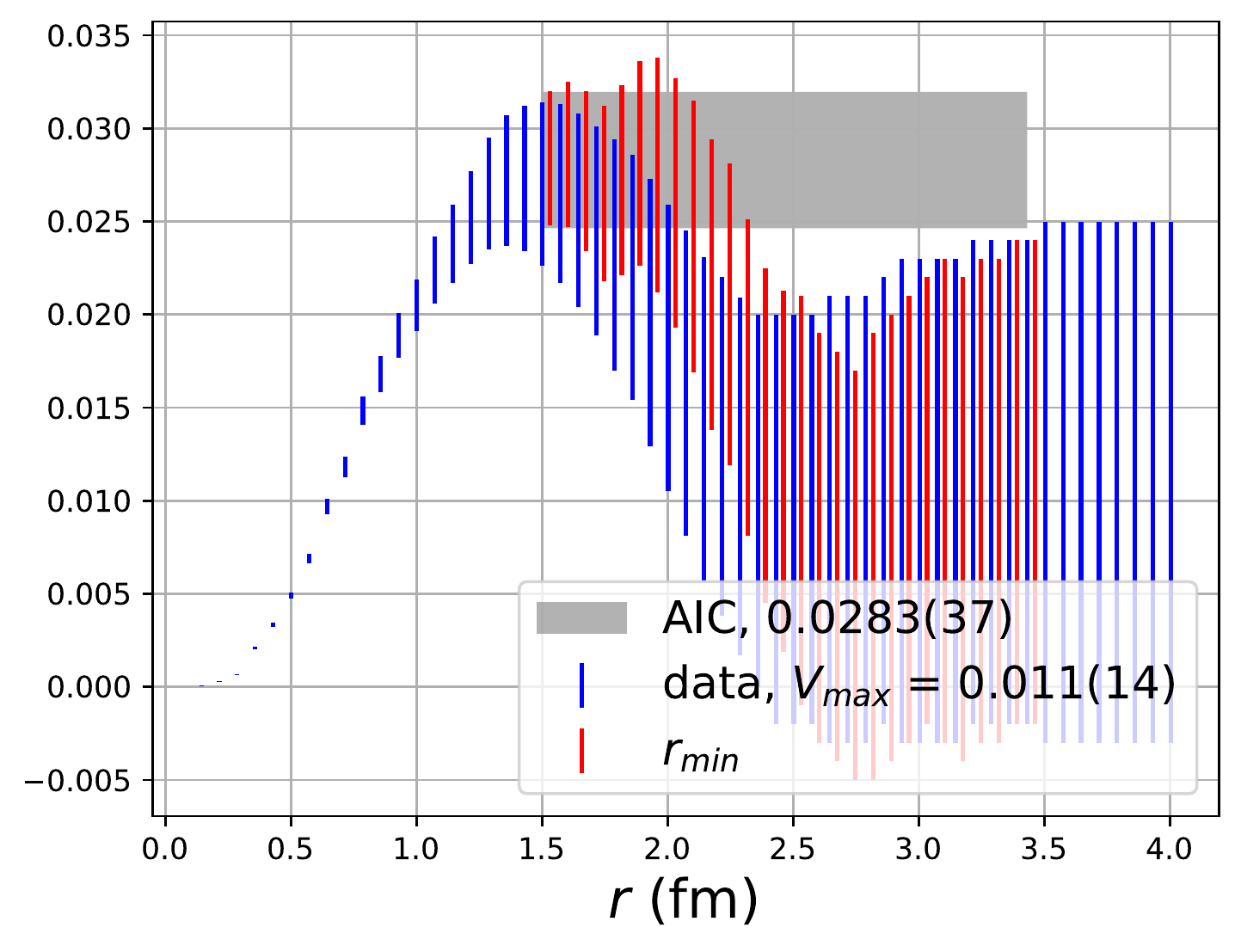} 
  \caption{Example plots for glue (upper panels) and strange quark (lower panels) 3pts at $t\ppf=4, \tau=2$ with valence pion mass 174 MeV. The left panels correspond to the correlation functions in Eq.~(\ref{eq:secDI_3pt_DI_cder}) as a function of the current-sink separation $r$. The middle panels correspond to the associated effective mass plots and the bands are the fit results using Eq.~(\ref{eq:secDI_cder_M}). The right panels correspond to the accumulated correlation functions as a function of $r$. The red lines are fits with errors explained in the text. The gray band is the result of AIC averages in Eq.~(\ref{eq:secDI_aic}).}
  \label{fig:DI_cder_basic_0}
\end{figure}

Let us first focus on the glue DI and the strange quark DI for the case with momenta setup $p\ppi = p\ppf$ and $q=0$ which gives $T_1(0)$.
Constant fits have been done with different $r_{\rm{min}}$ with fit range $[r_{\rm{min}},4.0\,{\rm{fm}}]$.
The fit results are shown in red points as a function of $r_{\rm{min}}$ in the right panels of Fig~\ref{fig:DI_cder_basic_0} and we have chosen $r_{\rm{min}} \geq r_{\rm{cut}} = 1.5\ {\rm{fm}}$.
We note that the fit errors for those $r_{\rm{min}}$ close to $r_{\rm{cut}}$ are much smaller than that of the total accumulated sum, the blue line at 4.0 fm.
The latter is the conventional approach with independent sums of $\vec{z}$ and $\vec{x}_{\rm{f}}$.
This is the essence of the CDER technique.
In order to estimate the systematic errors from different fits, we use the Akaike information criterion (AIC)~\cite{Akaike:1974aic} to average the fit results from different fit ranges with a weighting factor
\begin{eqnarray}\label{eq:secDI_aic}
\omega_{{\rm{AIC}}} = \exp \left[ - \frac{1}{2} (\chi^2 - 2 n_{\rm{dof}})\right],
\end{eqnarray} 
where $n_{\rm{dof}}$ is the number of degrees of freedom.
The prediction of the central value is
\begin{eqnarray}
\begin{aligned}
\bar{x} = \sum_i^P w_i x_i ;\ w_i = \frac{\omega_{{\rm{AIC}},i}}{\sum_i^P \omega_{{\rm{AIC}},i} },
\end{aligned}
\end{eqnarray}
with $P$ the total number of fits and $\omega_{{\rm{AIC}},i}$ the weight factor for fit $i$. The errors will be propagated with Jackknife resampling.
The AIC averaged values are shown in gray bands and given in legends in the right panels of Fig~\ref{fig:DI_cder_basic_0}.

\begin{figure}[htbp]
  \centering
  \includegraphics[page=1,width = 0.40 \textwidth]{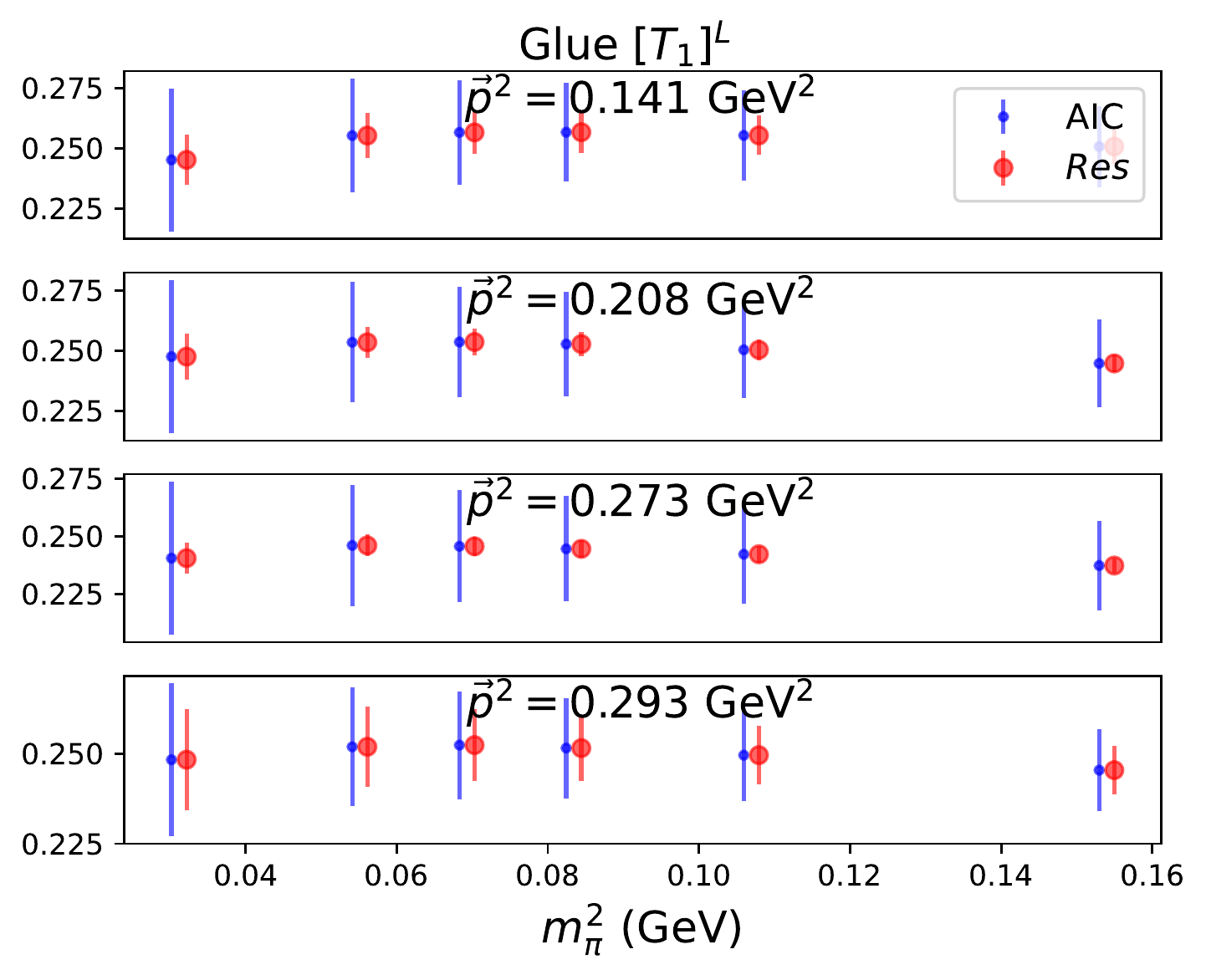}
  \includegraphics[page=1,width = 0.40 \textwidth]{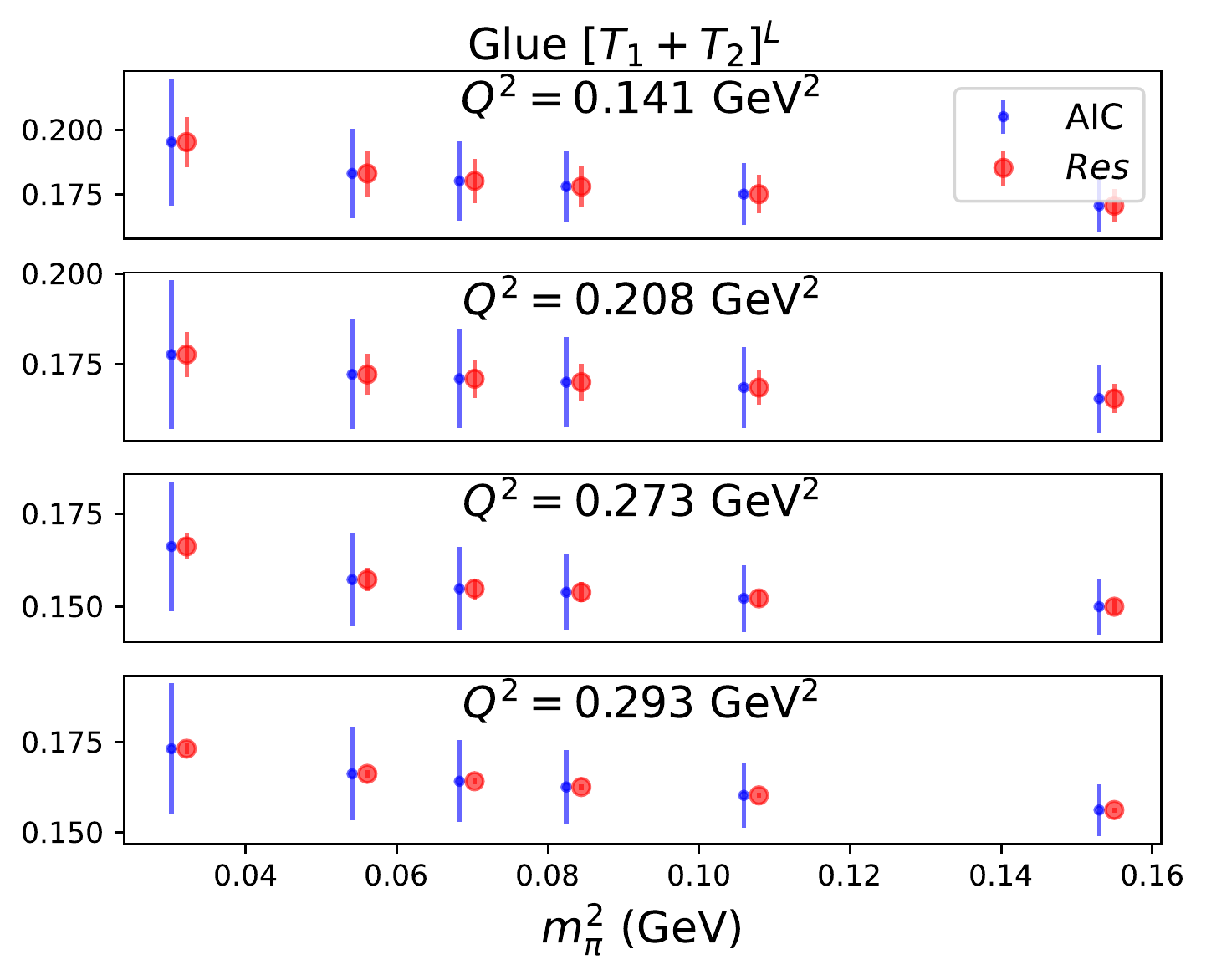}  \\
  \includegraphics[page=1,width = 0.40 \textwidth]{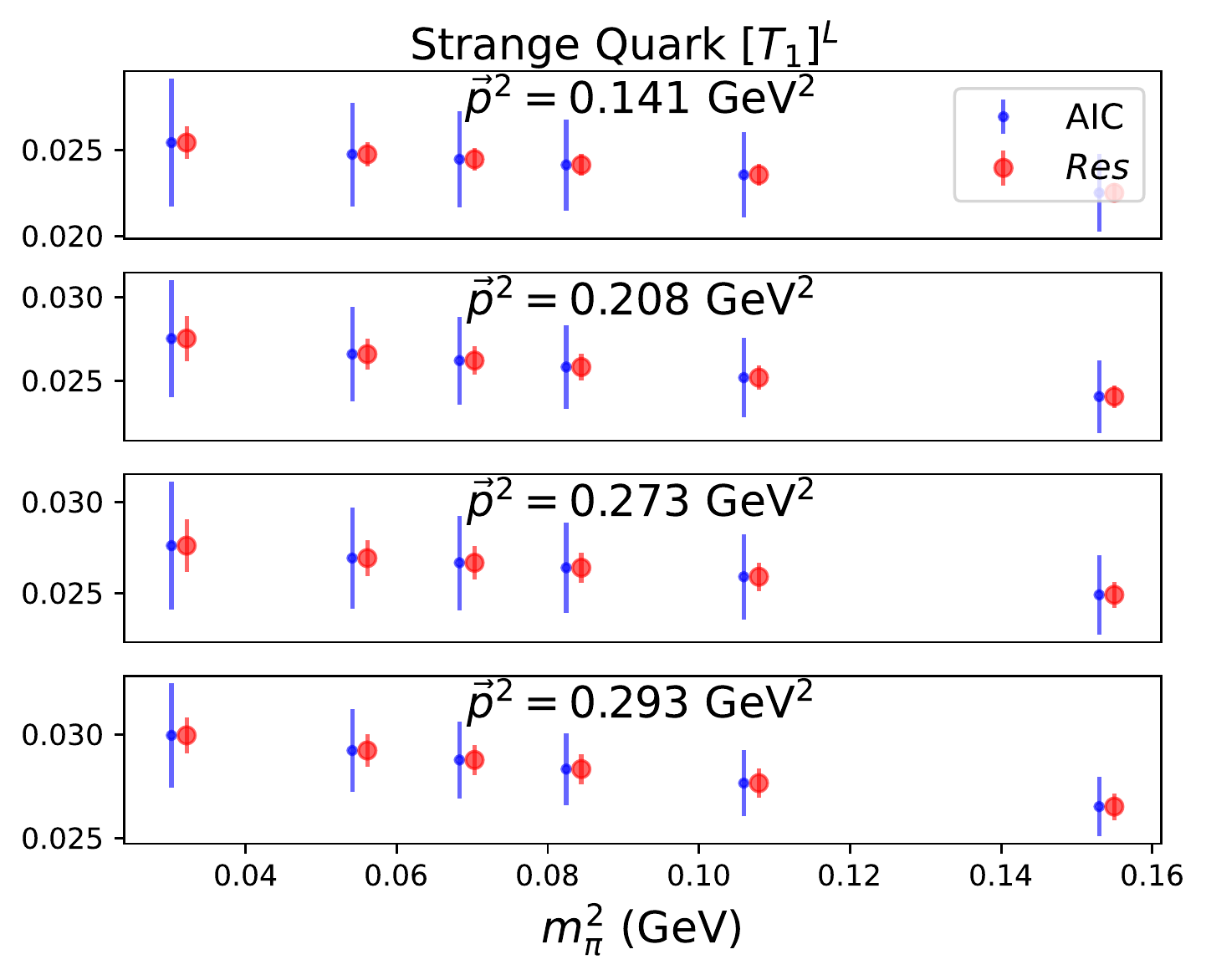}
  \includegraphics[page=1,width = 0.40 \textwidth]{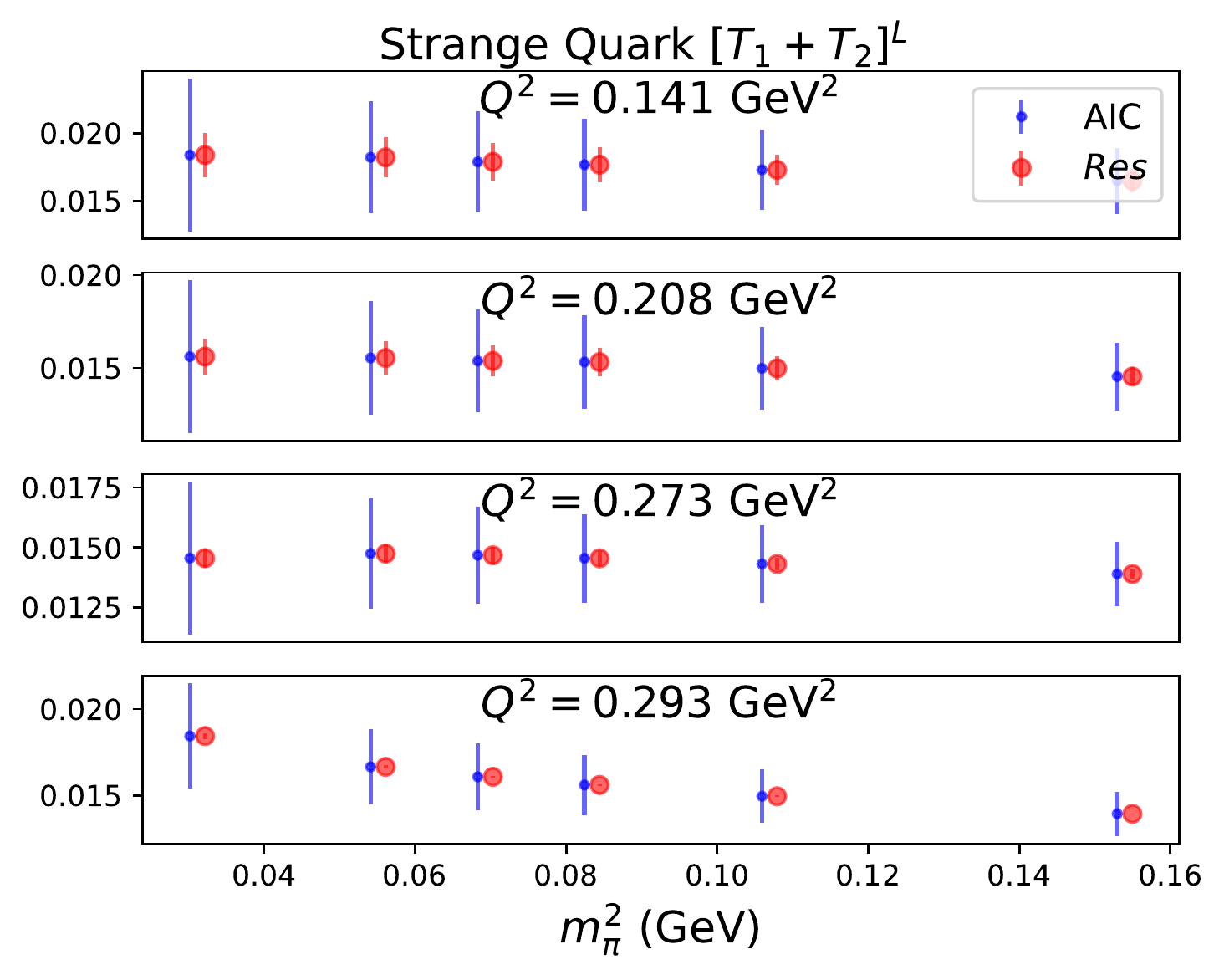}
  \caption{Comparison of AIC averaged values and $Res_{\rm{ max}}$ for $[T_1]^L$ and $[T_1+T_2]^L$ of glue (upper panels) and strange quark (lower panels) 3pts at $t\ppf=4, \tau=2$ at different valence pion masses. The different subplots correspond to different nucleon momenta $\vec{p}^2$ for $[T_1]^L$ and different $Q^2$ for $[T_1 + T_2]^L$. }
  \label{fig:cder_Res_1}
\end{figure}

Since we have gotten $M$ and $A$ from fitting the correlator, we could try to estimate a residue by the sum of the correlator after $r_{\rm{cut}}$ as 
\begin{eqnarray}
\begin{aligned}
Res = \sum_{r > r_{\rm{cut}}}^{r < r_{\rm{max}}} A r^{-\frac{3}{2}} e^{- M r},
\end{aligned}
\end{eqnarray}
in which the sum is over discrete points of the 3D volume and $r_{\rm{max}} \sim 4\, {\rm fm}$ for the current lattice.
For the glue DI, we get $Res = 0.0053(45)$ with $r_{\rm{cut}} = 1.5 \ {\rm{fm}}$ which gives the upper bound of the residue to be $Res_{\rm{max}} = 0.0053 + 0.0045 = 0.0098$.
This is smaller than the AIC error $0.025$.
And for the strange quark DI, we get $Res = 0.00068(95)$ which gives $Res_{\rm{max}} = 0.00068 + 0.00095 = 0.00163$.
This is also smaller than the AIC error $0.0035$.

We have gathered similar results for the glue DI and strange quark DI for both $T_1$ and $[T_1 + T_2]$ (light quarks DI have been omitted as they have similar behavior to the strange quark DI) at different pion masses, source momenta, and sink momenta in Fig.~\ref{fig:cder_Res_1}.
The blue points are the AIC averaged values and the red points are plotted with the same central values as the blue points with error bounds equal to $Res_{\rm{max}}$.
It can be seen that all of the residues are much smaller than the AIC errors except for a very few cases at small pion masses which are due to unstable fits of $M$ and $A$.
This confirms that our current way of estimating systematics for CDER with AIC is reliable.

}

{
\subsection{{\textit{z}}-Expansion fit }\label{sec3:sec_zexp}
In order to fit the $[T_1+T_2](Q^2)$ form factor and extrapolate it to $Q^2=0$, we have done a model-independent $z$-expansion~\cite{Lee:2015jqa} fit using the following equation with $k_{\rm{max}} \geq 2$:
\begin{eqnarray}\label{eq:z-exp}
\begin{aligned}
T(Q^2) &= \sum_{k=0}^{k_{\rm{max}}} a_k z^k \\
z(t,t_{\rm{cut}},t_0) &= \frac{\sqrt{t_{\rm{cut}}-t} - \sqrt{t_{\rm{cut}}-t_0}}{\sqrt{t_{\rm{cut}}-t} + \sqrt{t_{\rm{cut}}-t_0}} ,\\
\end{aligned}
\end{eqnarray}
where $T(Q^2)$ represents a nucleon form factor such as $T_1$, $T_2$, $D$ and their linear combinations such as $[T_1+T_2]$;
{$t = -Q^2$};
$t_{\rm{cut}} = 4 m_{\pi}^2$ correspond to the two-pion production threshold with $m_{\pi} = 172 \ {\rm{MeV}}$ chosen to be the sea pion mass;
and $t_0$ is chosen to be its ``optimal" value {$t_0^{\rm{opt}}(Q_{\rm{max}}^2) = t_{\rm{cut}}(1- \sqrt{1 + Q_{\rm{max}}^2/t_{\rm{cut}}})$} to minimize the maximum value of $|z|$, with $Q_{\rm{max}}^2$ the maximum $Q^2$ under consideration.
And we adopt the Gaussian prior proposed in~\cite{{Lee:2015jqa}} with $|a_k/a_0|_{\rm max}=5$ [use $a_k/a_0 = ``0(5)"$ (a Gaussian prior with central value $0$ and width $5$) for all $a_k (k>1)$ in the fits].
}

}

{
\section{Analysis and results}\label{sec:ana}

{
\subsection{Three-point correlation function fits}\label{sec4:3pt_fit}
We adopt the two-state fit formula to fit the quark/gluon ratio $R^{\mu \nu}_{\Gamma_\alpha}(t\ppf,\tau,\vec{p}\ppf,\vec{p}\ppi)$ in Eq.~(\ref{eq:sec5_case0})
\begin{eqnarray}\label{eq:secV_ratio_fit}
\begin{aligned}
R^{\mu \nu}_{\Gamma_\alpha}(t\ppf,\tau,\vec{p}\ppf,\vec{p}\ppi) = & A  + B_1 \, e^{-\Delta E_{p\ppf} (t\ppf - \tau)} \\
&  + B_2 \, e^{-\Delta E_{p\ppi} (\tau)}
  + B_3  \, e^{-\Delta E_{p\ppi} (\tau) - \Delta E_{p\ppf} (t\ppf - \tau)} ,
\end{aligned}
\end{eqnarray}
where $A$ is the ground-state matrix element, the terms with $B_1$, $B_2$, and $B_3$ are the contributions from the excited-state contamination, and $\Delta E_{p} = E^{1}_p-E_p$ is the energy difference between the nucleon ground-state energy $E_p$ and that of the first excited-state $E^{1}_p$. In order to stabilize the fit, we use $\Delta E_{p}$ from the fit of the 2pt as a prior for the 3pt fit with $\Delta E_{p} \in [300, 800] \ {\rm{MeV}}$.
The top panels of Fig.~\ref{fig:3pt_fit_CI_0} show sample plots for $T_1^{L}(0)$ for up quark CI, strange quark DI, and glue components.
We treat up and down quark DI contributions to be the same since we have exact isospin symmetry in the current simulation.
We have used the energy difference $\Delta E$ from 2pt to constrain our fits of Eq.~(\ref{eq:secV_ratio_fit}).
The source-sink separations $t'=7,8,9,10$ and $t'=4,5,6,7,8,9$ are used for CI and DI fits, respectively.
And $4$ points are dropped (2 points close to the source $t=0$ and 2 points close to the sink $t\ppf$) for each separation.
The gray bands are the fitted results of $T_1^L$.

To check the convergence of the ground-state matrix elements,
we also calculate the differential summed ratio as
\begin{eqnarray}\label{eq:secV_summed_ratio}
\begin{aligned}
\tilde{R}(t\ppf) \equiv \frac{SR(t\ppf) - SR(t\ppf - \Delta t)}{\Delta t}
\end{aligned}
\end{eqnarray}
with $SR(t\ppf) \equiv \sum_{t_c < \tau < t\ppf - t_c} R(t\ppf, \tau), t_c = 2$ and fit $\tilde{R}(t\ppf)$ with a constant, given current statistics.
It has been shown that excited-state contamination is better suppressed compared to the above two-state fit~\cite{Chang:2018uxx,Zhang:2021oja,He:2021yvm}.
The corresponding sample fits are shown in the lower panels of Fig.~\ref{fig:3pt_fit_CI_0}.
Simple linear fits are used to extrapolate ground-state matrix elements which are marked as gray bands.
We observe that the error of the gray band shown in the lower left panel for the up quark CI component is significantly smaller than that of each data point,
whereas for each of the strange quark DI and glue components (shown in the lower middle and right panels, respectively) the final fit error is similar to that of the smallest $t\ppf$, which has the best statistics compared with larger $t\ppf$.
The difference comes mainly from the fact that the data is negatively correlated for the lower left panel and positively correlated for the lower middle and right panels.
To understand this intuitively, consider a simple correlated data-averaging model in which
\begin{eqnarray}
\begin{aligned}
\bar{x} = \frac{1}{n} \sum_i^{n} x_i, \quad \sigma_{\bar{x}}^2 = \sigma_x^2   \frac{1}{n} (1+ \frac{1}{n-1} b),
\end{aligned}
\end{eqnarray}
where all $x_i$ have variance $\sigma_x^2$ and the data correlation between different $x_i$ is chosen to be $b$.
For a negative value of $b$, the final variance $\sigma_{\bar{x}}^2$ is smaller than $\sigma_x^2$ with an additional enhancement factor $1+ \frac{1}{n-1} b$ compared to the average of uncorrelated data.
On the other hand, a positive value of $b$ increases the final variance compared to the uncorrelated case.
This partially explains the observed difference, although we are using correlated fits instead of direct averaging.

The final predictions of the two-state fits and the differential summed-ratio fits are labeled on each panel and they agree very well with each other within errors.
Similar behaviors are observed with different valence pion masses for the down quark CI and up/down quark DI components and $[T_1 + T_2]^{L}(Q^2)$ form factors at different $Q^2$.
This confirms that we have good control of excited-state contamination under our current statistics.
We will focus on the two-state fits for the following discussions and estimate systematic uncertainty from excited-state contamination based on the difference between the final predictions from the two-state fit results and the differential summed-ratio fit results.

\begin{figure}[htbp]
  \centering
  \includegraphics[width = 0.32 \textwidth]{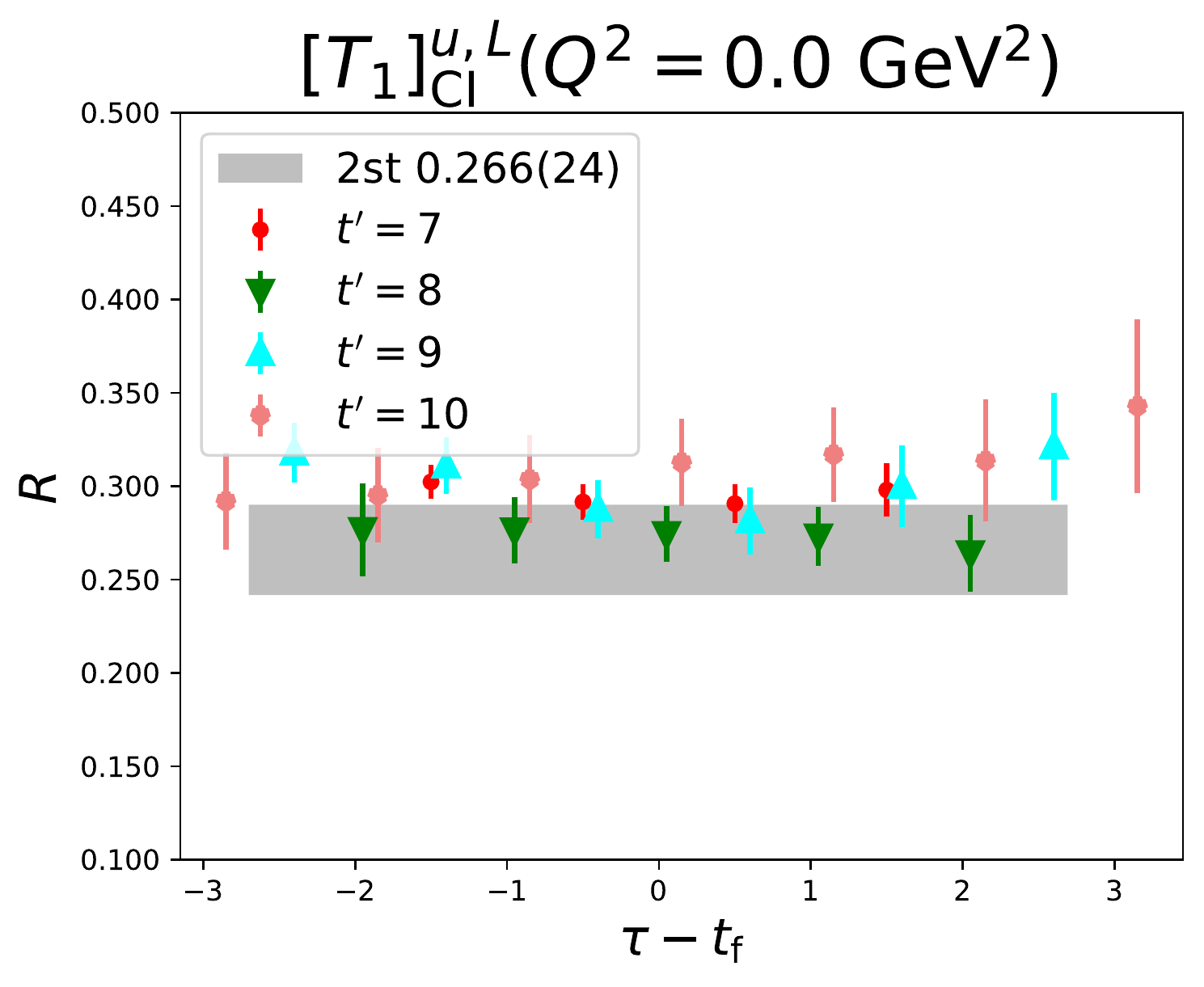}
  \includegraphics[width = 0.32 \textwidth]{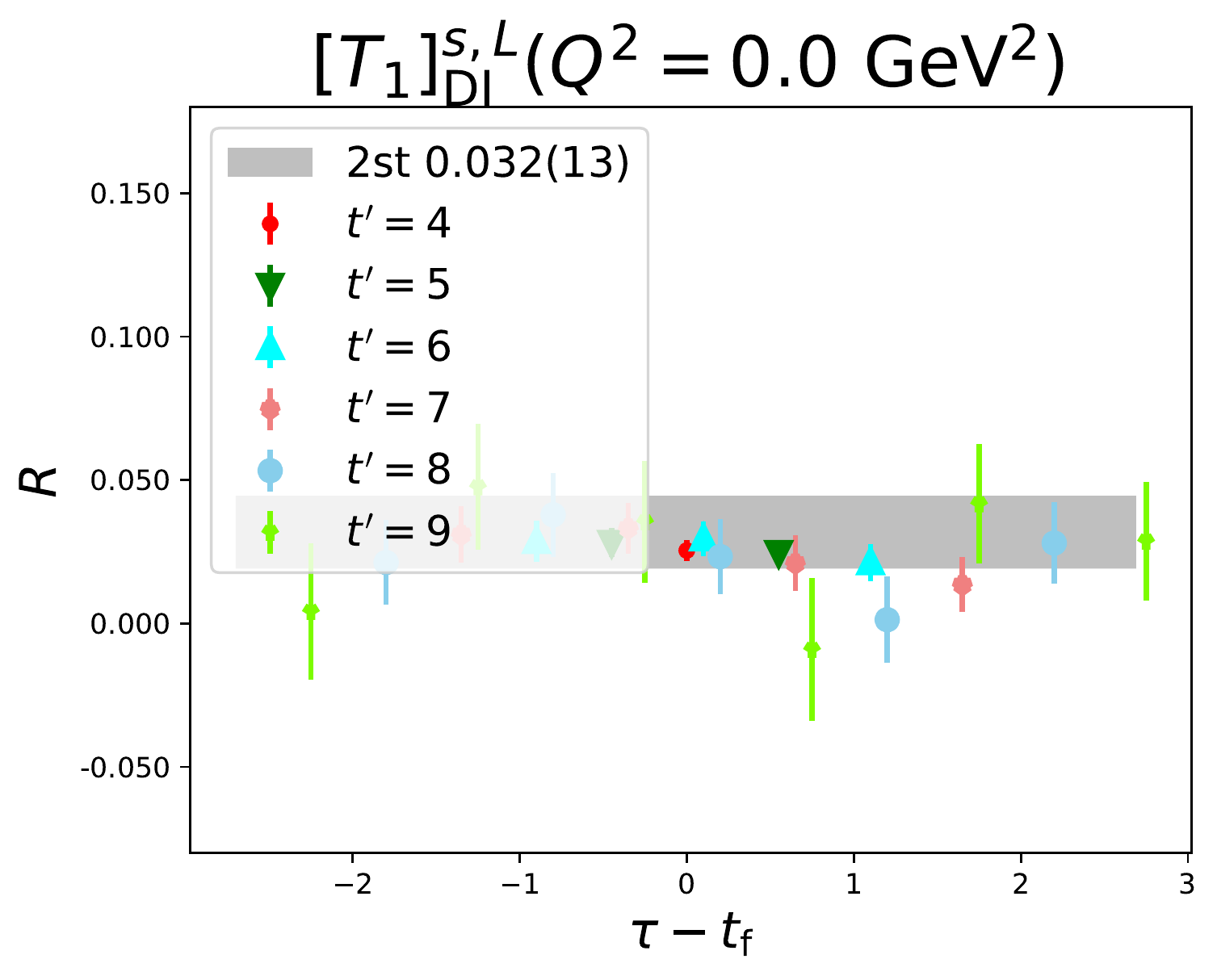}
  \includegraphics[width = 0.32 \textwidth]{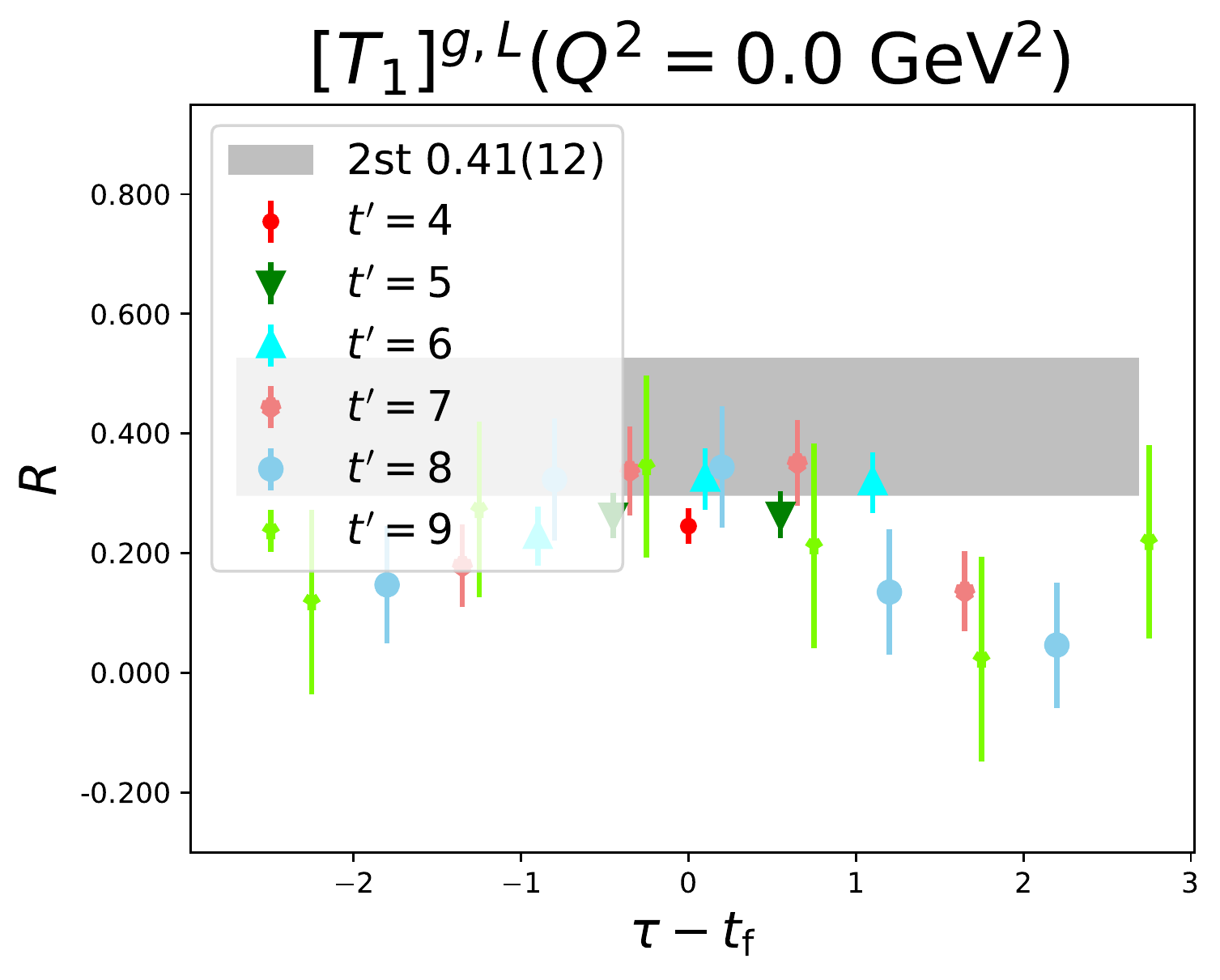}  \\
  \includegraphics[width = 0.32 \textwidth]{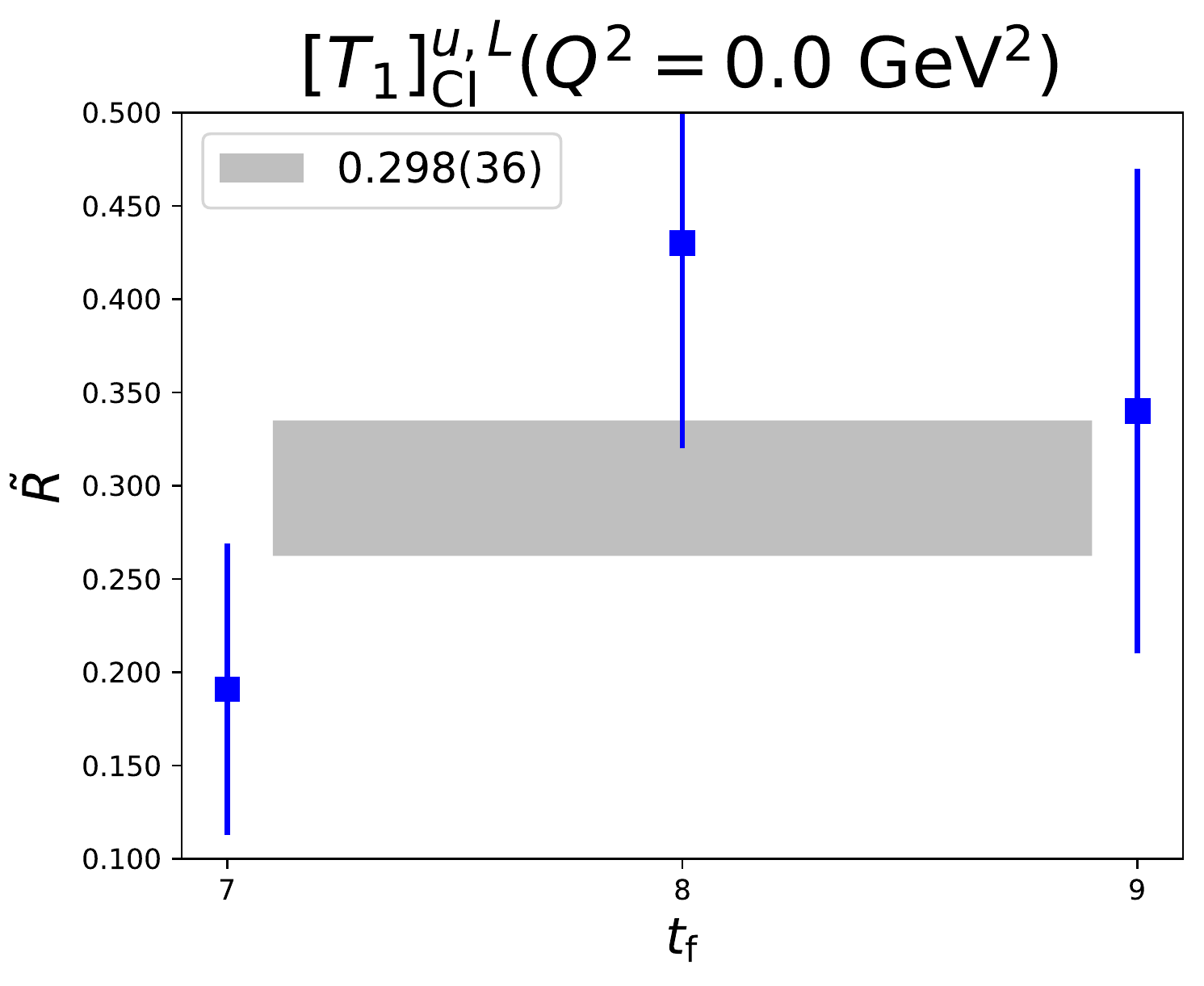}
  \includegraphics[width = 0.32 \textwidth]{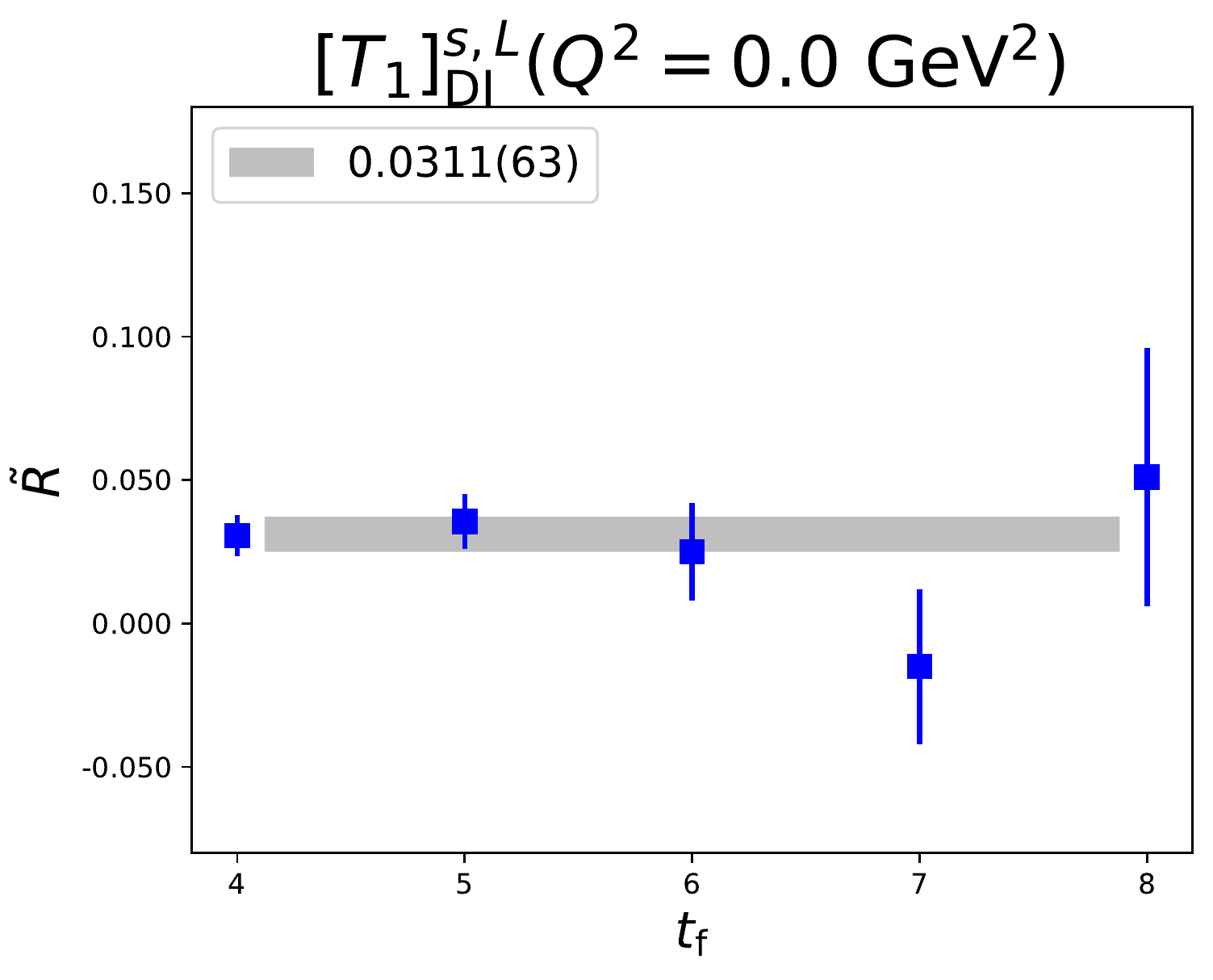}
  \includegraphics[width = 0.32 \textwidth]{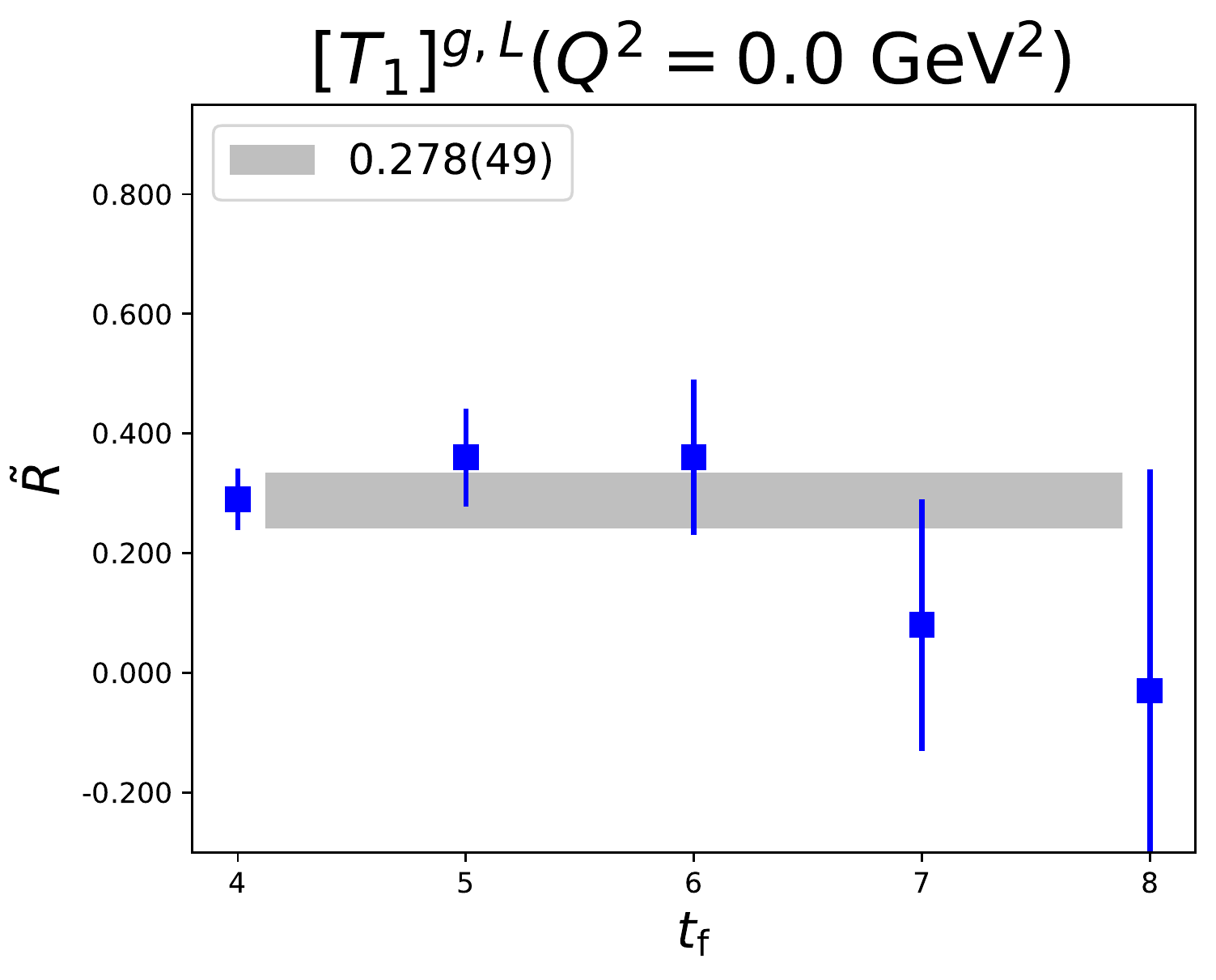}    
  \caption{The ratio $R(\tau,t\ppf)$ (top panels) and $\tilde{R}(t\ppf)$ (bottom panels) defined through Eq.~(\ref{eq:secV_ratio_fit}) and Eq.~(\ref{eq:secV_summed_ratio}), respectively.
The data of up quark CI, strange quark DI, and glue components with valence pion mass $174\ {\rm{MeV}}$ are shown in the left, middle, and right panels, respectively.
The gray bands are the fit predictions of the ground-state matrix elements $T_1(0)$ of each component.}
  \label{fig:3pt_fit_CI_0}
\end{figure}

}
{
\subsection{Form factor fits}\label{sec4:ff_fit}

We repeat the above procedure for all of the other cases.
The results of
$T_1^{L}(0)$ for up quark CI, down quark CI, $u$/$d$ quark DI, strange quark DI, and glue components as a function of nucleon momenta $\vec{p}^2$
are shown in Fig.~\ref{fig:ff_fit_T1}.
As shown in Eq.~(\ref{eq:sec5_case2}), the calculation of the $T_1^{L}$ form factor using the operator $\mathcal{T}_{4i}$ can only be done at $\vec{p} \neq \vec{0}$.
This is why there is not a point at $\vec{p}^2=0$ in each of the panels of Fig.~\ref{fig:ff_fit_T1}.
It can be seen that the results for $T_1^{L}(0)$ from different $\vec{p}^2$ are consistent with each other within uncertainty.
Thus we use a simple constant fit of the data points to give the final predictions which are marked as blue bands.
The fits of the $[T_1+T_2]^{L}(Q^2)$ form factors are shown in Fig.~\ref{fig:ff_fit_T12}.
As shown in Eq.~(\ref{eq:sec5_case0}), the calculation of the $[T_1+T_2]^{L}$ form factor using operator $\mathcal{T}_{4i}$ can also only be done at $\vec{p} \neq \vec{0}$.
Thus, we use the $z$-expansion defined in Eq.~(\ref{eq:z-exp}) to fit the data points and extrapolate to $Q^2=0$ to get $[T_1+T_2]^{L}(0)$ for each component.

\begin{figure}[htbp]
  \centering
  \includegraphics[page=1,width = 0.32 \textwidth]{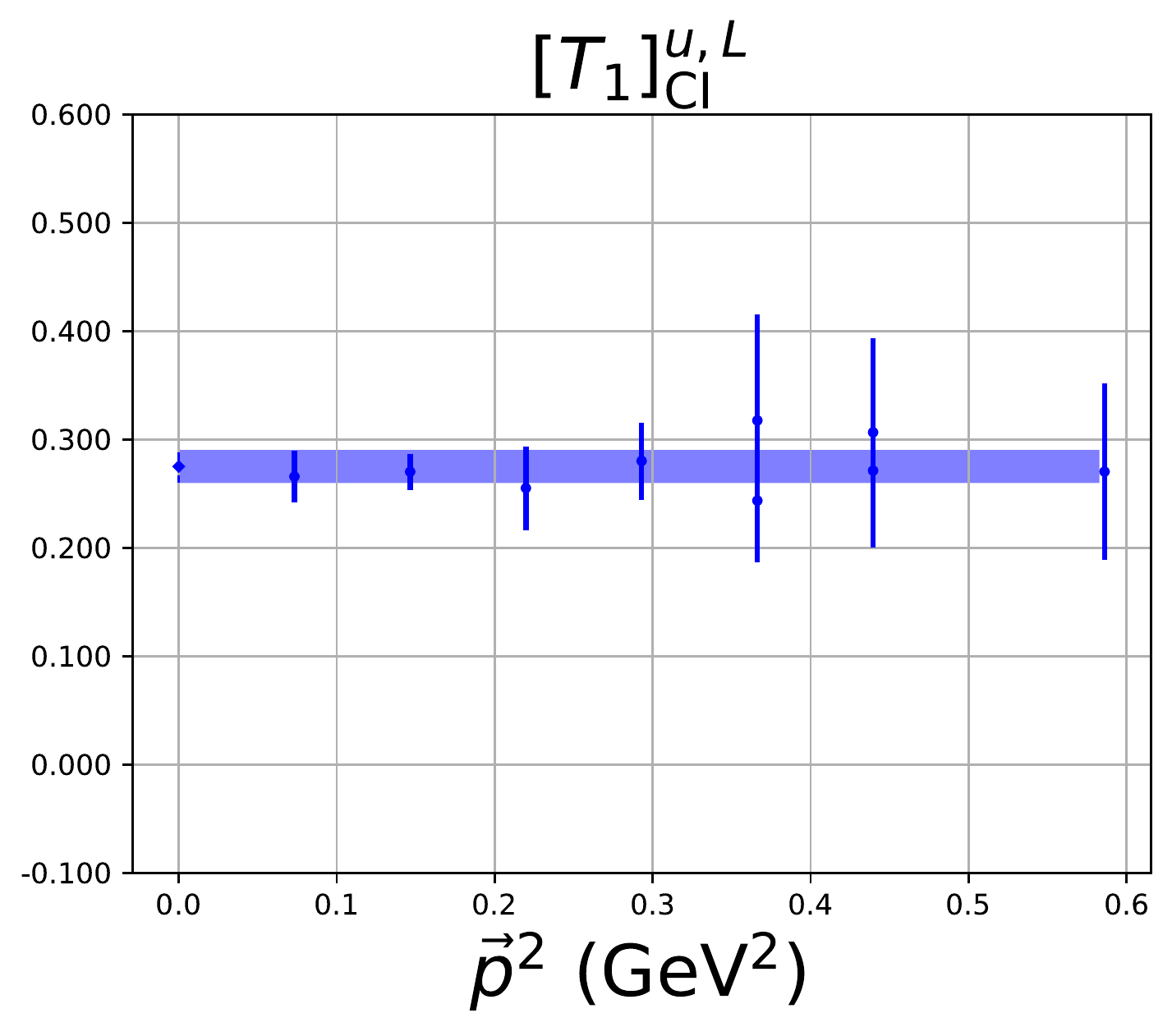} 
  \includegraphics[page=1,width = 0.32 \textwidth]{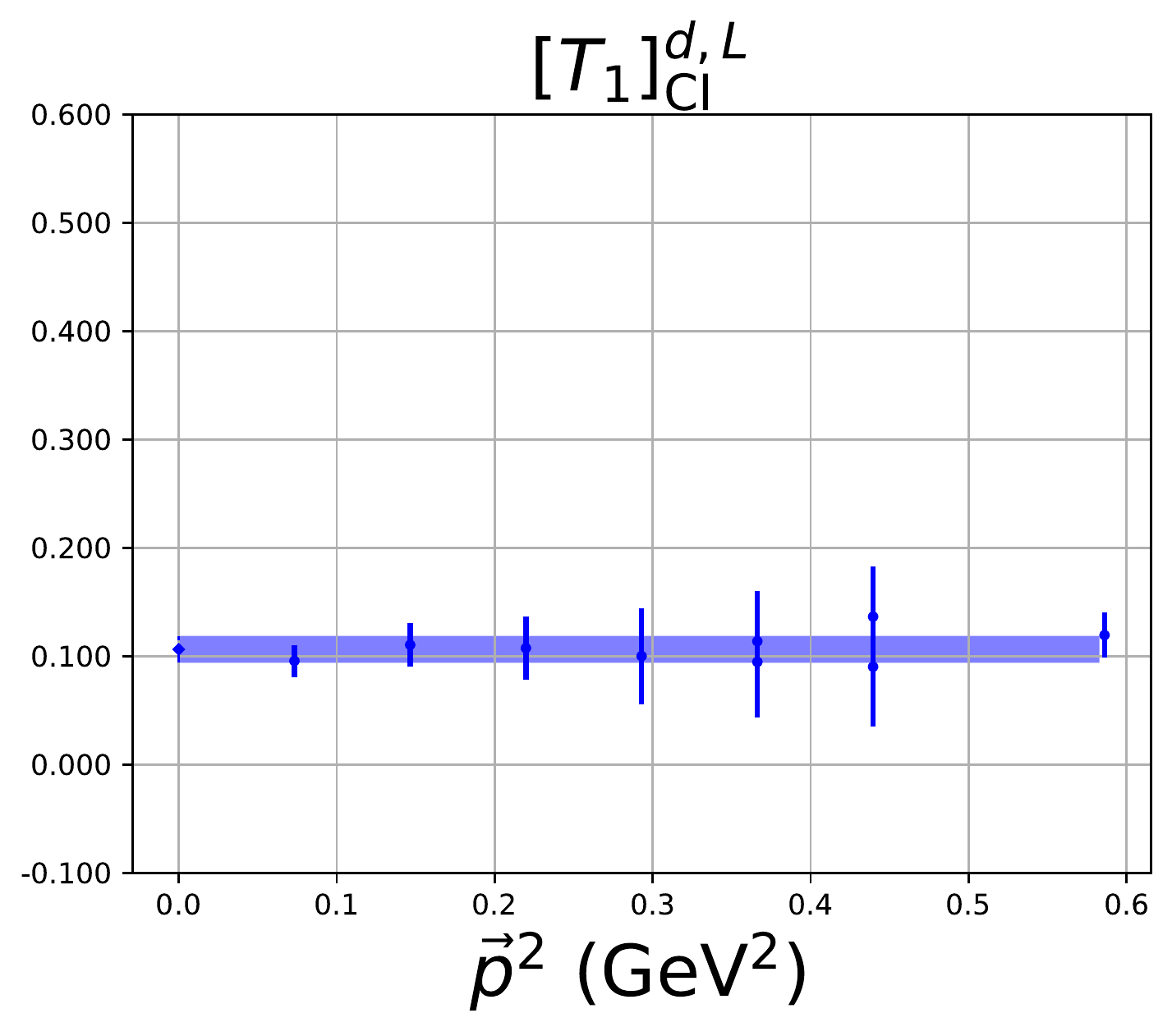} 
  \includegraphics[page=1,width = 0.32 \textwidth]{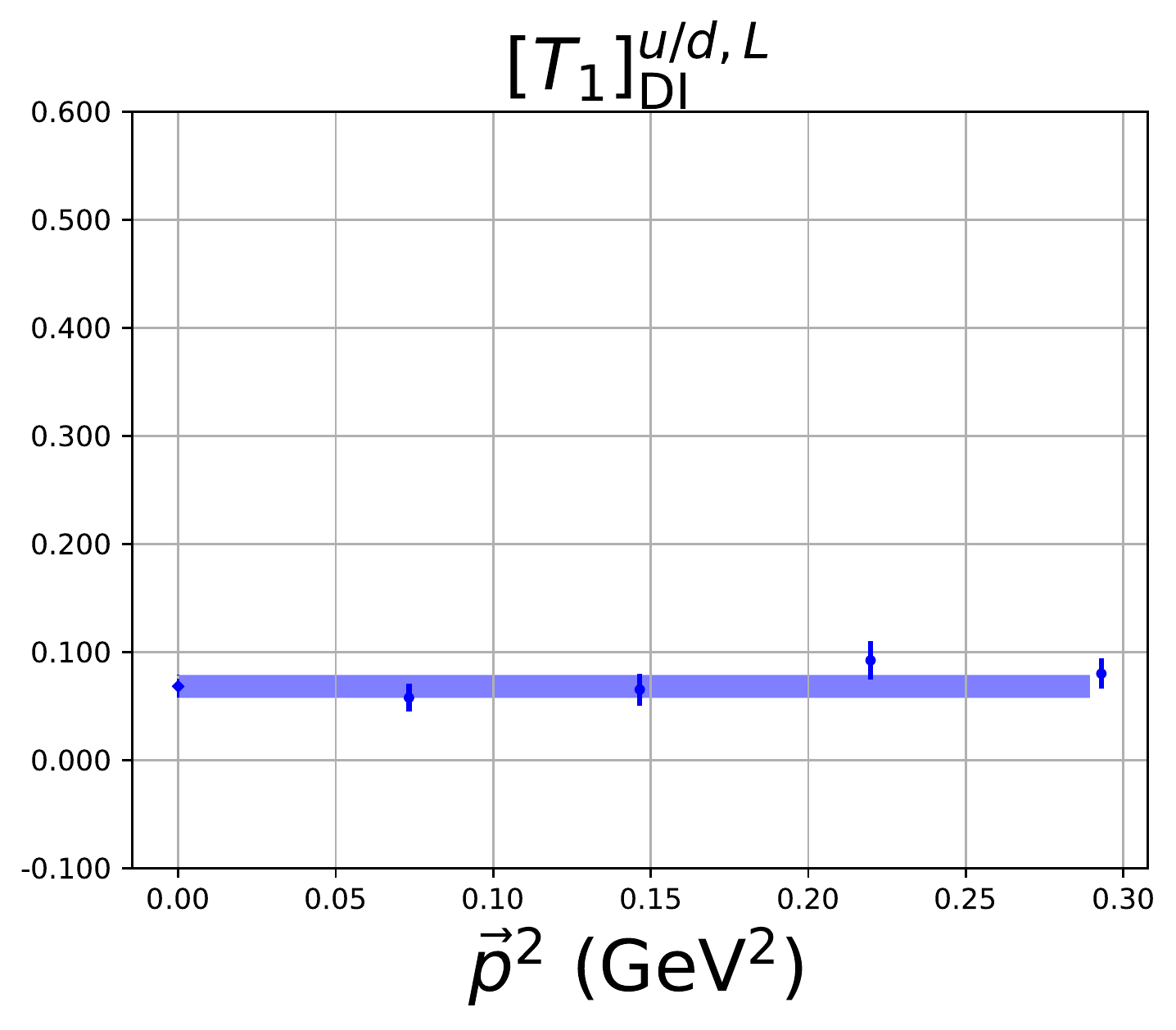}
  \includegraphics[page=1,width = 0.32 \textwidth]{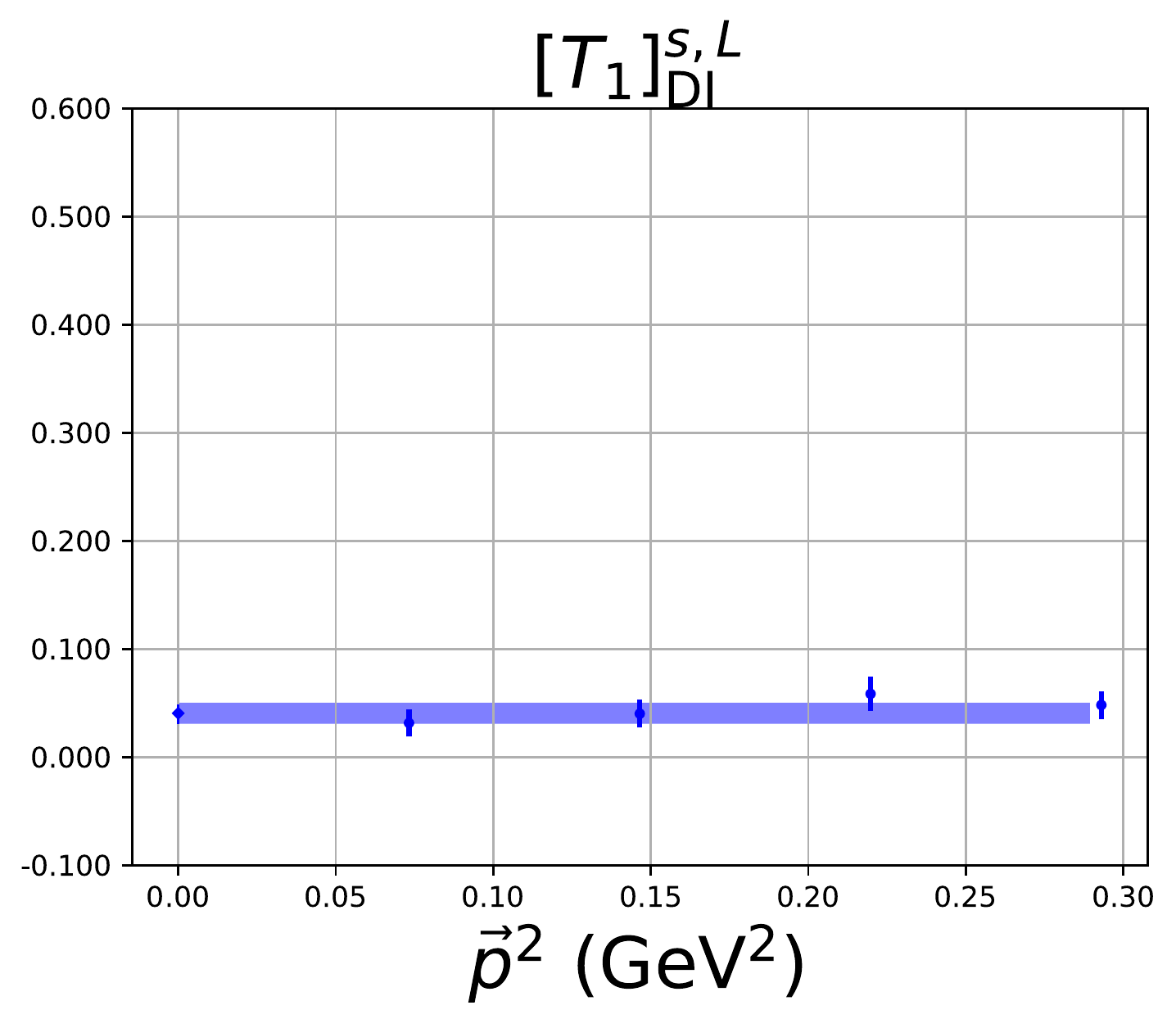} 
  \includegraphics[page=1,width = 0.32 \textwidth]{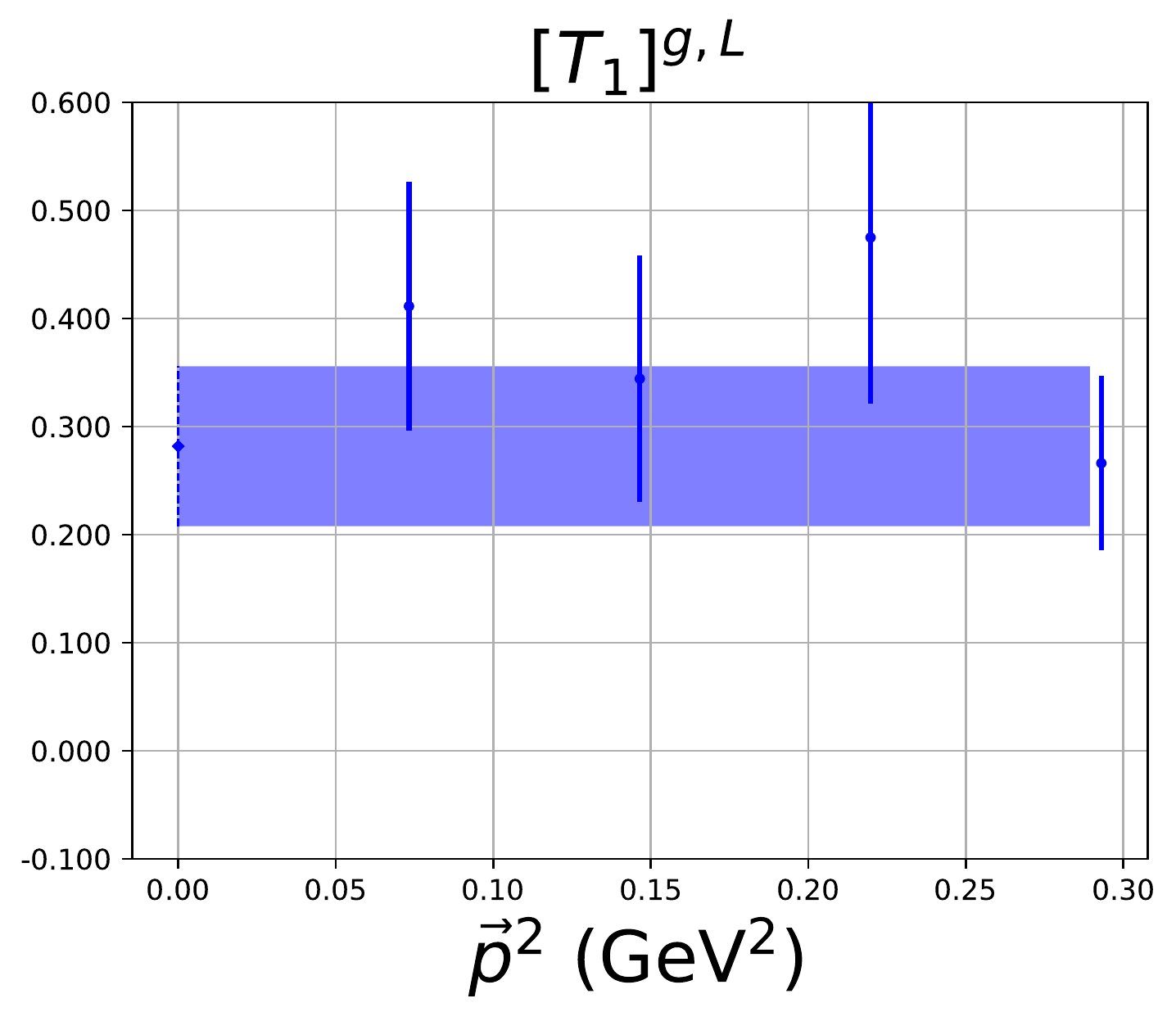} 
  \caption{Plots of $T_1^{L}(0)$ for the up quark CI, down quark CI, $u$/$d$ quark DI, strange quark DI, and glue components as a function of nucleon momenta $\vec{p}^2$ with valence pion mass $174\ {\rm{MeV}}$. In each plot, the blue band corresponds to a constant fit of the data points with each final result marked with a dashed line at $\vec{p}^2 = 0$.}
  \label{fig:ff_fit_T1}
\end{figure}

\begin{figure}[htbp]
  \centering
  \includegraphics[page=1,width = 0.32 \textwidth]{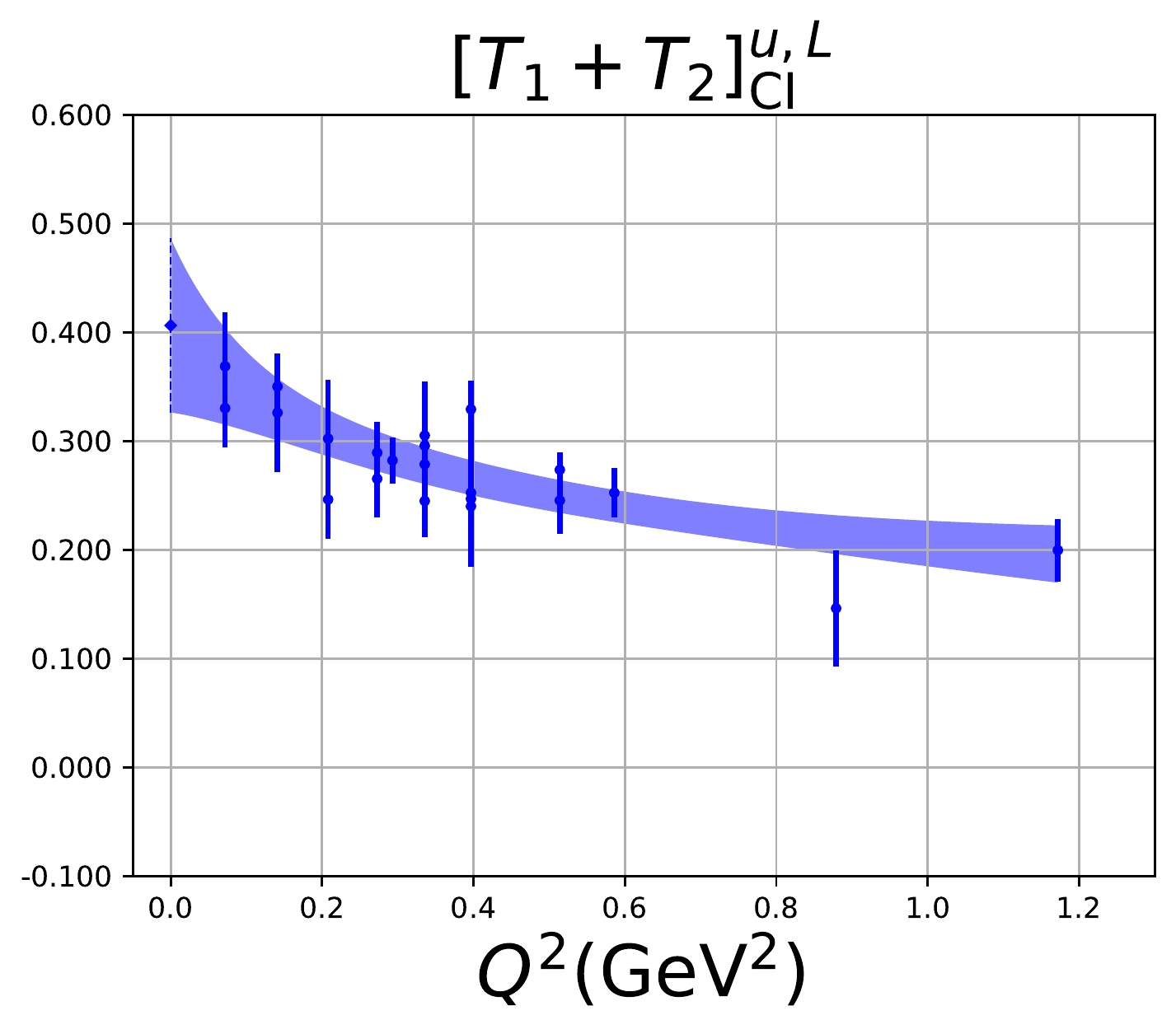} 
  \includegraphics[page=1,width = 0.32 \textwidth]{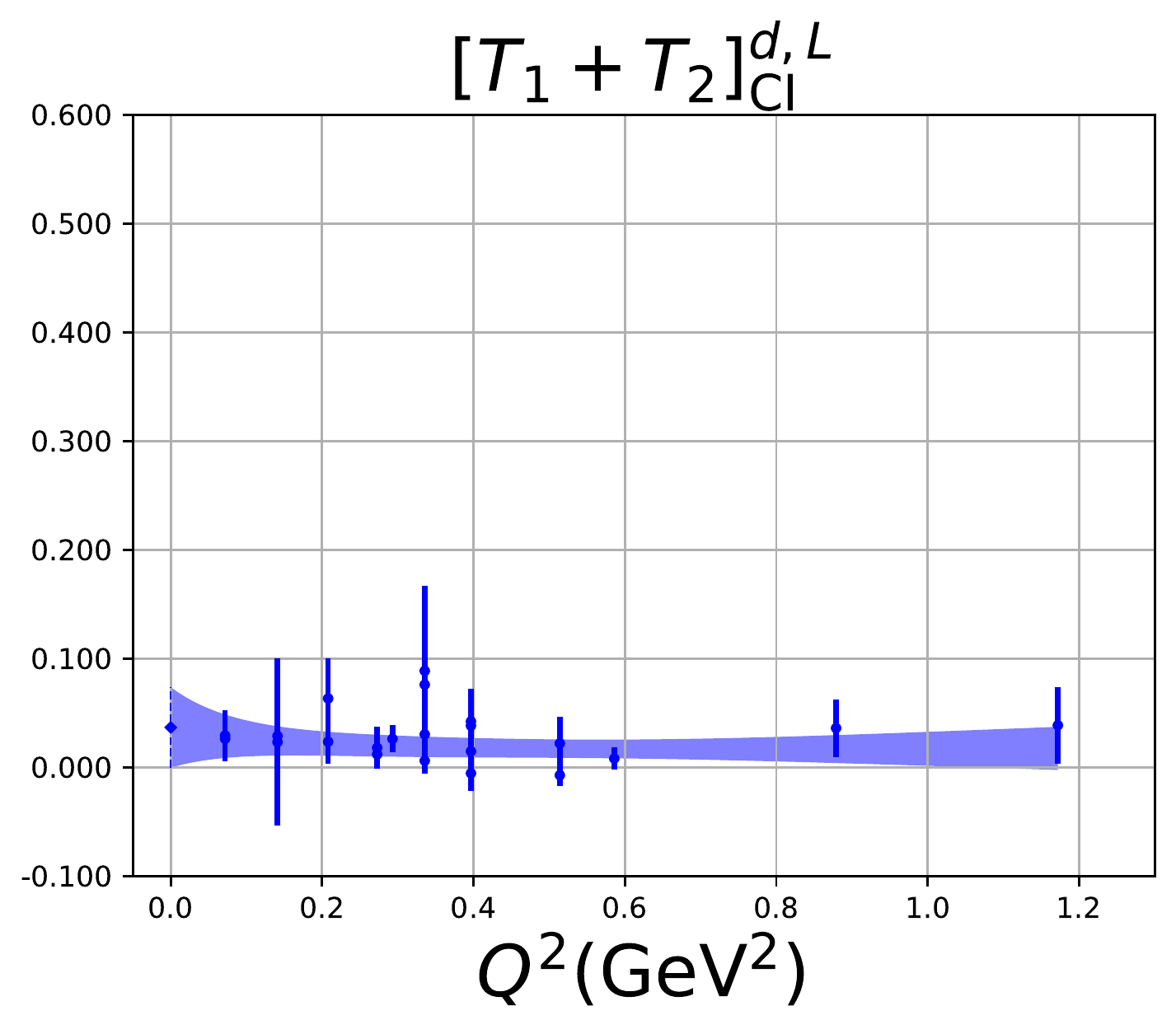} 
  \includegraphics[page=1,width = 0.32 \textwidth]{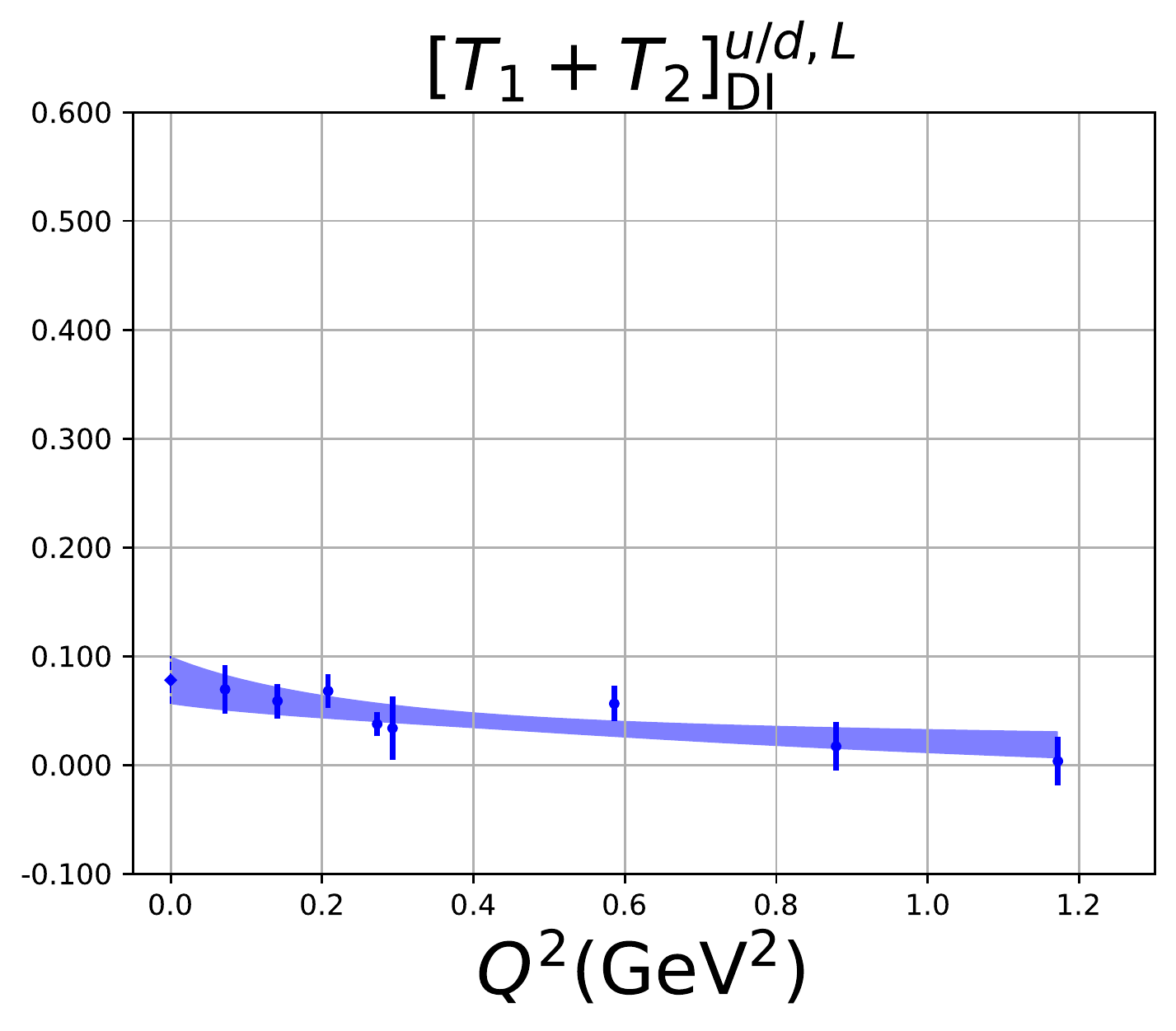}
  \includegraphics[page=1,width = 0.32 \textwidth]{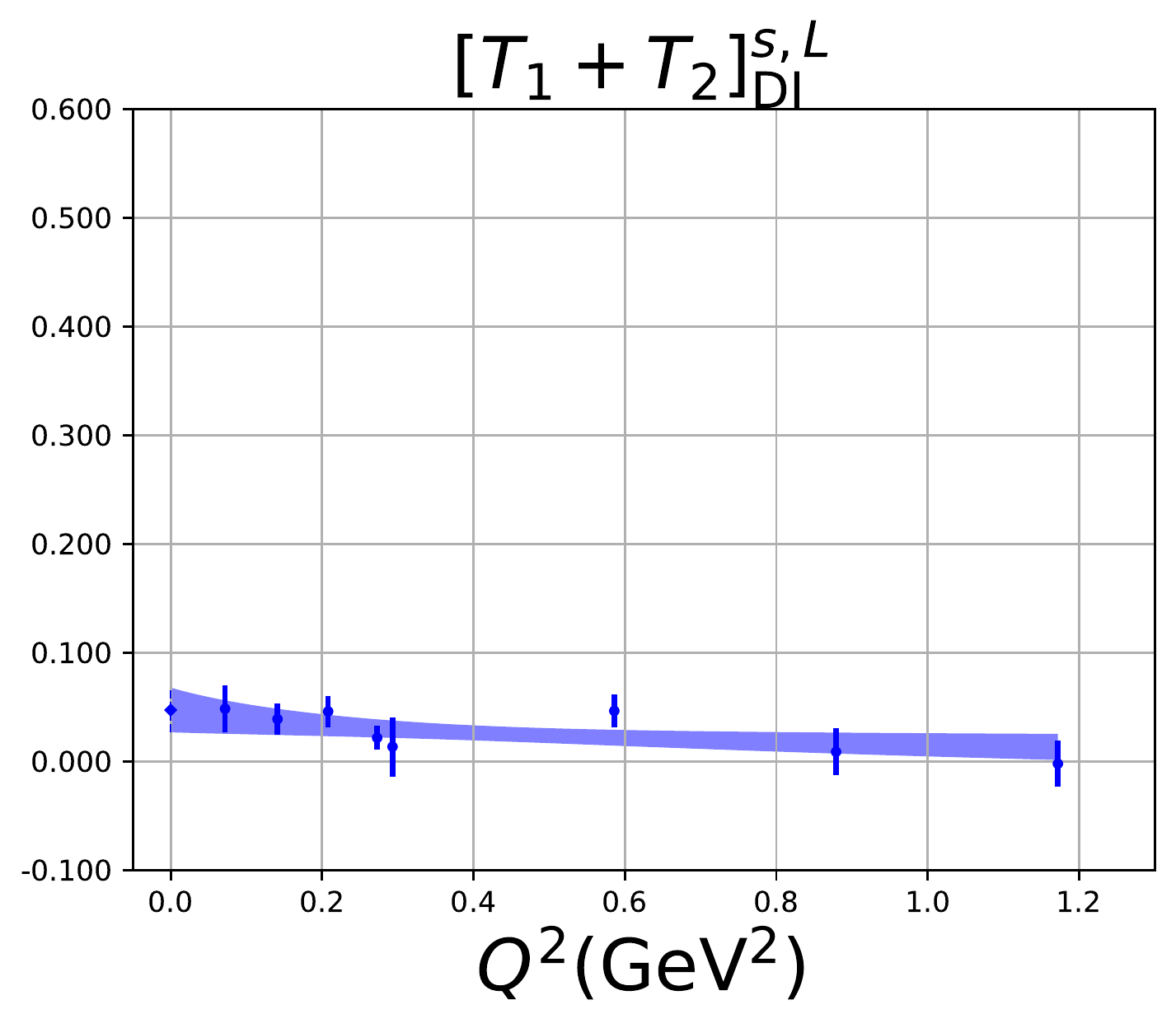} 
  \includegraphics[page=1,width = 0.32 \textwidth]{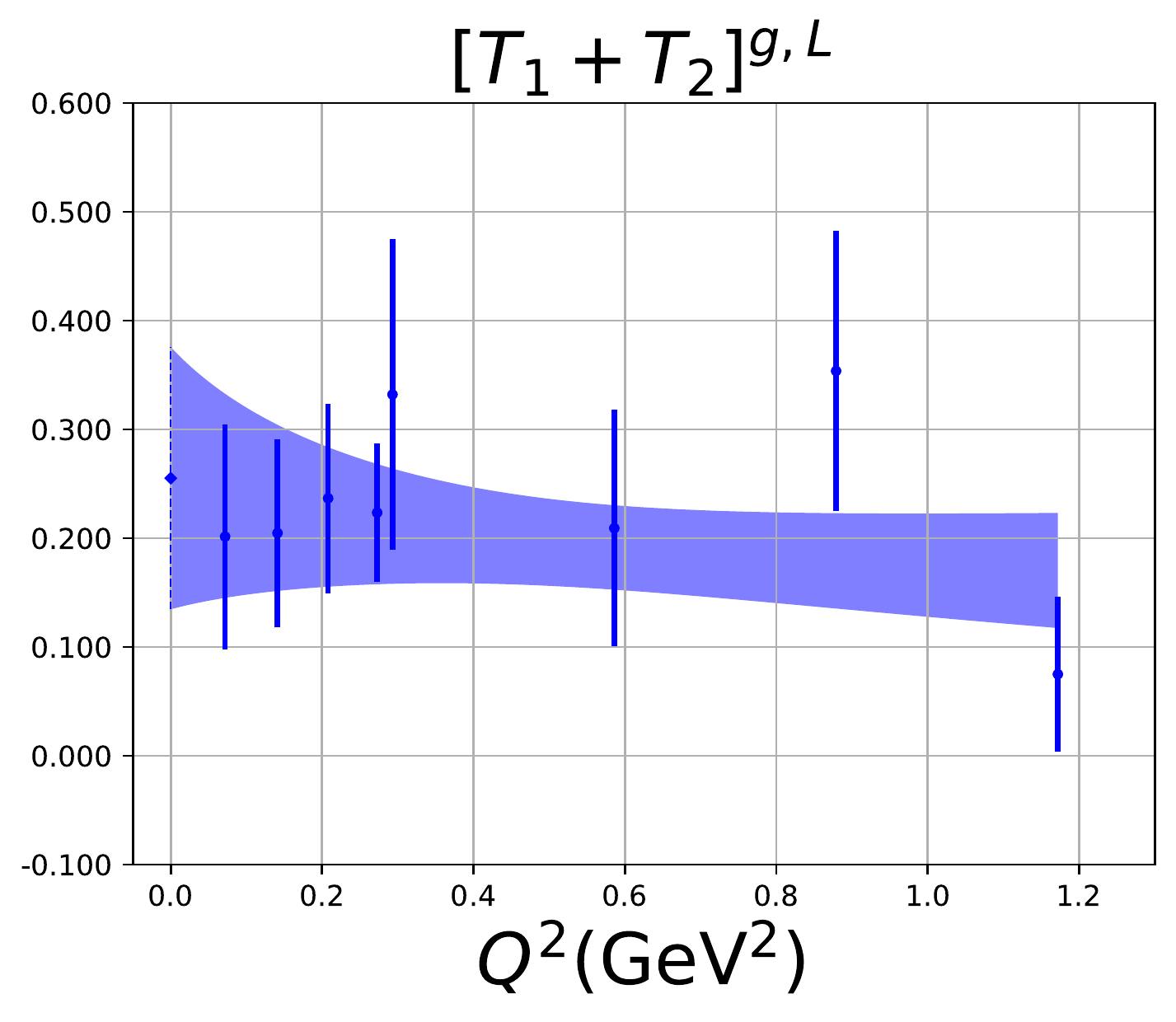} 
  \caption{Plots of $[T 1 + T 2]^{L}(Q^2)$ form factors for the up quark CI, down quark CI, $u$/$d$ quark DI, strange quark DI, and glue components as a function of $Q^2$ with valence pion mass $174\ {\rm{MeV}}$. In each plot, the band corresponds to the $z$-expansion fit with $k_{\rm{max}} = 2$ to extrapolate to $Q^2 = 0$ which is marked with a dashed line.}
  \label{fig:ff_fit_T12}
\end{figure}

}

{
\subsection{Final results}

Repeating the analysis for different valence pion masses, we gather the results of $T_1^{L}(0)$ and $[T_1+T_2]^{L}(0)$ at different valence pion masses without renormalization and normalization in Fig.~\ref{fig:3pt_fit_bar}. We see clear signals for all components. 
The renormalized results of $T_1^{R}(0)$ and $[T_1+T_2]^{R}(0)$ at $\overline{{\rm{MS}}}(\mu = 2 \ {\rm{GeV}})$ with Eq.~(\ref{eq:sec4_renorm}) are shown in Fig.~\ref{fig:3pt_fit_re_A}.
The $T_2^{R}(0)$ after renormalization calculated with $[T_1+T_2]^{R}(0) - T_1^{R}(0)$ are shown in Fig.~\ref{fig:3pt_fit_re_T2}.
It can been seen that $T_2^{q,R}$ and $T_2^{g,R}$ have almost no signals.
However, the total $T_2$ is consistent with zero without normalization.
Since the normalization condition Eq.~(\ref{eq:nor_condistion1}) is proportional to the $T_2^{R}$ form factor which will lead to unstable results, 
we choose to use the same normalization for the quarks and the glue in Eq.~(\ref{eq:nor_condition_Z}) to normalize our final results under current statistics.
Also, it can be seen that all components are quite linear in $m_{\pi}^2$; thus we perform a joint fit as
\begin{eqnarray}
\begin{aligned}
{T_1}^{i,R}   = a_1^i + a_2^i m_{\pi}^2, \quad
[T_1+T_2]^{i,R} = b_1^i + b_2^i m_{\pi}^2, \\
\end{aligned}
\end{eqnarray} 
with $i$ denoting each component (up quark CI, down quark CI, $u$/$d$ quark DI, strange quark DI and glue components) and the $a^i$s and $b^i$s are free parameters for fitting and the sum of the $T_1^R$ and $[T_1 + T_2]^R$ satisfy the constraints from Eq.~(\ref{eq:nor_condition_Z}) as 
\begin{eqnarray}
\begin{aligned}
\frac{\sum_i {T_1}^{i,R}}{N_1^L + N_2^L m_{\pi}^2} = 1 , \quad
\frac{\sum_i [T_1+T_2]^{i,R}}{N_1^L + N_2^L m_{\pi}^2} = 1, \\
\end{aligned}
\end{eqnarray} 
with $N_1^L$ and $N_2^L$ also free parameters for fitting.
The joint fit results are shown in Fig.~\ref{fig:3pt_fit_re_A} with $\chi^2/d.o.f.\sim 1.1$. 
It can be seen from the right panel of Fig.~\ref{fig:3pt_fit_re_B} that the sum of momentum fractions and angular momentum fractions are consistent with each other within current uncertainty which confirms our assumption in Eq.~(\ref{eq:nor_condition_Z}) of using one normalization constant $N^L$ at the present stage.

\begin{figure}[htbp]
  \centering
  \includegraphics[page=1,width = 0.45 \textwidth]{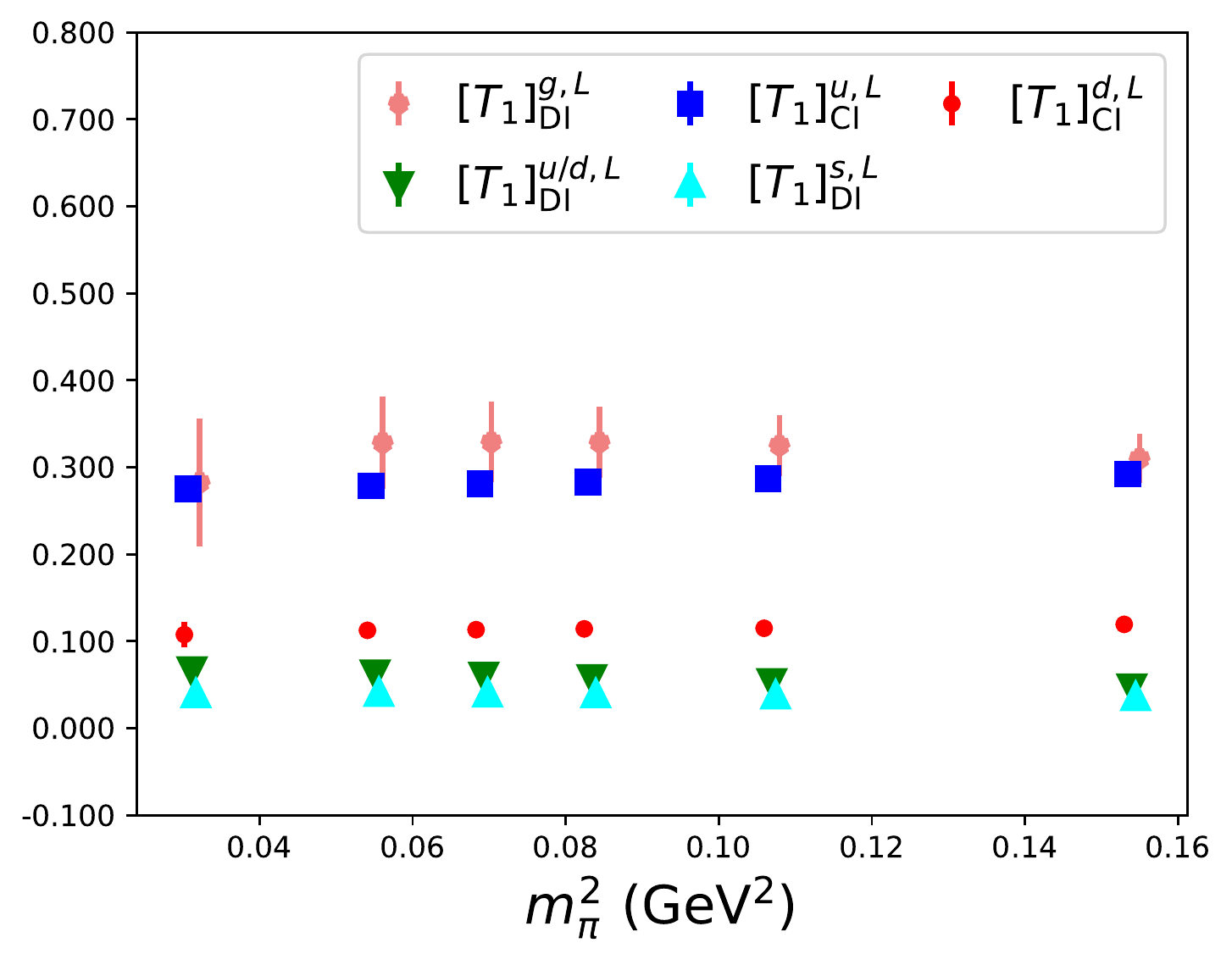} 
  \includegraphics[page=1,width = 0.45 \textwidth]{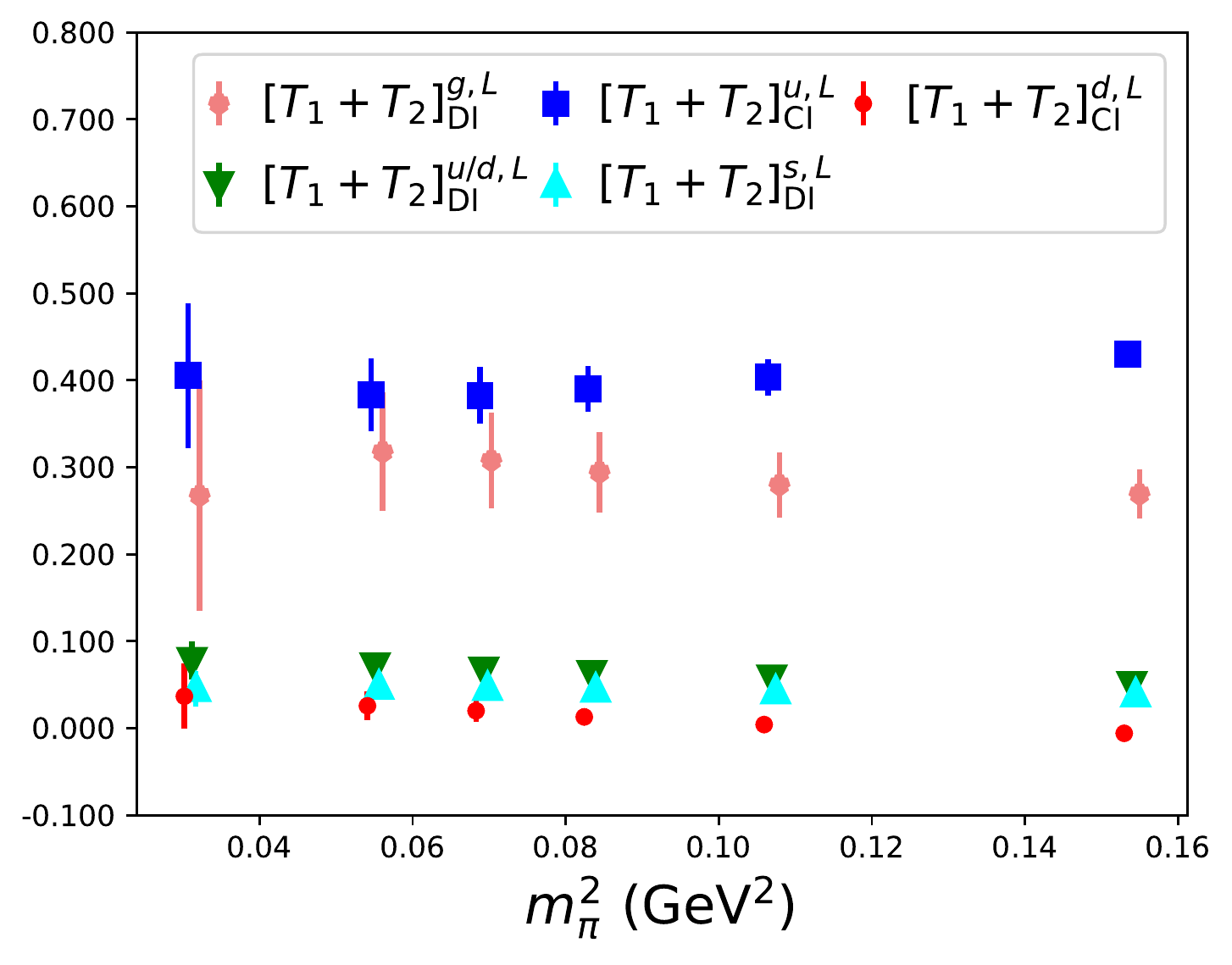} 
  \caption{Plots of the $T_1^{L}(0)$ (left panel) and $[T_1+T_2]^{L}(0)$ (right panel) at different valence pion masses without renormalization and normalization. Different colors correspond to up quark CI, down quark CI, up/down quark DI, strange quark DI and glue components.}
  \label{fig:3pt_fit_bar}
\end{figure}

\begin{figure}[htbp]
  \centering
  \includegraphics[page=1,width = 0.45 \textwidth]{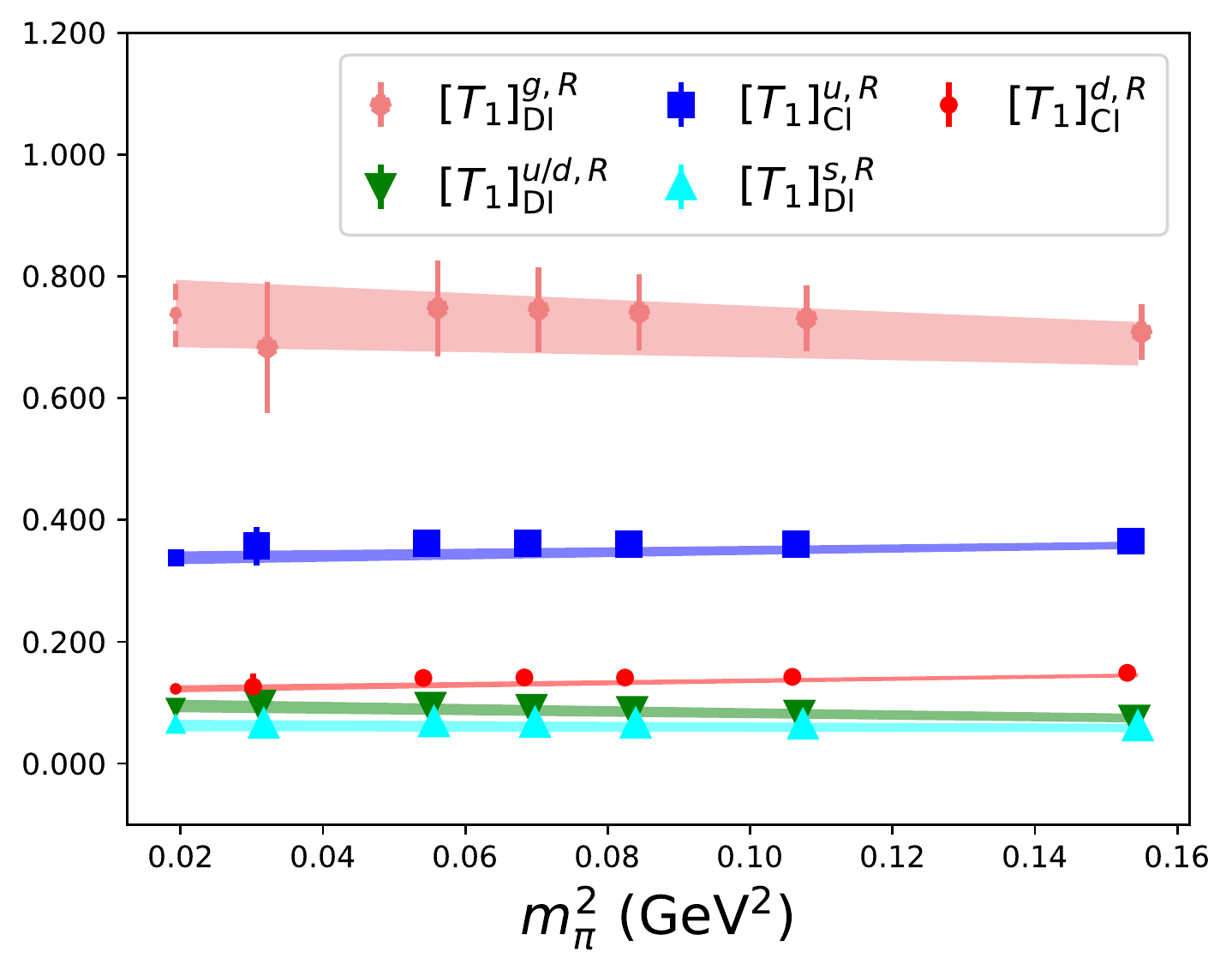} 
  \includegraphics[page=1,width = 0.45 \textwidth]{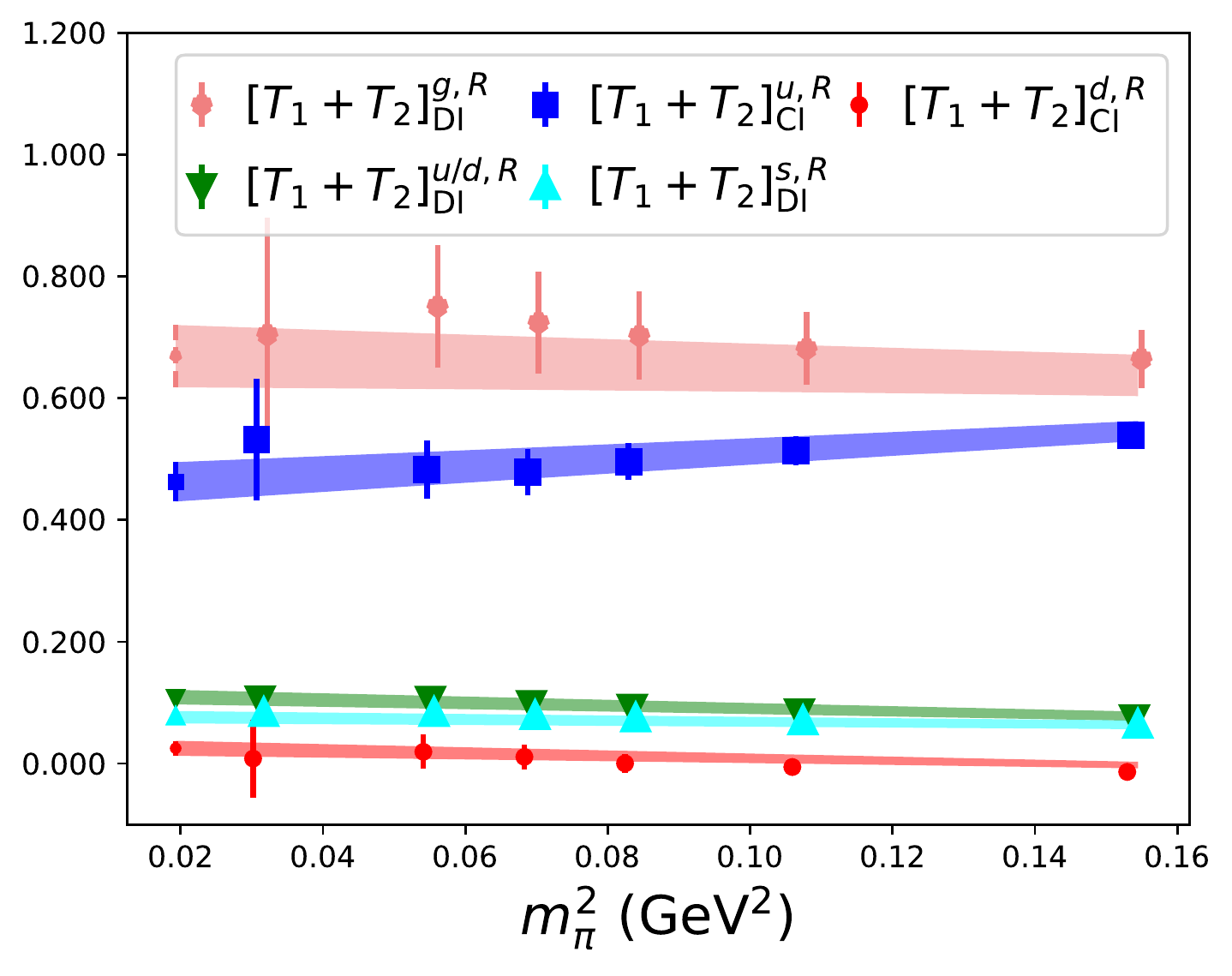} 
  \caption{Plots of the $T_1^{R}(0)$ (left panel) and $[T_1+T_2]^{R}(0)$ (right panel) at different valence pion masses after renormalization but without normalization.
Different colors correspond to the up quark CI and DI, down quark CI and DI, strange DI, and glue DI.
The bands are linear fit of the data points to extrapolate to the physical pion mass marked with a dashed line.
}
  \label{fig:3pt_fit_re_A}
\end{figure}

\begin{figure}[htbp]
  \centering
  \includegraphics[page=1,width = 0.45 \textwidth]{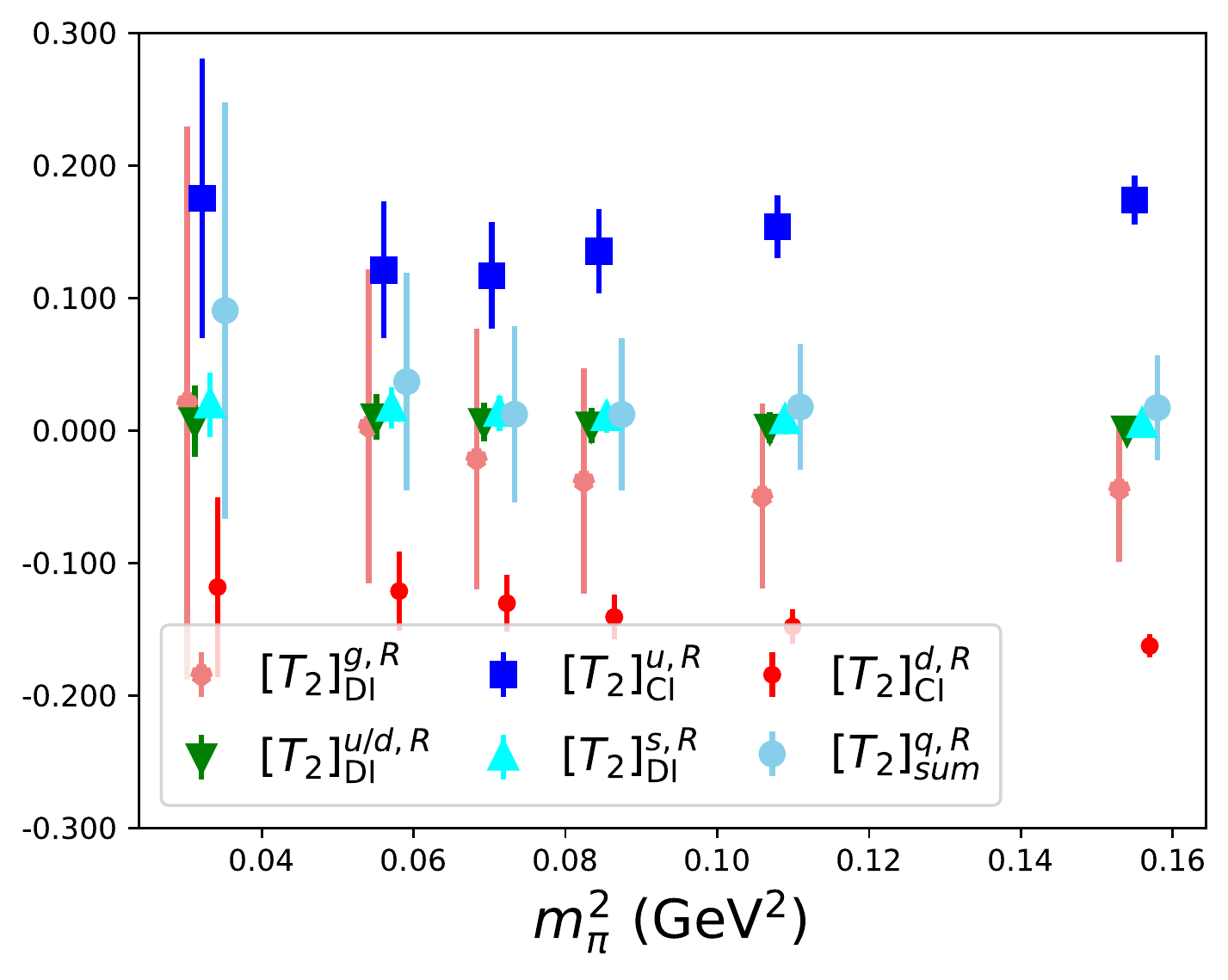} 
  \caption{Plot of the $T_2^{R}(0)$ at different valence pion masses after renormalization but without normalization. Different colors correspond to the up quark CI, down quark CI, $u$/$d$ quark DI, strange quark DI, and glue components.}
  \label{fig:3pt_fit_re_T2}
\end{figure}

\begin{figure}[htbp]
  \centering
  \includegraphics[page=1,width = 0.45 \textwidth]{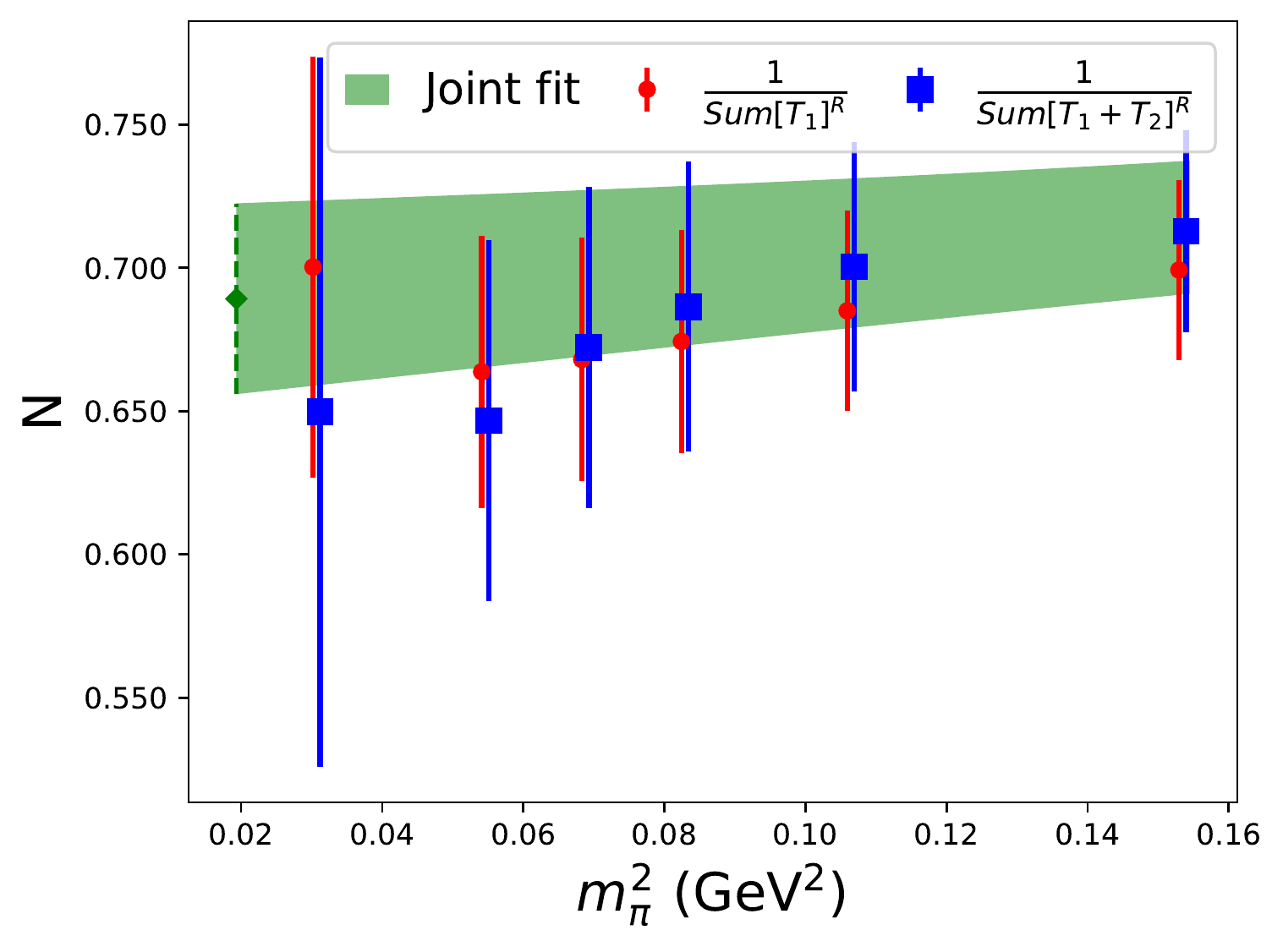} 
  \caption{Plot of the inverse of the sum of $T_1^{R}(0)$ and $[T_1+T_2]^{R}(0)$ which is the normalization factor that we apply to the $T_1^{R}(0)$ and $[T_1+T_2]^{R}(0)$. The bands are linear fit of the data points to extrapolate to the physical pion mass marked with a dashed line.}
  \label{fig:3pt_fit_re_B}
\end{figure}

\begin{table}[!htb]
  \footnotesize
  \centering{
  \begin{tabular}{| c | c | c | c | c | c | c| c| }
    \hline
                   & $u$(CI) & $d$(CI) & $u/d$(DI) & $s$(DI) & ${\rm{Sum}}^q$& glue  & ${\rm{Sum}}$\\
    \hline
    $\braket{x}$        &
  0.233(12)(26) &      0.085(5)(3) &    0.065(6)(2) &    0.043(6)(4) &  0.491(20)(23) &  0.509(20)(23) &     1.0 \\   
    \hline
    $2J$   &
  0.319(22)(63) &      0.017(9)(23) &      0.075(7)(16) &      0.052(6)(10) &     0.539(22)(44)  &  0.461(22)(44) &   1.0  \\
    \hline
    $T_2$   &
    0.086(22)(37) &     -0.067(9)(26) &      0.010(7)(14) &      0.010(6)(14) &   0.048(22)(21) &  -0.048(22)(21) &    0.0   \\  
    \hline
    $g_A $~\cite{Liang:2018pis}   &
    0.917(13)(28) & -0.337(10)(10) & -0.070(12)(15) & -0.035(6)(7) & 0.405(25)(37) &  $\cdots$ & $\cdots$ \\
    \hline
    $2L$   &
  -0.598(22)(63) &     0.354(9)(23) &      0.145(7)(16) &      0.087(6)(10) & 0.134(22)(44) &  $\cdots$  &   $\cdots$  \\          
    \hline
  \end{tabular}
  \caption{Renormalized and normalized values of momentum fractions $\braket{x}$ and angular momentum fractions $2J$ at $\overline{{\rm{MS}}}(\mu = 2 \ {\rm{GeV}})$ on a $32^3 \times 64$ domain wall lattice with lattice spacing $a=0.143$ fm and $m_{\pi} = 171$ MeV. ``${\rm{Sum}}^q$" in the table is the sum of all the quark CI and DI contributions. ``${\rm{Sum}}$" in the table is the sum of all the quark and glue contributions. The quark spin $g_A$ is from Ref.~\cite{Liang:2018pis} at $\overline{{\rm{MS}}}(\mu = 2 \ {\rm{GeV}})$. The orbital angular momentum fractions $2L$ are calculated with $2L = 2 J - g_A$.}
  \label{tab:final_renor_0}
  }
\end{table}

\begin{figure}[htbp]
  \centering
  \includegraphics[page=1,width = 0.50 \textwidth]{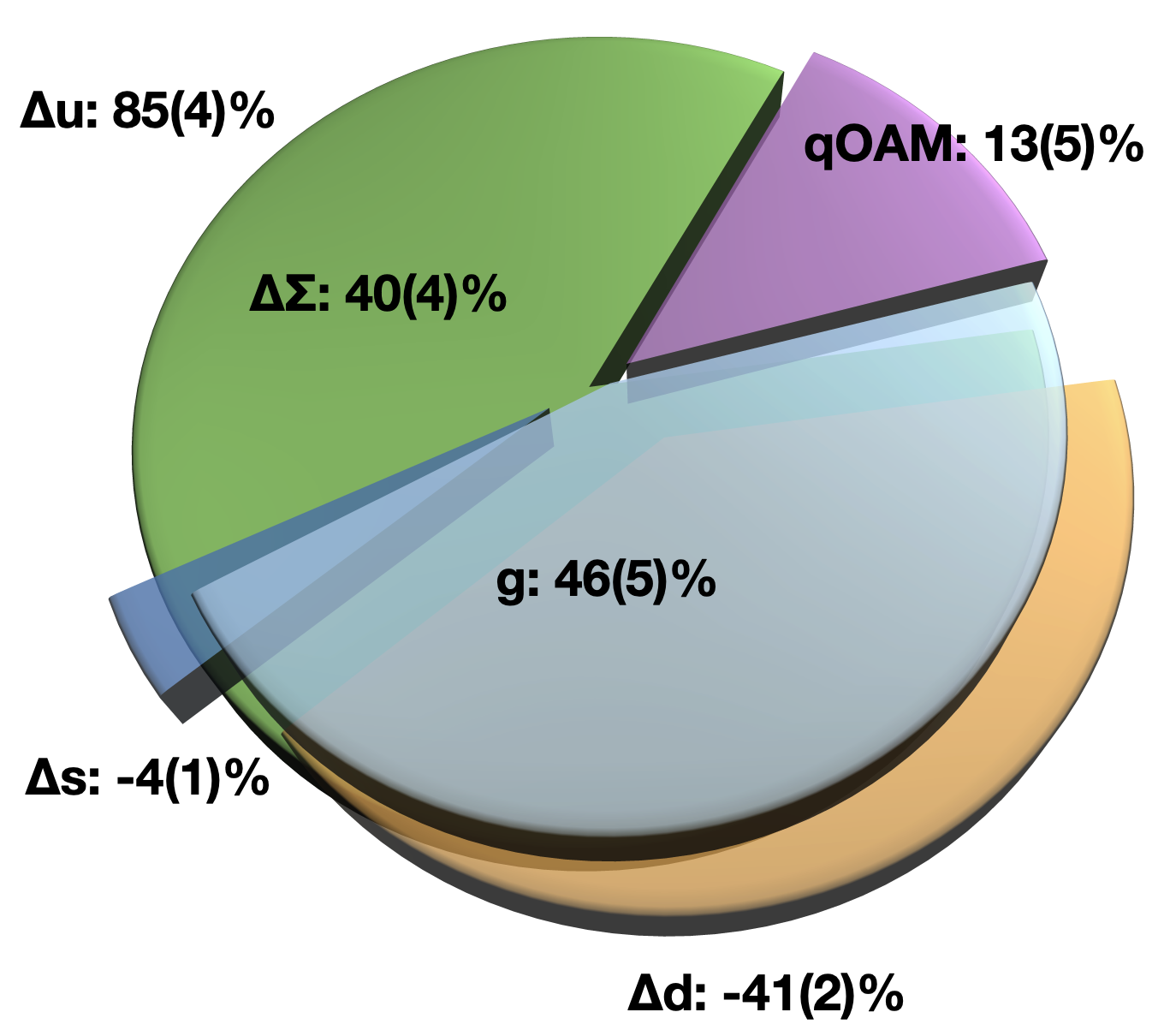} 
\caption{Summary plot of the quark spin $g_A$ from Ref.~\cite{Liang:2018pis}, the quark orbital angular momentum fraction and glue angular momentum fraction.}
  \label{fig:final_spin_decomp}
\end{figure}

The final renormalized and normalized momentum fractions $\braket{x}$ and angular momentum fractions $2J$ are listed in Table~\ref{tab:final_renor_0}.
We have performed fits using the two-state fits in Eq.~(\ref{eq:secV_ratio_fit}) and the differential summed-ratio fits in Eq.~(\ref{eq:secV_summed_ratio}).
They give similar statistical errors and agree with each other within uncertainty.
Thus, we choose the central value and statistical error to be given by the results from two-state fits in Eq.~(\ref{eq:secV_ratio_fit}) which we believe to have better control of excited-state contamination under current statistics.
The systematic uncertainties only include the contributions from the excited-state contamination estimated by taking the central value difference of the results from the two different fits.
Our predictions of the momentum fractions $\langle x \rangle^R_{u,d,s,g}$ and iso-vector momentum fraction $\langle x \rangle^R_{u-d}$ are 0.298(12)(24), 0.150(7)(5), 0.043(6)(4), 0.509(20)(23), and 0.148(10)(29), respectively, which are consistent with the preliminary results from Ref.~\cite{Yang:2018bft} on the same ensemble, but with much smaller errors due to the application of CDER for the DI.
Extrapolations of all the predictions of the momentum fractions to the continuum and infinite-volume limits are needed in order to be compared to 
phenomenological global fits at $\overline{{\rm{MS}}}(\mu = 2 \ {\rm{GeV}})$ such as the CT14~\cite{Dulat:2015mca} values
$\langle x \rangle^R_{u} = 0.348(5)$, $\langle x \rangle^R_{d} = 0.190(5)$, $\langle x \rangle^R_{s} = 0.035(9)$, $\langle x \rangle^R_{g} = 0.416(9)$, and $\langle x \rangle^R_{u-d} = 0.158(6)$.
Our predictions of the angular momentum percentage fractions $\langle 2J \rangle^R_{u,d,s,g}$ are
0.394(20)(47), 0.092(10)(7), 0.052(6)(10), and 0.461(22)(44), respectively.
We have also listed the quark spin $g_A$ from Ref.~\cite{Liang:2018pis} at $\overline{{\rm{MS}}}(\mu = 2 \ {\rm{GeV}})$
along with the orbital angular momentum percentage fractions $2L$ calculated with $2L = 2 J - g_A$ which are summarized in Fig.~\ref{fig:final_spin_decomp}.
We see that the quark orbital angular momentum fraction at $0.134(22)(44)$ has a relatively small error and is not negligible.

}

}

{
\section{Summary}\label{sec:summary}
In summary, we have carried out a complete calculation of proton momentum and angular momentum fractions at several overlap valence pion masses on a $32^3 \times 64$ domain-wall lattice with lattice spacing $a=0.143$ fm and $m_{\pi} = 171$ MeV.
We report the renormalized, mixed, and normalized momentum fractions for the quarks and glue to be $0.491(20)(23)$ and $0.509(20)(23)$, respectively,
and the renormalized and normalized total angular momentum percentage fractions for quarks and glue to be $0.539(22)(44)$ and $0.461(22)(44)$, respectively.
The energy-momentum tensor three-point function (3pt) calculations include both the connected insertions for up and down quarks and the disconnected insertions for up/down quark, strange quark and glue.
We have used complex $Z_3$ grid sources to increase signals of the nucleon correlation functions and $Z_4$ noise to estimate the quark loops.
We have also used FFT on CI 3pts along with low-mode substitution on both the source and sink nucleon.
The new sandwich method of constructing the 2pts and 3pts with LMS has direct projection of nucleon momentum for the source, and FFT helps the statistics by averaging different kinematic configurations having the same $Q^2$.
The errors of DI 3pts for up/down quark, strange quark, and glue are greatly reduced through the use of the cluster-decomposition error reduction technique ~\cite{Liu:2017man,Yang:2018bft}, especially for the unnormalized glue improved by a factor of 3.
With the full nonperturbative renormalization, mixing, and normalization using momentum and angular momentum sum rules, we find the momentum fractions and angular momentum percentage fractions listed in Table~\ref{tab:final_renor_0} at $\overline{{\rm{MS}}}(\mu = 2 \ {\rm{GeV}})$.
Finally, we should note that this work should be extended to include other lattices with different volumes and lattice spacings to control systematic errors from finite volume and lattice spacing, and the mixed action effects in our current result can also be eliminated during the continuum extrapolation.}

\begin{acknowledgments}
We thank the RBC/UKQCD Collaborations for providing their domain-wall gauge configurations.
This work is supported in part by the U.S. DOE Grant No.\ DE-SC0013065 and DOE Grant No.\ DE-AC05-06OR23177 which is within the framework of the TMD Topical Collaboration.
Y.Y.\ is supported by the Strategic Priority Research Program of Chinese Academy of Sciences, Grants No.\ XDC01040100, No.\ XDB34030300, and No.\ XDPB15.
Y.Y.\ is also supported in part by a National Natural Science Foundation of China (NSFC) and the Deutsche Forschungsgemeinschaft (DFG, German Research Foundation) joint Grant No.\ 12061131006 and SCHA~458/22.
J.L.\ is supported by the Science and Technology Program of Guangzhou (No.\ 2019050001).
This research used resources of the Oak Ridge Leadership Computing Facility at the Oak Ridge National Laboratory, which is supported by the Office of Science of the U.S. Department of Energy under Contract No.\ DE-AC05-00OR22725. This work used Stampede time under the Extreme Science and Engineering Discovery Environment (XSEDE), which is supported by National Science Foundation Grant No.\ ACI-1053575.
We also thank the National Energy Research Scientific Computing Center (NERSC) for providing HPC resources that have contributed to the research results reported within this paper.
We acknowledge the facilities of the USQCD Collaboration used for this research in part, which are funded by the Office of Science of the U.S. Department of Energy.
\end{acknowledgments}

\bibliography{./Proton_AM.bib}

\end{document}